\documentclass[acmtog]{acmart}

\usepackage{booktabs} 
\usepackage[linesnumbered,lined,ruled,commentsnumbered]{algorithm2e}
\acmPrice{15.00}

\setcopyright{acmcopyright}
\acmJournal{TOG}
\acmYear{2020}\acmVolume{39}\acmNumber{4}\acmArticle{1}\acmMonth{7} \acmDOI{10.1145/3386569.3392473}

\setcitestyle{square}

\usepackage[normalem]{ulem}
\usepackage{stackengine}
\usepackage{subfigure}
\usepackage{graphicx}
\usepackage{float}

\renewcommand{\vec}[1]{\mathbf{#1}}

\newcommand{\todo}[1]{}

\newcommand\hide[1]{}

\newcommand\add[1]{{#1}}

\definecolor{bkcolor}{RGB}{210,10,210}
\definecolor{vcacolor}{RGB}{123,50,210}
\definecolor{barbaracolor}{RGB}{246,150,70}

\newcommand\ie{i.e.,~}
\newcommand\eg{e.g.,~}
\newcommand\Fig[1]{Figure~\ref{fig:#1}}
\newcommand\Sec[1]{Section~\ref{sec:#1}}
\newcommand\Eq[1]{Equation~(\ref{eq:#1})}

\newcommand\Algo[1]{Algorithm~(\ref{algo:#1})}
\newcommand\Tab[1]{Table~(\ref{tab:#1})}

\newcommand\twoD{2D}
\newcommand\threeD{3D}
\newcommand\particleToGrid{particle-to-grid}
\newcommand\gridToParticle{grid-to-particle}

\DeclareMathOperator*{\argmin}{arg\,min}

\newcommand{\gridSuper}{\fbox{\small{+}}}
\newcommand{\particleSuper}{\circ}
\newcommand{\grid}[1]{#1^{\setlength{\fboxsep}{0pt}\gridSuper}}
\newcommand{\particle}[1]{#1^{\particleSuper}}

\begin{document}

\title{Lagrangian Neural Style Transfer for Fluids}

\author{Byungsoo Kim}
\affiliation{%
   \institution{ETH Zurich}}
\email{kimby@inf.ethz.ch}

\author{Vinicius C. Azevedo}
\affiliation{%
  \institution{ETH Zurich}}
\email{vinicius.azevedo@inf.ethz.ch}

\author{Markus Gross}
\affiliation{%
   \institution{ETH Zurich}}
\email{grossm@inf.ethz.ch}

\author{Barbara Solenthaler}
\affiliation{%
   \institution{ETH Zurich}}
\email{solenthaler@inf.ethz.ch}

\acmSubmissionID{517}
\renewcommand{\shortauthors}{B. Kim, V. C. Azevedo, M. Gross, B. Solenthaler}

\begin{teaserfigure}
\centering
\includegraphics[trim={310 0 250 0},clip,height=180px]{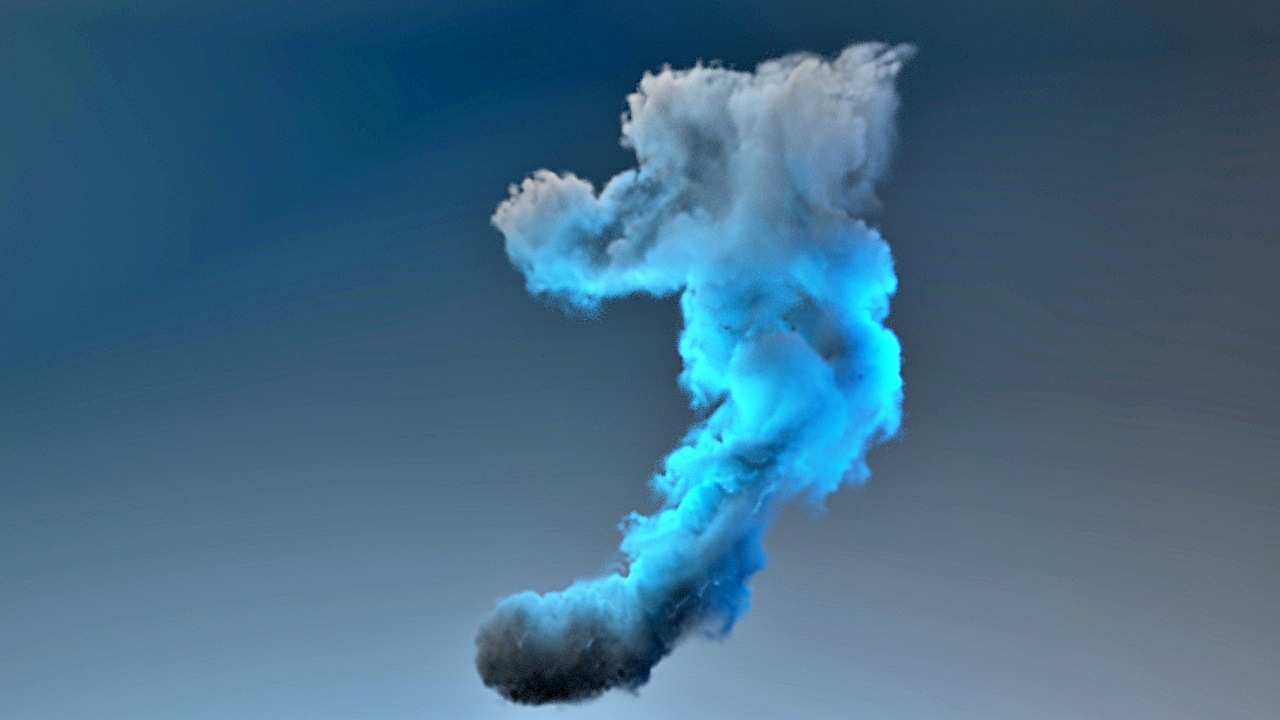}
\includegraphics[trim={190 0 200 0},clip,height=180px]{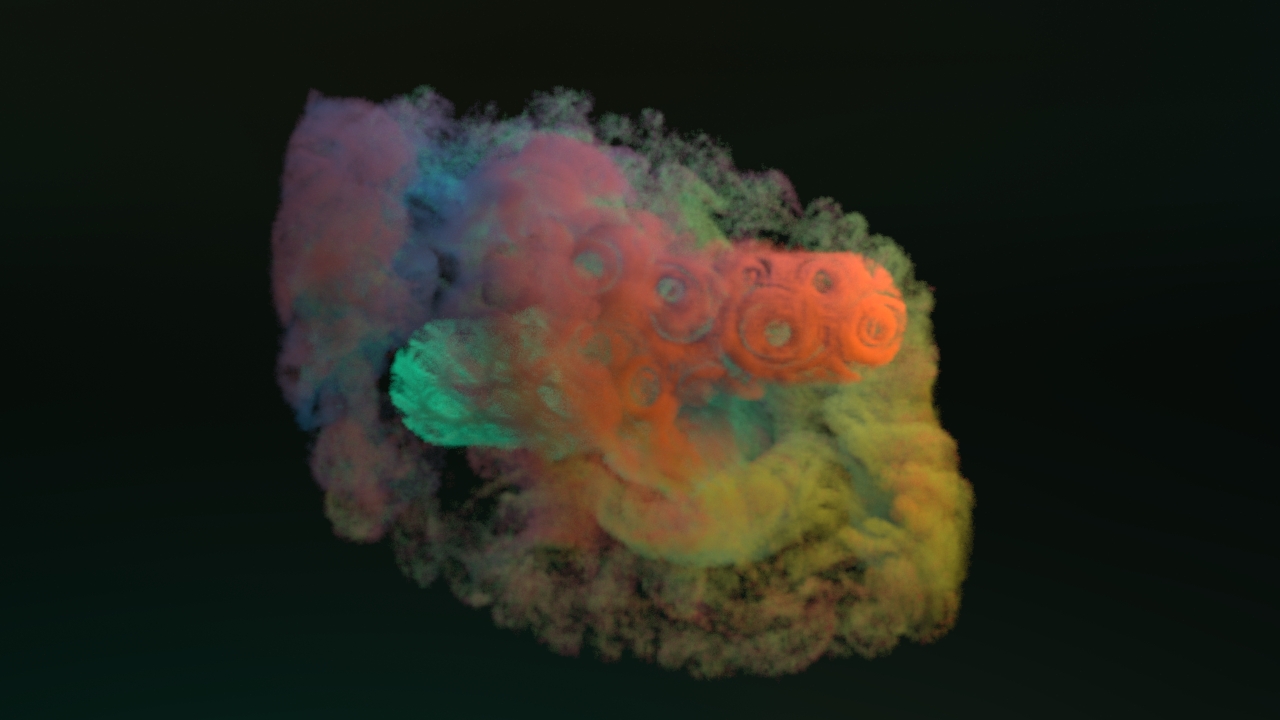} 
\includegraphics[trim={290 0 280 0}, clip,height=180px]{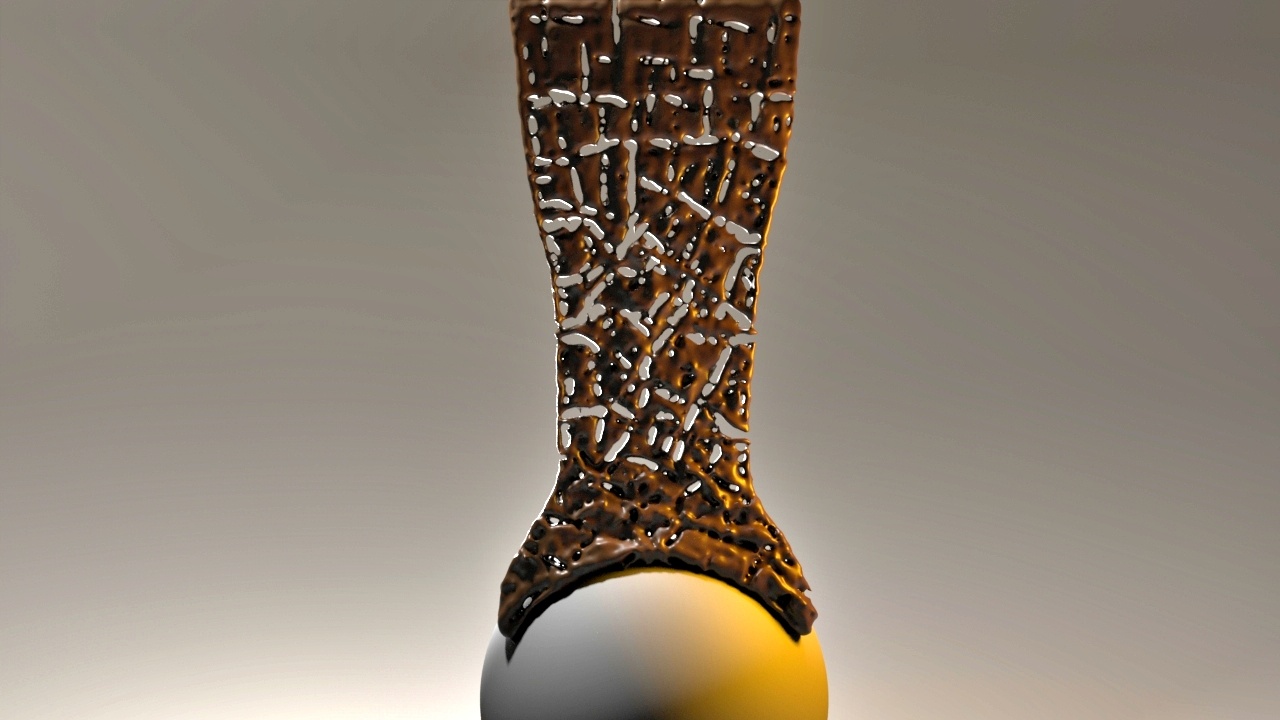} 
\caption{Our Lagrangian neural style transfer enables novel artistic manipulations, such as time-coherent stylization of smoke, multiple fluids and liquids.}
\label{fig:teaser}
\end{teaserfigure}

\begin{abstract}
Artistically controlling the shape, motion and appearance of fluid simulations pose major challenges in visual effects production. In this paper, we present a neural style transfer approach from images to 3D fluids formulated in a Lagrangian viewpoint. Using particles for style transfer has unique benefits compared to grid-based techniques. Attributes are stored on the particles and hence are trivially transported by the particle motion. This intrinsically ensures temporal consistency of the optimized stylized structure and notably improves the resulting quality.  Simultaneously, the expensive, recursive alignment of stylization velocity fields of grid approaches is unnecessary, reducing the computation time to less than an hour and rendering neural flow stylization practical in production settings. Moreover, the Lagrangian representation improves artistic control as it allows for multi-fluid stylization and consistent color transfer from images, and the generality of the method enables stylization of smoke and liquids likewise.  

\end{abstract}



\maketitle
\section{Introduction}

\begin{figure*}[ht!]
    \centering
    \includegraphics[width=0.75\textwidth]{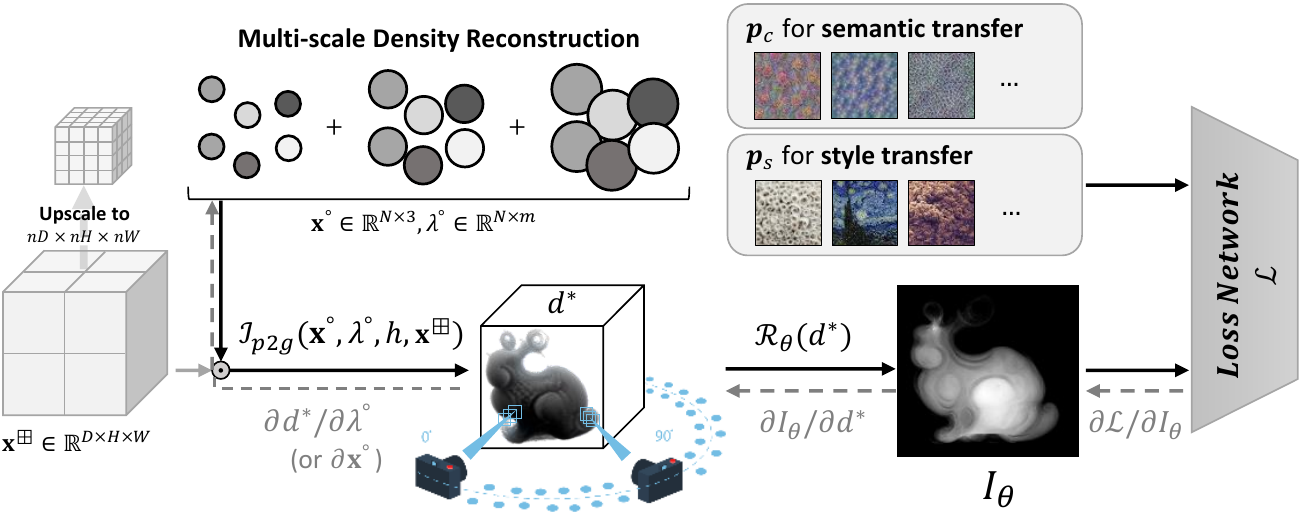}
    \caption{
    \add{Overview of our LNST method. We optimize particle positions $\particle{\vec{x}}$ and attributes $\particle{\lambda}$ to stylize a given density field $d^*$. We transfer information from particles to the grid with the splatting operation $\mathcal{I}_{p2g}$, and jointly update loss functions and attributes. 
    The black arrows show the direction of the feed-forward pass to the loss network $L$, and the gray arrows indicate backpropagation for computing gradients.
    For grid-based simulation inputs, we sample and re-simulate particles in a multi-scale manner (\Algo{densitySample}).}
    }
    \label{fig:overview}
    \vspace{-5px}
\end{figure*}

In visual effects production, physics-based simulations are not only used to realistically re-create natural phenomena, but also as a tool to convey stories and trigger emotions. Hence, artistically controlling the shape, motion and the appearance of simulations is essential for providing directability for physics. Specifically to fluids, the major challenge is the non-linearity of the underlying fluid motion equations, which makes optimizations towards a desired target difficult. Keyframe matching either through expensive fully-optimized simulations \cite{Treuille2003, McNamara2004, Pan2016} or simpler distance-based forces \cite{Nielsen2011,Raveendran2012} provide control over the shape of fluids. The fluid motion can be enhanced with turbulence synthesis approaches \cite{kim2008wavelet,sato2018example} or guided by coarse grid simulations \cite{Nielsen2011}, while patch-based texture composition \cite{Gagnon2019, jamrivska2015lazyfluids} enables manipulation over appearance by automatic transfer of input 2D image patterns. 


The recently introduced Transport-based Neural Style Transfer (TNST) \cite{Kim2019} takes flow appearance and motion control to a new level: arbitrary styles and semantic structures given by 2D input images are automatically transferred to 3D smoke simulations. The achieved effects range from natural turbulent structures to complex artistic patterns and intricate motifs. The method extends traditional image-based Neural Style Transfer \cite{gatys2016} by reformulating it as a transport-based optimization. Thus, TNST is physically inspired, as it computes the density transport from a source input smoke to a desired target configuration, allowing control over the amount of dissipated smoke during the stylization process.
 %
However, TNST faces challenges when dealing with time coherency due to its grid-based discretization. The velocity field computed for the stylization is performed independently for each time step, and the individually computed velocities are recursively aligned for a given window size. Large window sizes are required, rendering the recursive computation expensive while still accumulating inaccuracies in the alignment that can manifest as discontinuities. 
Moreover, transport-based style transfer is only able to advect density values that are present in the original simulation, and therefore it does not inherently support color information or stylizations that undergo heavy structural changes.

\begin{figure}[h!]
\newcommand*{\mywidth}{0.14}
 \newcommand*{\triml}{70.0}
 \newcommand*{\trimr}{250.0}
 \newcommand*{\trimb}{20.0}
 \newcommand*{\trimt}{300}
 \centering
     \includegraphics[width=\mywidth\textwidth]{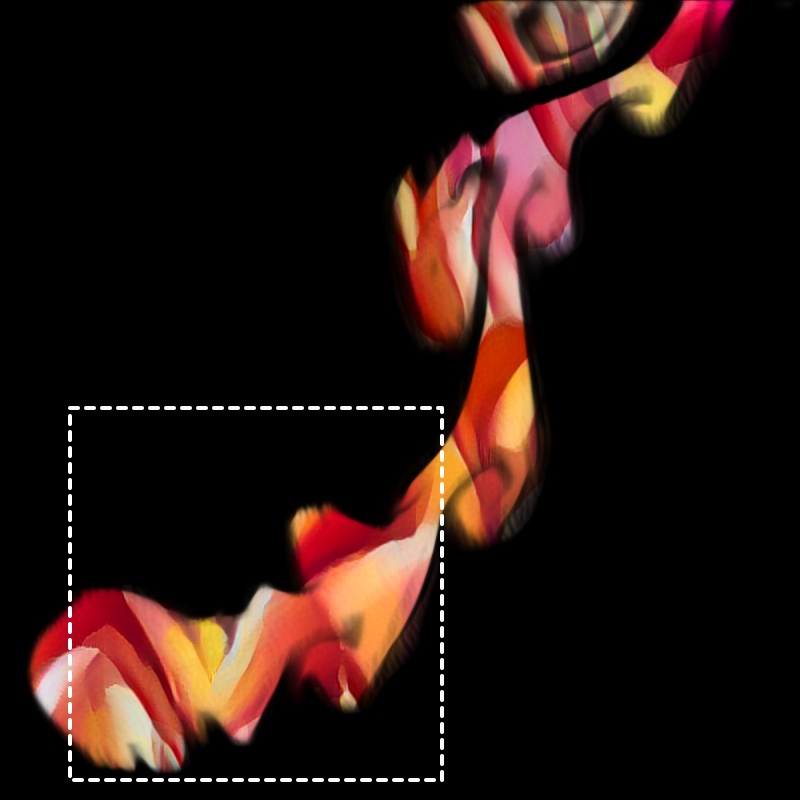} 
    \hspace{-26.5px}\includegraphics[width=0.1\linewidth]{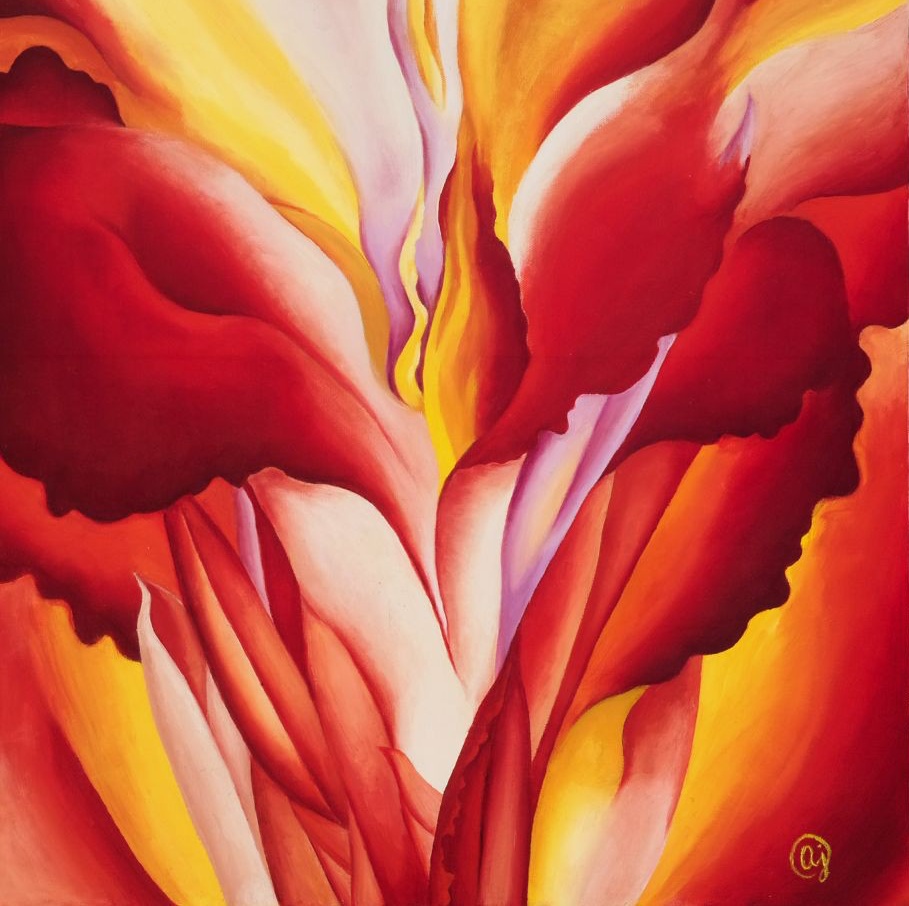}
    \includegraphics[trim=\triml px \trimb px \trimr px \trimt px, clip, width=\mywidth\textwidth]{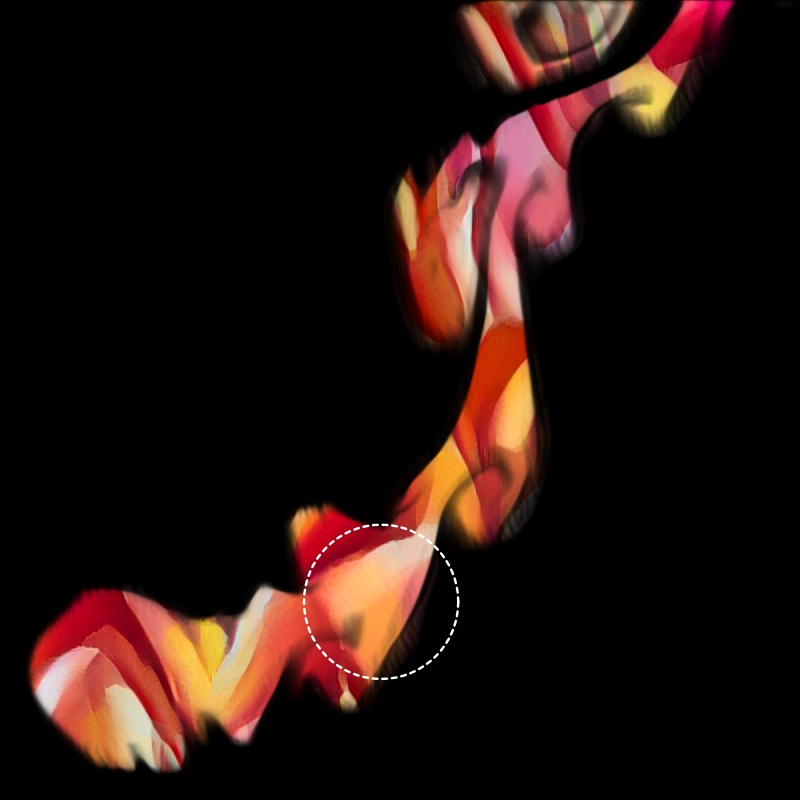} 
    \includegraphics[trim=\triml px \trimb px \trimr px \trimt px, clip, width=\mywidth\textwidth]{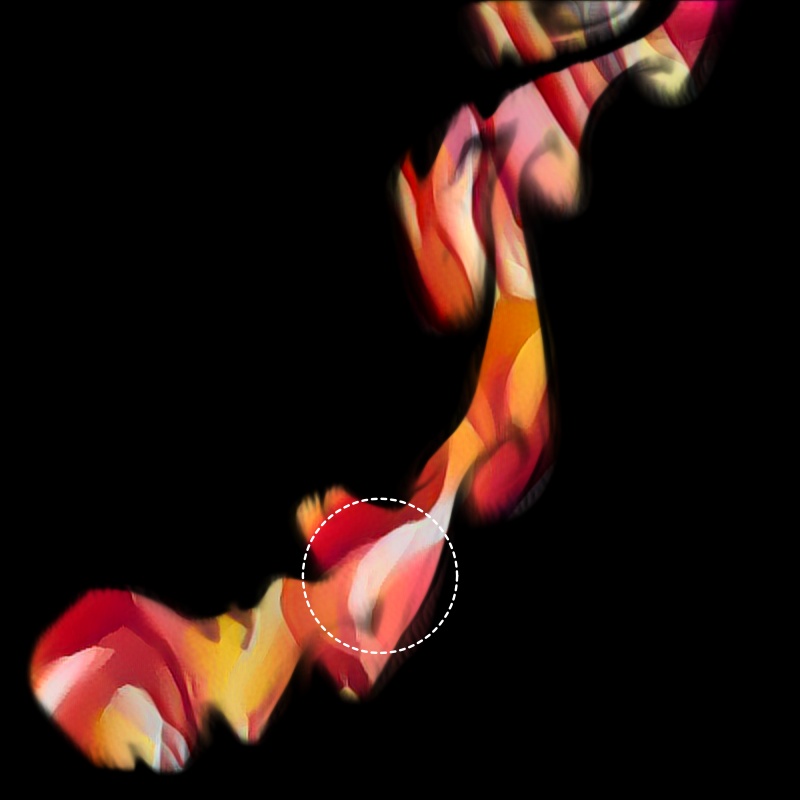} 
    \\
    \includegraphics[width=\mywidth\textwidth]{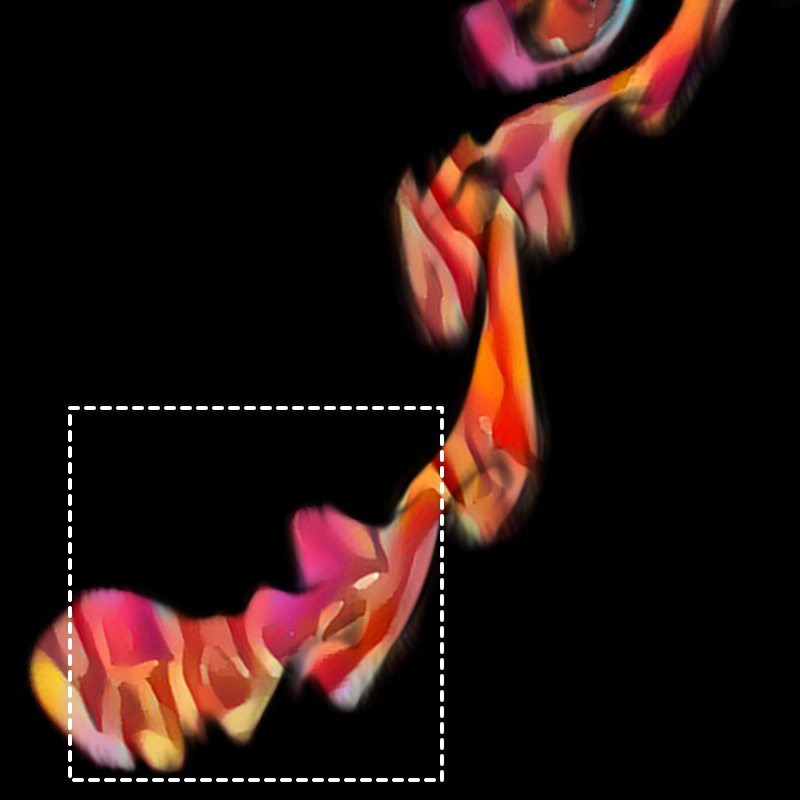}
    \hspace{-26.5px}\includegraphics[width=0.1\linewidth]{fig/results/color/cokeffe}
    \includegraphics[trim=\triml px \trimb px \trimr px \trimt px, clip, width=\mywidth\textwidth]{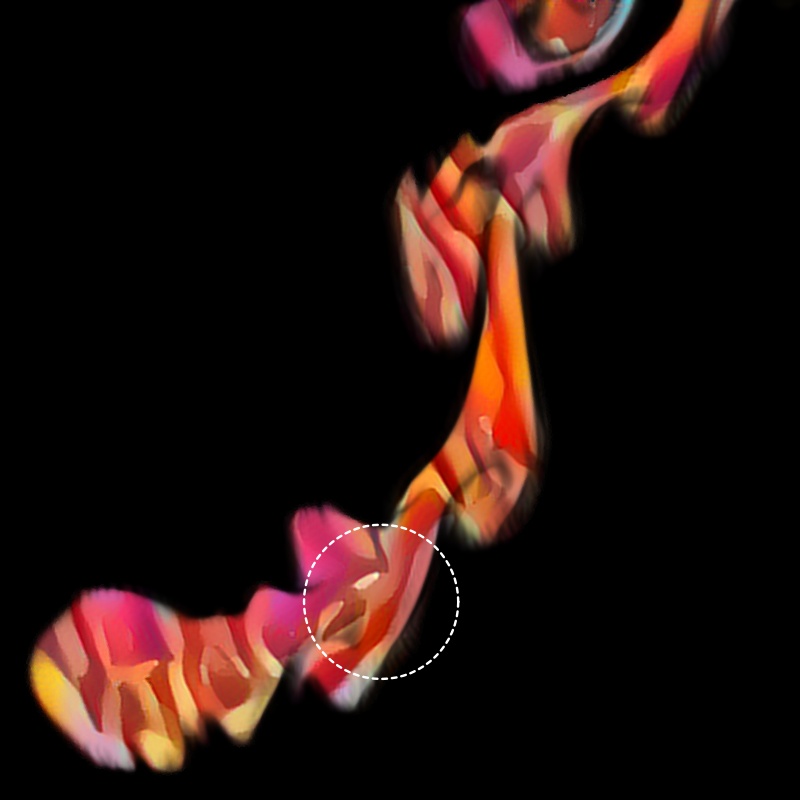} 
    \includegraphics[trim=\triml px \trimb px \trimr px \trimt px, clip, width=\mywidth\textwidth]{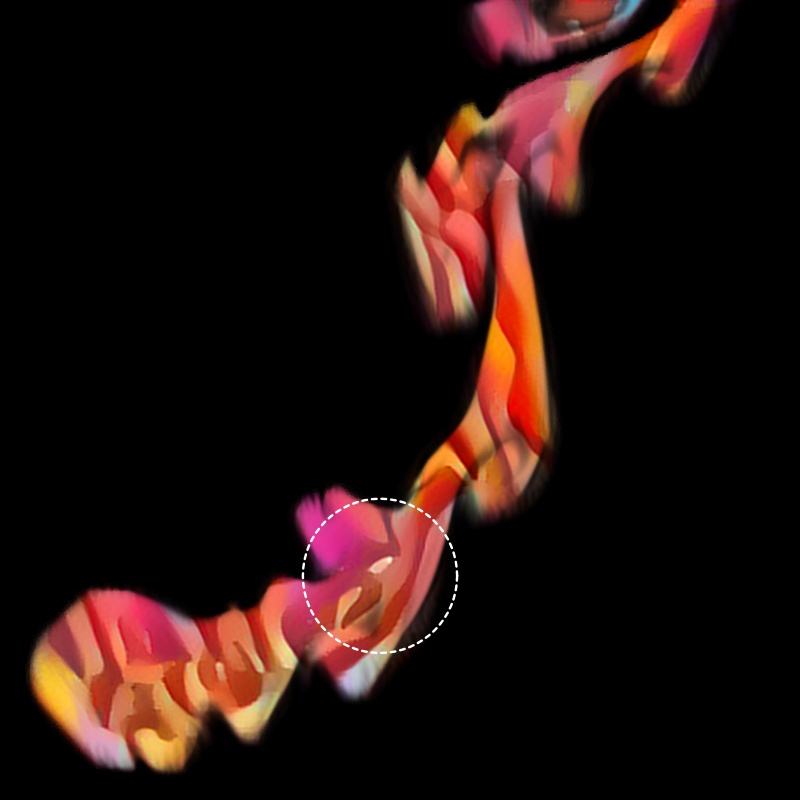} 
    %
    %
    \caption{Neural color stylization \cite{christen19color} using the input \emph{Red Canna} applied to a smoke scene with TNST (top) and LNST (bottom). The close-up views (\add{dashed box}, frames 60 and 66) reveal that LNST is more time-coherent than TNST (\add{dashed circle}).}
\label{fig:colorSmoke}
\end{figure}

Thus, in this work, we reformulate Neural Style Transfer in a Lagrangian setting \add{(see Figure~\ref{fig:overview})}, demonstrating its superior properties compared to its Eulerian counterpart. 
In our Lagrangian formulation, we optimize per-particle attributes such as positions, densities and color. This intrinsically ensures better temporal consistency as shown for example in \Fig{colorSmoke}, eliminating the need for the expensive recursive alignment of stylization velocity fields. 
The Lagrangian approach reduces the computational cost to enforce time coherency, increasing the speed of results from one day to a single hour.
The Lagrangian Style transfer framework is completely oblivious to the underlying fluid solver type. Since the loss function is based on filter activations from pre-trained classification networks, we transfer the information back and forth from particles to the grids, where loss functions and attributes can be jointly updated.
We propose regularization strategies that help to conserve the mass of the underlying simulations, avoiding oversampling of stylization particles. 
Our results demonstrate novel artistic manipulations, such as stylization of liquids, color stylization, stylization of multiple fluids, and time-varying stylization.

\section{Related Work}

\emph{Lagrangian Fluids} have become popular for simulating incompressible fluids and interactions with various materials. 
Since the introduction of SPH to computer graphics \cite{Desbrun1996,Muller2003}, various extensions have been presented that made it possible to efficiently simulate millions of particles on a single desktop computer. Accordingly, particle methods reached an unprecedented level of visual quality, where fine-scale surface effects and flow details are reliably captured. To enforce incompressibility, the original state equation based method \cite{Monaghan2005,Becker2007} has been replaced by pressure Poisson equation (PPE) solvers using either a single source term for density invariance \cite{Solenthaler2009,Ihmsen2014} or two PPEs to additionally account for divergence-free velocities \cite{Bender2015}. Solvers closely related to PPE have been presented, such as Local Poisson SPH \cite{He2012}, Constraint Fluids \cite{Bodin2012} and Position-based Fluids \cite{Macklin2013}. Boundary handling is computed with particle-based approaches that sample boundary geometry (e.g. \cite{Gissler2019}) or implicit methods that typically use a signed distance field (e.g. \cite{Koschier2017}). Extensions include highly viscous fluids (e.g. \cite{Peer2015}), and multiple phases and fluid mixing (e.g. \cite{Ren2014}).
An overview of recent developments in SPH can be found in the course notes of Koschier et al. \shortcite{Koschier2019}.

\vspace{0.1cm}

\emph{Hybrid Lagrangian-Eulerian Fluids} combine the versatility of the particles representation to track transported quantities with the capacity of grids to enforce incompressibility. Among popular approaches, the Fluid Implicit Particle Method (FLIP) \cite{Brackbill1988} was first employed in graphics to animate sand and water \cite{Zhu2005}. Due to its ability to accurately capture sub-grid details it has been widely adopted for liquid simulations, being extended to animation of turbulent water \cite{Kim2007}, coupled with SPH for modelling small scale splashes \cite{Losasso2008}, improved for efficiency \cite{Ando2013, Ferstl2016}, used in fluid control \cite{Pan2013}, and enhanced with better particle distribution \cite{Ando2011, Um2014}. 
\add{The Material Point Method (MPM) \cite{stomakhin2013} was used to simulate a wide class of solid materials \cite{jiang2016material}.} 
Recent work on hybrid approaches extended the information tracked by the particles by affine \cite{Jiang2015} and polynomial \cite{Fu2017} transformations. For a thorough discussion of hybrid continuum models, we refer to Hu et al. \shortcite{Hu2019}.


\vspace{0.1cm}

\emph{Patch-based Appearance Transfer} methods compute similarities between source and target datasets in local neighborhoods, modifying the appearance of the source by transferring best-matched features from the target dataset. Kwatra et al. \shortcite{Kwatra2005} employ local similarity measures in an energy-based optimization, enabling texture patches animated by flow fields. This approach was further extended to liquid surfaces \cite{Kwatra2006, Bargteil2006}, and improved by modifying the texture based on visually salient features of the liquid mesh \cite{Narain2007}. Jamri\v{s}ka et al. \shortcite{jamrivska2015lazyfluids} improved previous work with better temporal coherency and matching precision for obtaining high-quality 2D textured fluids. Texturing liquid simulations was also implemented in a Lagrangian framework by using individually tracked surface patches \cite{Yu2011, Gagnon2016, Gagnon2019}. Image and video-based approaches also take inspiration from fluid transport. Bousseau et al. \shortcite{Bousseau2007} proposed a bidirectional advection scheme to reduce patch distortions. Regenerative morphing and image melding techniques were combined with patch-based tracking to produce in-betweens for artist-stylized keyframes \cite{browning2014stylized}. Recent advances in patch-based appearance transfer often rely on evaluating the underlying 3D geometric information; examples include improving template matching by a novel similarity measure \cite{Talmi2016}, patch matching for illumination effects \cite{Fiser2016}, extensions to texture mapping \cite{Bi2017} and intricate texture motifs \cite{Diamanti2015}. While these approaches were successful in 2D settings and for texturing liquids, they cannot inherently support 3D volumetric data.

\vspace{0.1cm}

\emph{Velocity Synthesis} methods augment flow simulations with velocity fields, which manipulate or enhance volumetric data. Due to the inability of pressure-velocity formulations to properly conserve different energy scales of flow phenomena, sub-grid turbulence~\cite{kim2008wavelet, Schechter2008, Narain2008} was modelled for better energy conservation. These approaches were extended to model turbulence in the wake of solid boundaries \cite{Pfaff2009}, liquid surfaces~\cite{kim2013closest} and example-based turbulence synthesis \cite{sato2018example}. In order to merge fluids of different simulation instances \cite{Thuerey2016} or separated by void regions \cite{sato2018example}, velocity fields where synthesized by solving an unconstrained energy minimization problem. Lastly, the Transport-based Neural Style Transfer (TNST) \cite{Kim2019} can also be seen as a velocity synthesis method: at each time-step, the method optimizes a velocity field that transports the smoke towards a desired stylization.


\vspace{0.1cm}

\emph{Machine Learning \& Fluids} 
was first introduced to graphics by Ladick\'{y} et al. \shortcite{Ladicky2015}. They used Regression Forests to predict positions of fluid particles over time, resulting in a substantial performance gain compared to traditional Lagrangian solvers.
CNN-based architectures were employed in Eulerian-based solvers to substitute the pressure projection step \cite{Tompson2016, Yang2016} and to synthesize flow simulations from a set of reduced parameters \cite{Kim2019deep}. An LSTM architecture \cite{Wiewel2018} predicted changes on pressure fields for multiple subsequent time-steps, speeding up the pressure projection step. Differentiable fluid solvers \cite{Schenck2018, Hu2018, Hu2020DiffTaichi, Holl2020} have been introduced that can be automatically coupled with \add{deep learning} architectures and provide a natural interface for image-based applications. Patch-based \cite{Chu2017} and GAN-based \cite{xie2018tempogan} fluid super-resolution enhance coarse simulations with rich turbulence details, while also being computationally inexpensive. While these approaches produce detailed, high-quality results, they do not support transfer of arbitrary smoke styles.

\vspace{0.1cm}

\emph{Differentiable Rendering and Stylization} is used in Neural Style Transfer algorithms to transfer the style of a source image to a target image by matching features of a pre-trained classified network \cite{gatys2016}. However, stylizing 3D data requires a differentiable renderer to map the representation to image space. Loper and Black \shortcite{Loper2014} proposed the first fully differentiable renderer with automatically computed derivatives, while a novel differentiable volume sampling was implemented by Yan et al. \shortcite{Yan2016}. Raster-based differentiable rendering for meshes for stylization with approximate \cite{kato2018neural} and analytic \cite{liu2018paparazzi} derivatives was proposed to approximate visibility changes and mesh filters, respectively. A cubic stylization algorithm \cite{Liu2019} was implemented by minimizing a constrained energy formulation and employed to mesh stylization. Closer to our work, Kim et al. \shortcite{Kim2019} defines an Eulerian framework for a transport-based neural style transfer of smoke. Their approach computes individually stylized velocity fields per-frame, and temporal coherence is enforced by aligning subsequent stylization velocity fields and performing smoothing. 
We compare the Eulerian approach with our method in the subsequent sections. 
For an overview on differentiable rendering and neural style transfer we refer to Yifan et al. \shortcite{Yifan2019} and Jing et al. \shortcite{jing1705neural}, respectively.

\section{Eulerian Transport-Based NST}
\label{sec:Preliminaries}
We briefly review previous Eulerian-based TNST \cite{Kim2019} for completeness and to better compare against our novel Lagrangian approach. Transport-Based Neural Style Transfer (TNST) extends the original NST algorithm to transfer the style of a given image to a flow-based 3D smoke density. As opposed to NST where individual pixels of the target image are optimized, TNST optimizes a velocity field that modifies density values through indirect smoke transport. The velocity field $\hat{\vec{v}}$ that stylizes the input density $d$ is defined by a loss function $\mathcal{L}$ computed from a pre-trained image classification CNN by 
\begin{equation}
\hat{\vec{v}} = \argmin_{\vec{v}} \sum_{\theta \in \Theta} \mathcal{L}(\mathcal{R}_\theta(\mathcal{T}(d,\vec{v}))\,, \vec{p}),
\label{eq:tnstSingle}
\end{equation}
where $\mathcal{T}$ is a transport function that advects $d$ with $\vec{v}$, generating the stylized density $\hat{d}=\mathcal{T}(d,\hat{\vec{v}})$; $\mathcal{R}$ is a differentiable renderer converting the density field to image-space for a specific view $\theta$ by $I = \mathcal{R}_{\theta}(\hat{d})$, and $\vec{p}$ denotes the set of user-defined parameters used in the stylization process. The velocity field contributions are individually computed per view, resulting in a \threeD~volumetric smoke stylization. While the authors separate the velocity field into its irrotational and incompressible parts which can be optimized independently, we omit this here for simplicity.

The loss function is subdivided into semantic and style losses for additional control over artistic stylization given a rendered density field. Style transfer considers an input image and user-selected activation layers (levels of features), while semantic transfer selects a CNN layer with desirable attributes that will be transferred to the target stylized smoke. Since the smoke is advected towards a target objective, this guarantees that the original smoke shape and semantics is enforced without matching its original content loss, as in traditional NST algorithms \cite{gatys2015neural}. For simplicity, we restrict our discussion to the style loss, which is given by
%
%
%
%
%
%
\begin{equation}
\mathcal{L}_s(I, \vec{p}_s) = \sum_l^{L} \left [ \frac{1}{4C_l^2 (H_l\times W_l)^2} \sum_{m, n}^{C_l} \left( G^l_{mn}(I) - G^l_{mn}(I_s) \right)^2 \right ],
\label{eq:styleTransferFinal}
\end{equation}
where the Gram matrix $G$ computes correlations between different filter responses. The Gram matrix is calculated for a given layer $l$ and two channels $m$ and $n$, by iterating over all pixels of the flattened 1-D \add{feature map} $\mathcal{\hat{F}}^l(I)$ as
\begin{equation}
G^l_{mn}(I) = \sum_{i}^{H_l\times W_l} \mathcal{\hat{F}}_{mi}^l(I) \; \mathcal{\hat{F}}_{ni}^l(I).
\end{equation}
%

\begin{figure} [h!]
  \centering
  \includegraphics[width=0.435\textwidth]{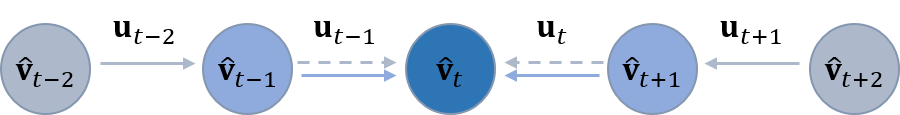}
  \caption{Recursive temporal alignment in TNST. For a window size $w$, $(w^2-1)/4$ recursive temporal alignment steps are performed for each stylization velocity $\hat{\vec{v}}$.
  \add{Colors indicate the distance to frame $t$, and arrows refer to advection steps (with recursive steps shown as dashed lines). 
  }
  }
  \label{fig:tnstalign}
\end{figure}

Extending the single frame stylization in a time-coherent fashion is expensive and inaccurate when computed in an Eulerian framework.
TNST aligns stylization velocities by recursively advecting them with the simulation velocities for a given window size as shown in \Fig{tnstalign}. The recursive nature renders this computation inefficient time- and memory-wise, especially when large window sizes are employed to enable smooth transitions between consecutive frames. Due to the large memory requirement, this operation often has to be computed on the CPU, which generates additional overhead by the use of expensive data transfer operations. 
%
%

%
%

\section{Lagrangian NST}

In contrast to its Eulerian counterpart, the Lagrangian representation uses particles that carry quantities such as the position, density and color value.
Neural style transfer methods compute loss functions based on filter activations from pre-trained classification networks, which are trained on image datasets. Thus, we have to transfer the information back and forth from particles to the grids, where loss functions and attributes can be jointly updated. We take inspiration from hybrid Lagrangian-Eulerian fluid simulation pipelines that use grid-to-particle $\mathcal{I}_{g2p}$ and particle-to-grid $\mathcal{I}_{p2g}$ transfers as
\begin{equation}
\particle{\lambda} = \mathcal{I}_{g2p}(\particle{\vec{x}}, \grid{\lambda}) \;\;\; \text{and} \;\;\; \grid{\lambda} = \mathcal{I}_{p2g}(\particle{\vec{x}}, \particle{\lambda}, h, \grid{\vec{x}}),
\label{eq:peaceAmongWorlds}
\end{equation}
where $\particle{\lambda}$ and $\grid{\lambda}$ are attributes defined on the particle and grid, respectively, $\particle{\vec{x}}$ refers to all particle positions, $\grid{\vec{x}}$ are grid nodes to which values are transferred, and $h$ is the support size of the particle-to-grid transfer. 

Our \gridToParticle~transfer employs a regular grid cubic interpolant, while the particle-to-grid transfer uses standard radial basis functions. Regular Cartesian grids facilitate finding grid vertices around an arbitrary particle position. For this, we extended a differentiable point cloud projector~\cite{insafutdinov2018pointclouds} to arbitrary grid resolution, neighborhood size and custom kernel functions. Given all the neighboring particles \add{$j \in \partial \Omega_{\vec{x}}$} around a grid node \add{$\vec{x}$}, a grid attribute $\grid{\lambda}$ is computed by summing up weighted particle contributions as
\begin{equation}
\grid{\lambda}(\vec{x}) = \frac{\sum_{j \in \partial \Omega_{\vec{x}}} \particle{\lambda_j} \; W(||\vec{x}-\particle{\vec{x_j}}||,h)}{\sum_{j \in \partial \Omega_{\vec{x}}}W(||\vec{x}-\particle{\vec{x_j}}||,h)},
\end{equation}
where we chose $W$ to be the cubic B-spline kernel, which is also often used in SPH simulations~\cite{Monaghan2005}:

\begin{equation}
W(r,h)_{\text{cubic}} =
\begin{cases}
\frac{2}{3} - r^2 + \frac{1}{2} r^3, & 0 \leq r \leq 1,\\
\frac{1}{6} (2 - r)^3, & 1 \leq r \leq 2, \\
0, & r > 2.
\end{cases}
\end{equation}
%



We now have all the necessary elements to convert the previous Eulerian style transfer (\Eq{tnstSingle}) into a Lagrangian framework. Given a set of Lagrangian attributes $\particle{\vec{\Lambda}}$, the optimization objective for a single frame is
\begin{align}
\particle{\hat{\vec{\Lambda}}} = \argmin_{\particle{\vec{\Lambda}}} \:    \sum_{\theta \in \Theta} \sum_{\particle{\lambda} \in \particle{\Lambda}} w_{\particle{\lambda}} \, \mathcal{L}(\mathcal{R}_\theta(\mathcal{I}_{p2g}(\particle{\vec{x}}, \particle{\lambda}), \, \vec{p}),
\label{eq:LNSTloss}
\end{align}
where $w_{\particle{\lambda}}$ are weights for the losses that include Lagrangian attributes. \add{In case of particle position $\particle{\vec{x}}$ given as the target quantity $\particle{\lambda}$, we use the SPH density $\mathcal{I}_{p2g}(\particle{\vec{x}}) = \sum_{j \in \partial \Omega_{\vec{x}}}{ m_j W(||\vec{x}-\particle{\vec{x}_j}||,h)}$, where $m_j$ represents the mass of the $j$-th particle \cite{splishsplash}.}
Note that our losses are evaluated similarly as in the original Eulerian method, since the gradients computed in image-space also modify grid values ($\grid{\lambda}$). However, these gradients are automatically propagated back to the particles by auto-differentiating the \particleToGrid~ $\mathcal{I}_{p2g}$ function. Thus, our method only reformulates the domain of the optimization, sharing the same stylization possibilities (semantic and content transfers) as in the original TNST.

\begin{figure}[h!]
\newcommand*{\mywidth}{0.1}
 \newcommand*{\triml}{370.0}
 \newcommand*{\trimr}{340.0}
 \newcommand*{\trimb}{30.0}
 \newcommand*{\trimt}{80}
 \centering
    \includegraphics[trim=\triml px \trimb px \trimr px \trimt px, clip,
    width=\mywidth\textwidth]{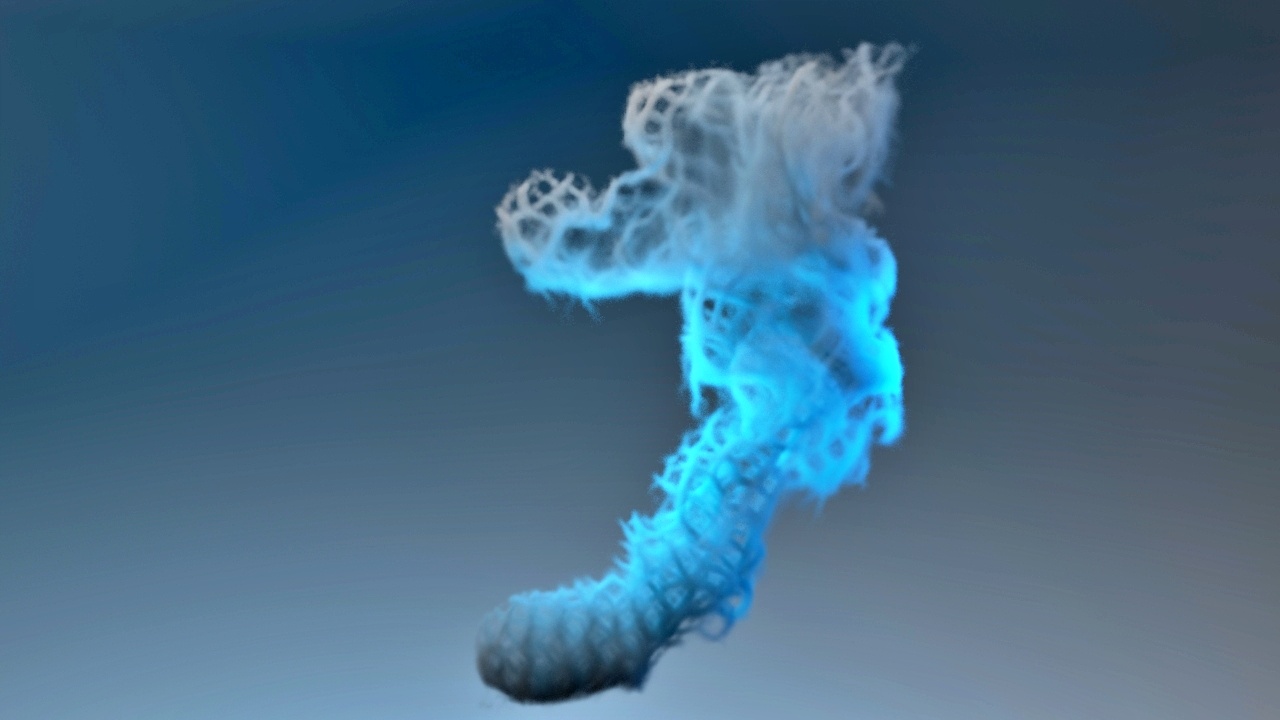}
    \includegraphics[trim=\triml px \trimb px \trimr px \trimt px, clip, 
    width=\mywidth\textwidth]{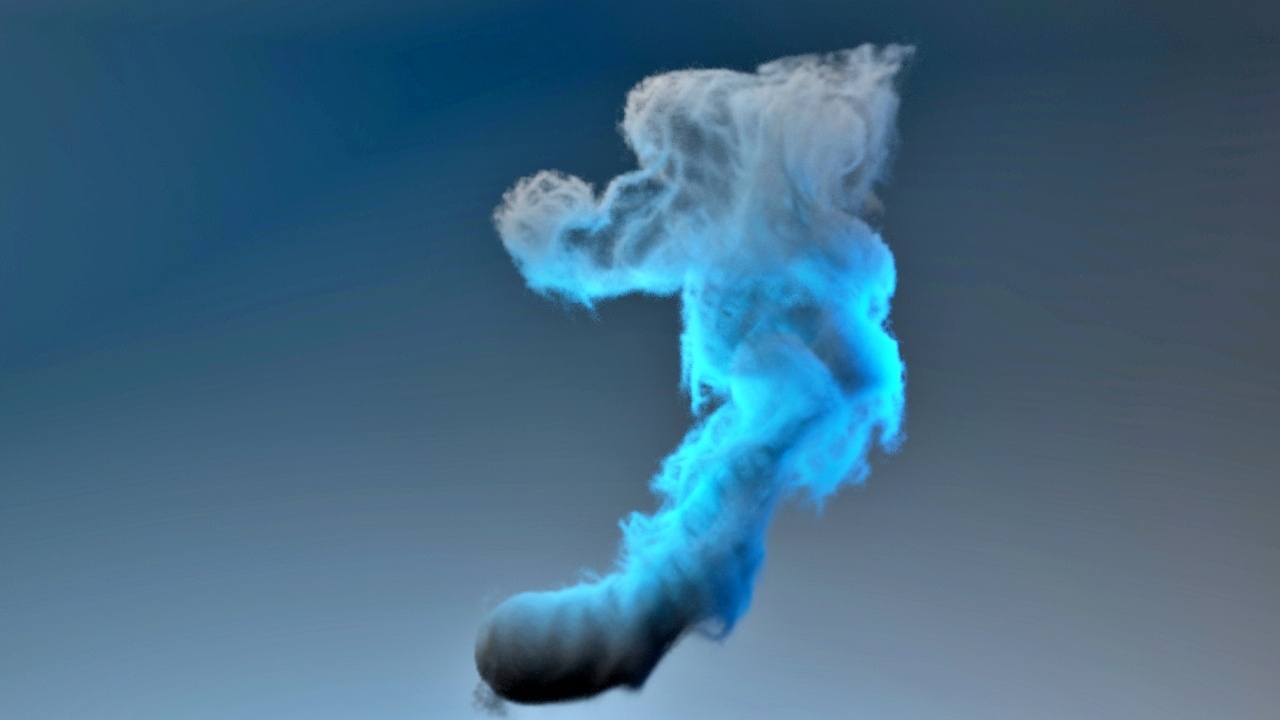} 
    \includegraphics[trim=\triml px \trimb px \trimr px \trimt px, clip, width=\mywidth\textwidth]{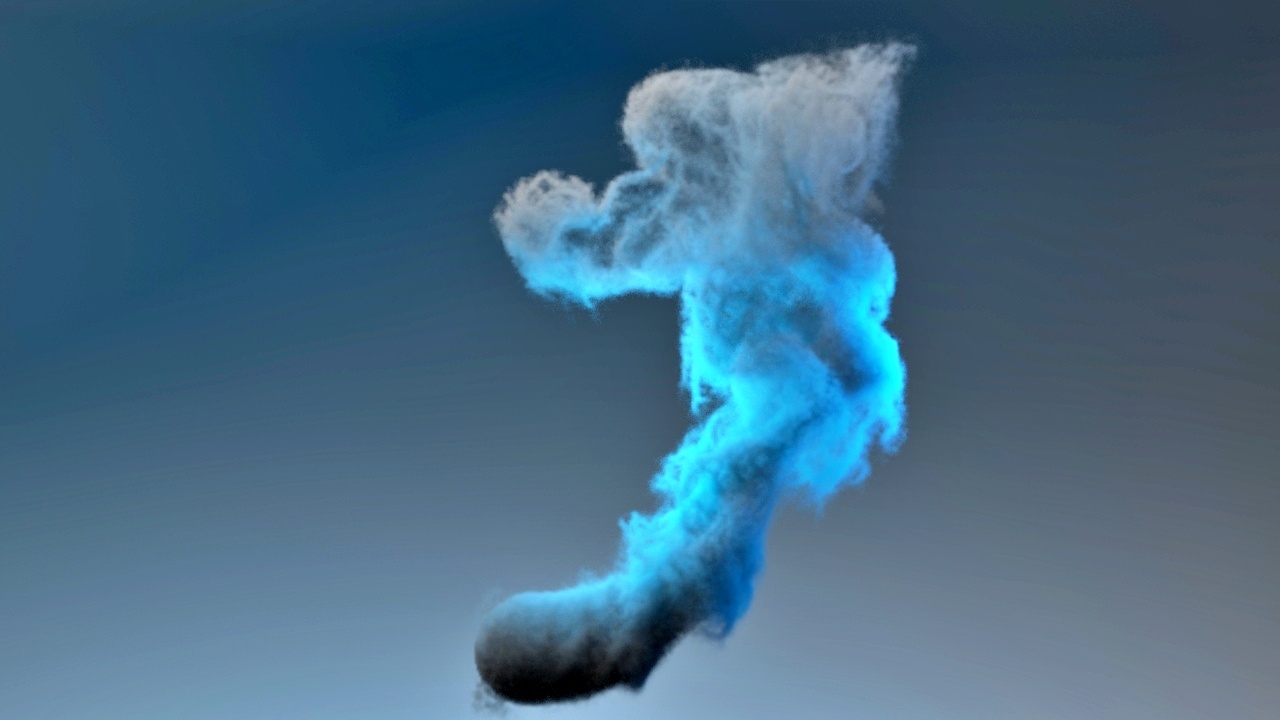} 
    \includegraphics[trim=\triml px \trimb px \trimr px \trimt px, clip, width=\mywidth\textwidth]{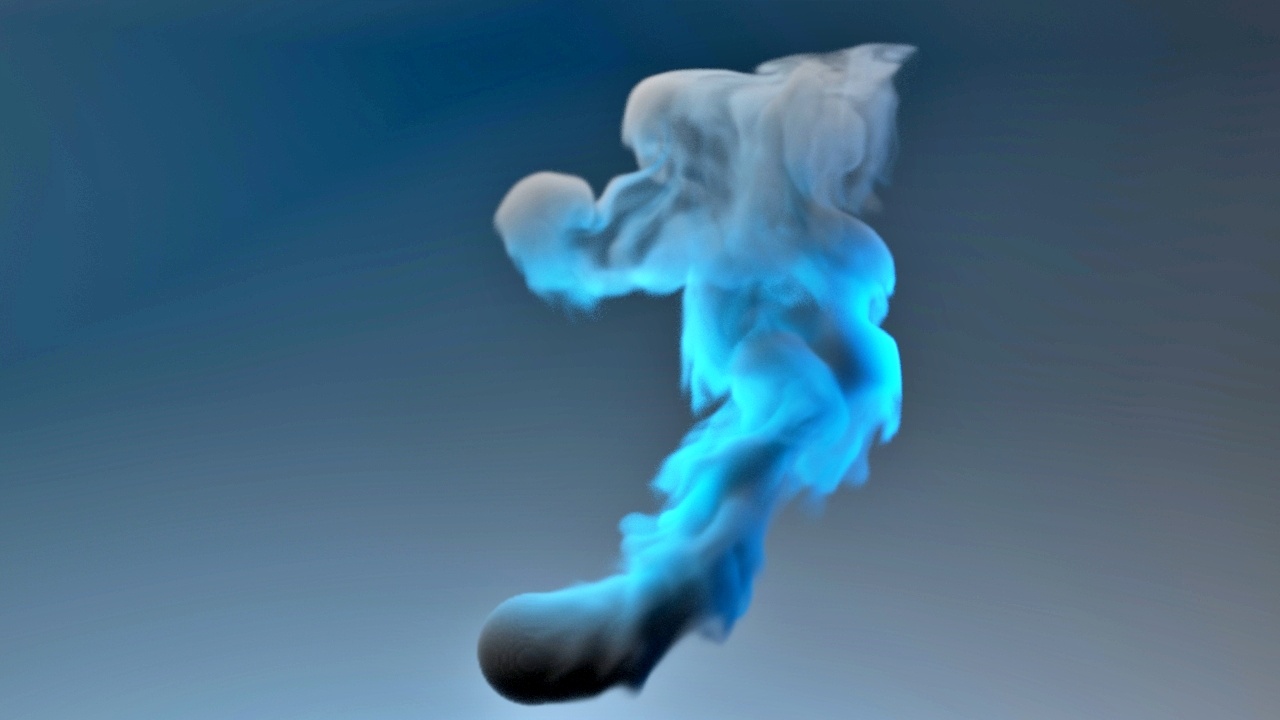}
    \caption{
    Different weights for the density regularization show the trade-off between pronounced structures and conservation of mass. The images on the left show results with zero, low, and high weights, respectively, and the right image is the ground truth.
    }
\label{fig:regularizer_density}
\end{figure}

Since the Lagrangian optimization is completely oblivious to the underlying solver type, the chosen attributes for creating stylizations can be arbitrarily combined, enabling a wide range of artistic manipulations in different scene setups. 
We outline two strategies and demonstrate their impact on the stylization. The first one is particularly suitable for participating volumetric data, which are often simulated with grid-based solvers. It involves optimizing a scalar value carried by the Lagrangian stylization particles by \Eq{LNSTloss}. For most of our smoke scenes, this scalar value is the density, though it can also be the color or emission. The regularization term
%
%
\begin{equation}
\mathcal{L}({\particle{\lambda}})_\rho =  (\sum {\Delta \particle{\lambda}})^2 - \sum \log {||\Delta \particle{\lambda}||_1}
\label{eq:densityConserv}
\end{equation}
reinforces the conservation of the original amount of smoke. It minimizes the total net smoke change, preventing the stylization to undesirably fade out particles and keeping changes non-zero by minimizing cross-entropy loss at the same time. \Fig{regularizer_density} demonstrates the impact of different regularizer weights.

\begin{figure}[h!]
\newcommand*{\mywidth}{0.095}
 \newcommand*{\triml}{370.0}
 \newcommand*{\trimr}{370.0}
 \newcommand*{\trimb}{0.0}
 \newcommand*{\trimt}{70}
 \centering
    \includegraphics[trim=\triml px \trimb px \trimr px \trimt px, clip,
    width=\mywidth\textwidth]{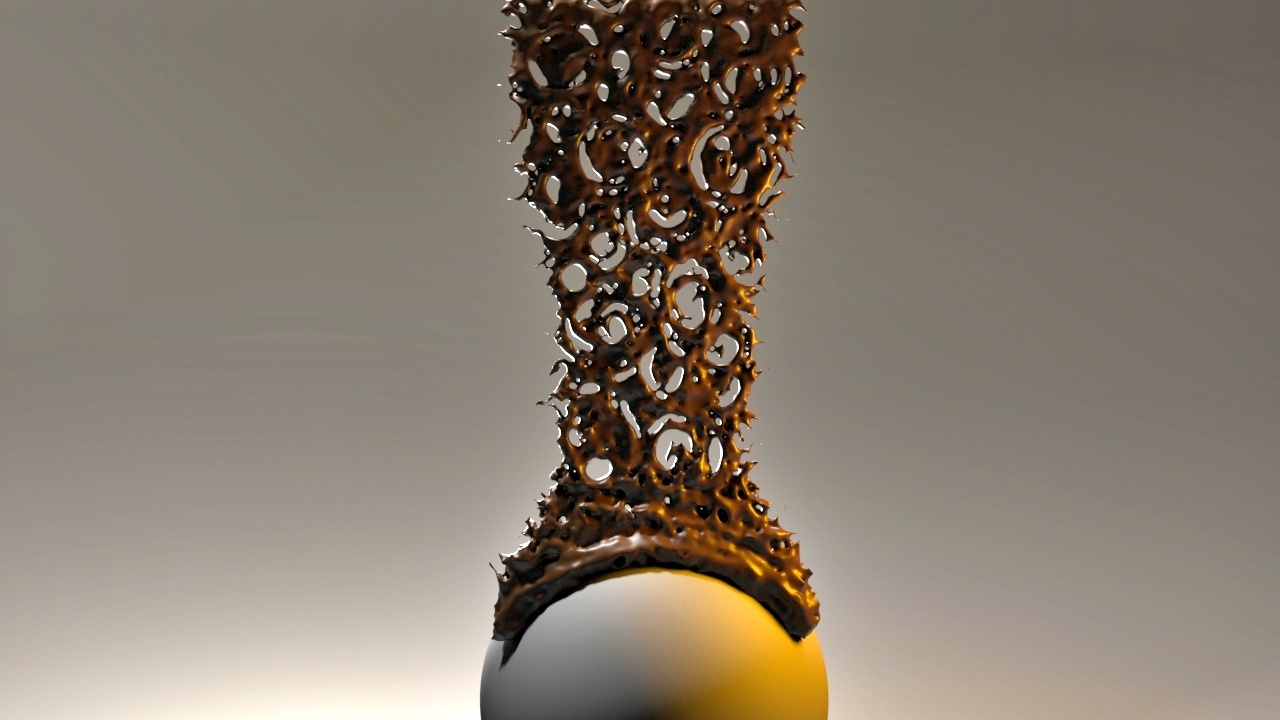}
    \includegraphics[trim=\triml px \trimb px \trimr px \trimt px, clip, 
    width=\mywidth\textwidth]{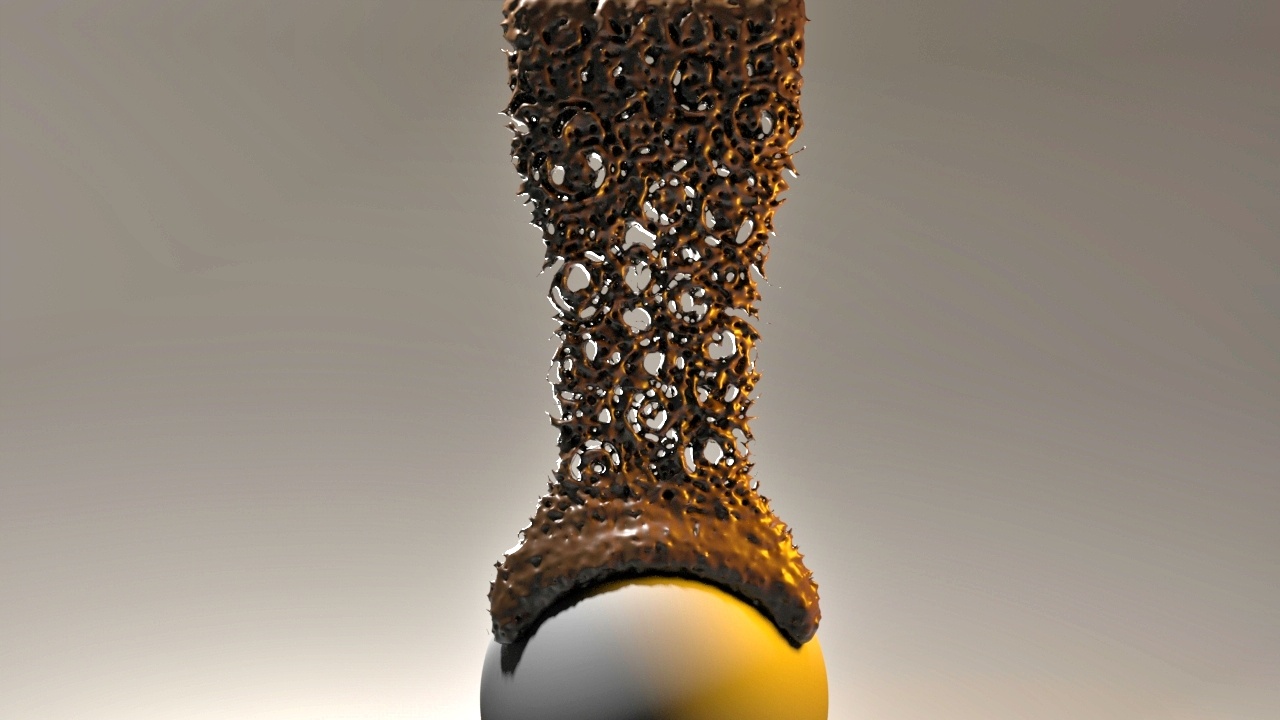} 
    \includegraphics[trim=\triml px \trimb px \trimr px \trimt px, clip, width=\mywidth\textwidth]{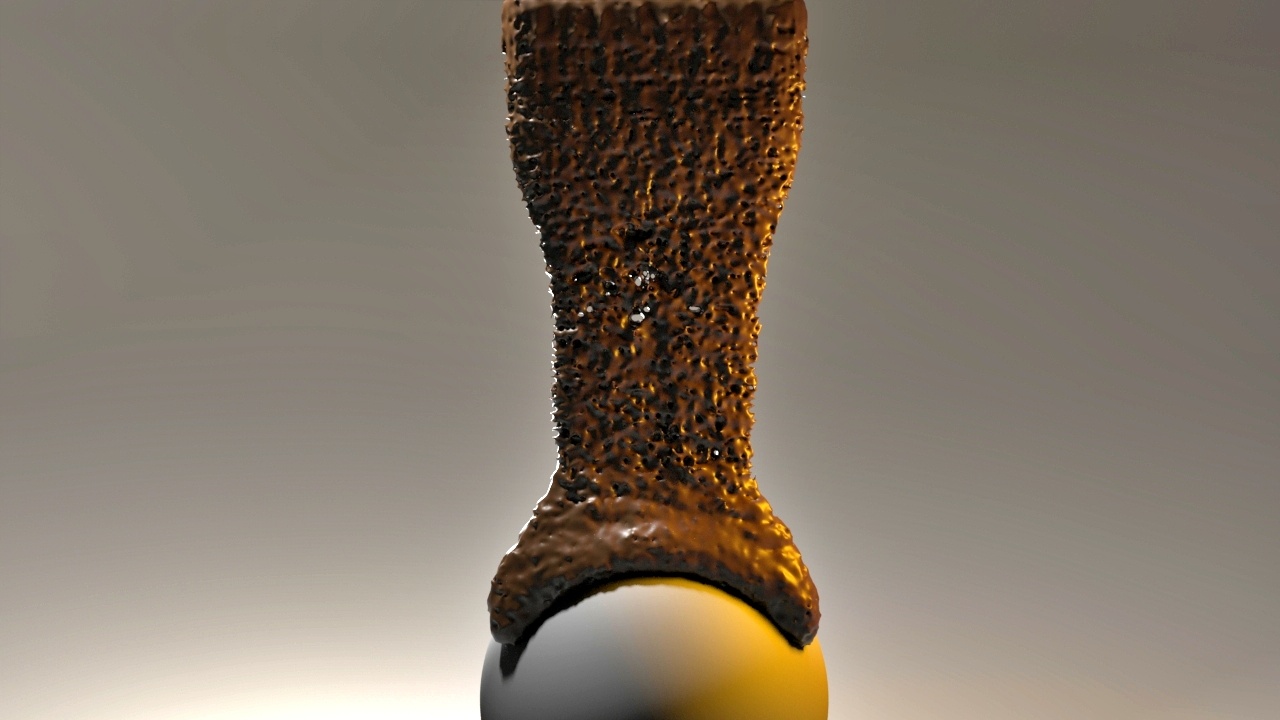} 
    \includegraphics[trim=\triml px \trimb px \trimr px \trimt px, clip, width=\mywidth\textwidth]{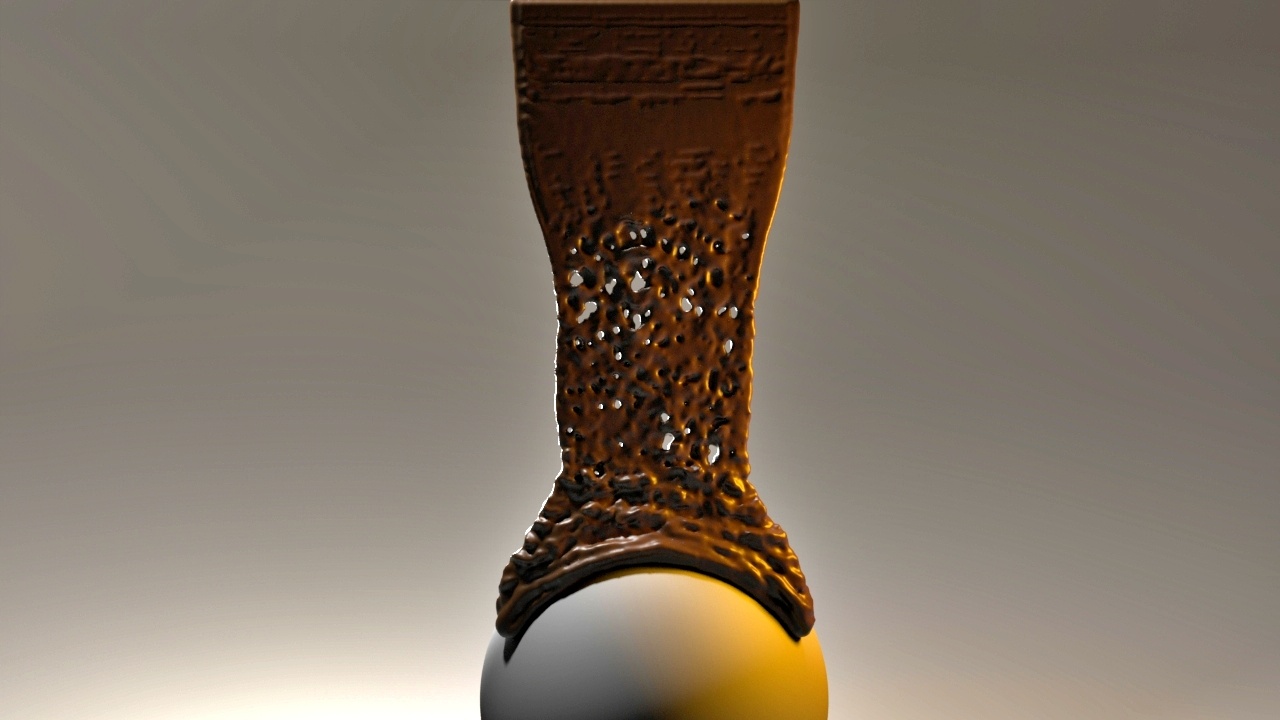}
    \caption{
    Different weights for the position regularization show the trade-off between pronounced structures and uniform sampling. The images on the left show results with zero, low, and high weights, respectively, and the right image is the ground truth.
    }
\label{fig:regularizer_position}
\end{figure}

The second strategy is suitable if the underlying fluid solver is particle-based or hybrid, which is often the case for liquids.
For these simulations, we can define particle position displacements as the optimized Lagrangian attributes. However, generating stylizations by modifying particle displacements may cause cluttering or regions with insufficient particles. The regularization penalizes irregular distribution of particle positions 
and is defined as 
\begin{align}
\mathcal{L}(\particle{\vec{x}})_{\Delta x} = ||\mathcal{I}_{p2g}(\particle{\vec{x}}) - \grid{\rho_0}||_2 ^ 2,
\label{eq:positionUpdato}
\end{align}
where $\grid{\rho_0}$ corresponds to the rest density for cells that contain particles, and is zero otherwise. 
Note that \Eq{positionUpdato} does not account for the particle deficiency near fluid surfaces. This could be addressed by adding virtual particles \cite{Schechter2008} or applying (variants of) the Shepard correction to the kernel function \cite{Reinhardt2019}.
We show the impact of this regularizer on the particle sampling in \Fig{regularizer_position}, highlighting the trade-off between uniform distribution and stylization strength.

We notice that both regularizations in \Eq{densityConserv} and \Eq{positionUpdato} are different incarnations of the mass conservation property commonly used in fluid simulations. In TNST, mass conservation is enforced by decomposing the stylization velocities into their irrotational and incompressible parts, which can be optimized independently. 
Both techniques enable a high degree of artistic control over the content manipulation.




\subsection{An Efficient Particle-Based Smoke Re-Simulation}

{
\add{If the input is a grid-based simulation,} we have to sample and re-simulate particles. We can use a sparse representation with only one particle per voxel, in constrast to hybrid liquid simulations that usually sample 8 particles per voxel to properly capture momentum conservation \cite{Zhu2005}.
}
Combining a low number of particles with a position integration algorithm that accumulates errors over time will yield irregularly distributed particles \cite{Ando2011}. This manifests in a rendered image as smoke with overly dense or void regions. 
We therfore solve the following optimization problem
\begin{align}
\particle{\hat{\vec{x}}}, \particle{\hat{\rho}} = \argmin_{{\particle{\vec{x}}, \particle{\rho}}} \sum_t ||\mathcal{I}_{p2g}(\particle{\vec{x}}_t, \particle{\rho_t}) - \grid{\rho_t}||_2 ^ 2.
\label{eq:resimulationOptimization}
\end{align}

%
The optimization problem presented above is not only severely under-constrained but also has a time-varying objective term, and optimizing for \Eq{resimulationOptimization} is challenging if tackled jointly for both particle positions $\particle{\vec{x}}$ and densities $\particle{\rho}$. Thus, we use a heuristic approach for solving this optimization, subdividing it into two steps, position optimization and multi-scale density update (\Sec{multiscale}). Firstly, we minimize the irregular distribution of particle positions by employing a position-based update, optimizing particle distributions using \Eq{positionUpdato} as objective.
%
 The distribution of the particles is optimized per frame and serves as an input for optimizing subsequent frames, enabling temporally coherent position updates. \Eq{positionUpdato} can be automatically computed by our fully differentiable pipeline.

\subsubsection{Multi-scale Density Representation}
\label{sec:multiscale}

\begin{figure}[t!]
    \centering
    \stackunder{\includegraphics[trim = 0px 0px 0px 0px, clip, width=0.128\textwidth]{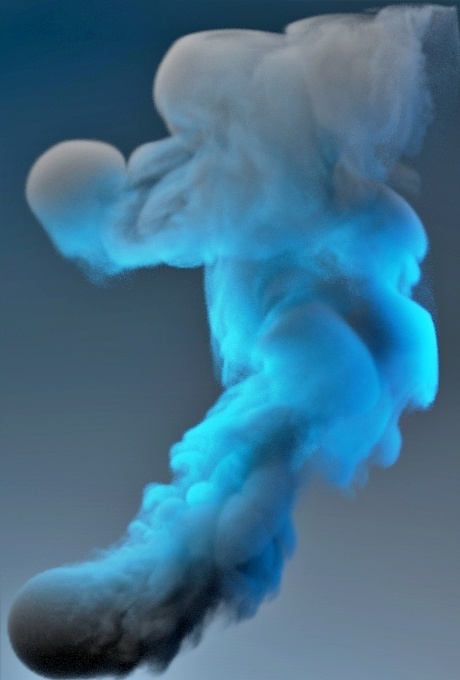}}{\scriptsize{(a)}}
    \stackunder{\includegraphics[trim = 0px 0px 0px 0px, clip, width=0.128\textwidth]{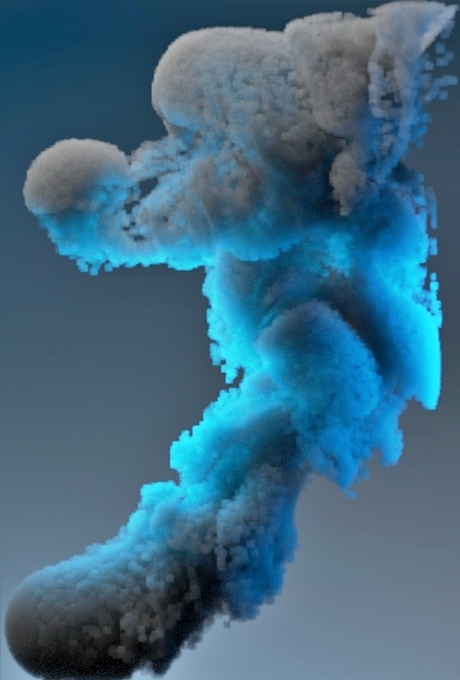}}{\scriptsize{(b)}}
    \stackunder{\includegraphics[trim = 0px 0px 0px 0px, clip, width=0.128\textwidth]{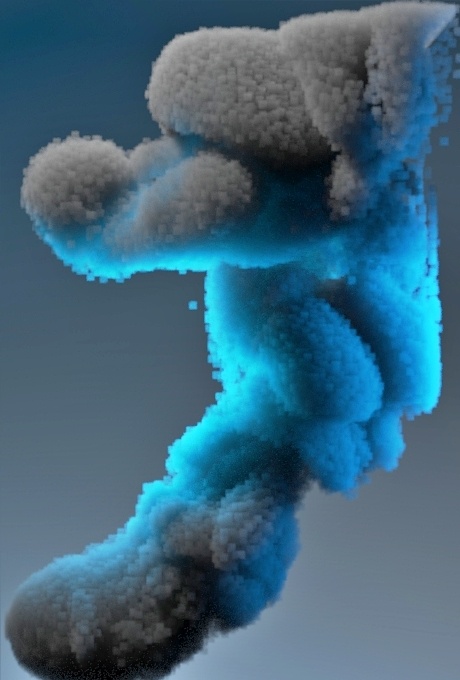}}{\scriptsize{(c)}}
    \\
    \stackunder{\includegraphics[trim = 0px 0px 0px 0px, clip, width=0.128\textwidth]{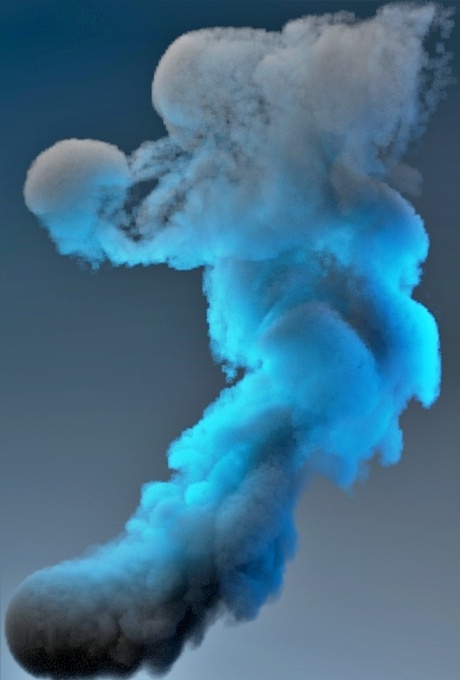}}{\scriptsize{(d)}}
    \stackunder{\includegraphics[trim = 0px 0px 0px 0px, clip, 
    width=0.128\textwidth]{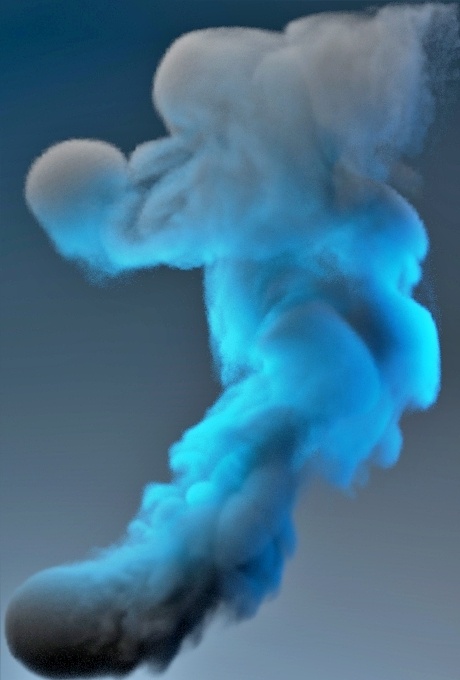}}{\scriptsize{(e)}}
    \stackunder{\includegraphics[trim = 0px 0px 0px 0px, clip, width=0.128\textwidth]{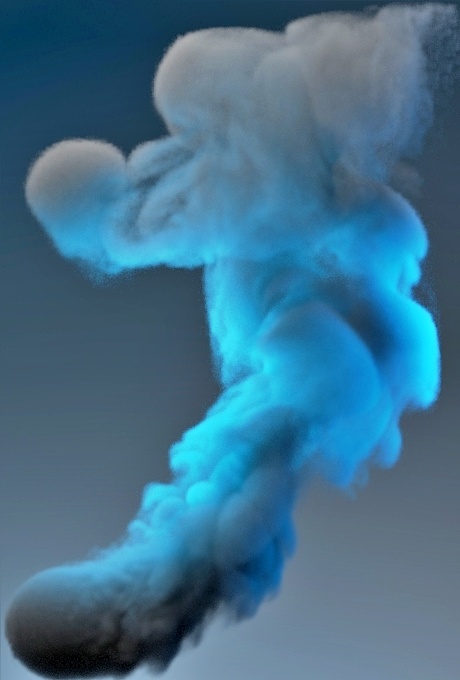}}{\scriptsize{(f)}}
    \vspace{-3mm}
    \caption{
    Comparison of different re-simulation strategies. (a): ground truth density, (b): constant density carried by particles, (c): (b) with redistribution by \Eq{positionUpdato}, (d): single-scale sampled density, (e): (d) with redistribution, (f): multi-scale \add{($n_s=3$)} sampled density with redistribution (final method).
    }
    \label{fig:resim_compare}
    \vspace{-5px}
\end{figure}

In addition to the position update, we also compute smoke densities individually carried by the particles to further eliminate small gaps that may appear due to the sparse discretization, further enhancing the solution of \Eq{resimulationOptimization}. Owing to the low number of sampled particles and the mismatches between grid and particle transfers, carrying a constant density will either produce grainy (\Fig{resim_compare}, (b)) or diffuse (\Fig{resim_compare}, (c)) volumetric representations, depending on if particle re-distribution (\Eq{positionUpdato}) is applied or not.
A simple approach is to interpolate density values directly from the grid over time. Larger kernel sizes could be used to remedy sparse sampling, but would excessively smooth structures and degrade quality.
%
%
%


We take inspiration from Laplacian pyramids, where distinct grid resolution levels are treated separately. In our case, we compute residuals of different support kernel sizes of the \particleToGrid~transfer. This efficiently captures both low- and high-frequency information, covering potentially empty smoke regions while also providing sharp reconstruction results. The residual computation of kernels of varying support sizes is synergistically coupled with matching grid resolutions, which creates an efficient multi-scale representation of the smoke.


The multi-scale reconstruction works as follows: we first sample grid densities to the particles. This represents the smoke low-frequency information, which we interpolate to the particle variables $\particle{\rho_0}$. The variables above the first level (\eg $\particle{\rho_1}, \particle{\rho_2}$) will carry residual information computed between subsequent levels. The Lagrangian representations vary between each level because they perform \gridToParticle~transfers with progressively reduced kernel support sizes. To compare residuals between Lagrangian representations, we make use of \particleToGrid~transfers, which act as a low-pass filter, similarly to blurring operations of Laplacian pyramids. This process is performed until the original grid resolution is matched. Our multi-scale density representation is summarized in \Algo{densitySample}.
\Fig{resim_compare} illustrates the impact of using a single scale without (d) and with (e) particle re-distribition (\Eq{positionUpdato}). The multi-scale result with re-distribution (f), which corresponds to our final method, has a higher PSNR (31.89) than its single-scale counterpart (31.39) and is very close to the ground truth (a).

\begin{algorithm}
\SetAlgoLined
\KwData{Particle positions \(\particle{\vec{x}} \) optimized by \Eq{positionUpdato}\\
        Original grid-based smoke simulation \(\grid{\rho}\) \\
        Grid node positions $\grid{\vec{x}}$ \\
        Coarsest support kernel radius $r$ \\
        Number of pyramid subdivisions $n_s$}
\KwResult{Multi-scale residual density \(\particle{\rho}\) stored on particles}
\BlankLine
\(\particle{\rho_0} \leftarrow \mathcal{I}_{g2p} (\particle{\vec{x}}, \grid{\rho}) \) \\
\(\grid{\rho_*} \leftarrow \mathcal{I}_{p2g} (\particle{\vec{x}}, \particle{\rho_0}, r, \grid{\vec{x}}) \)\\
\For{\(i \leftarrow 1 \) \KwTo \(n_{s}\)}{
    \(\grid{\rho_*} \leftarrow \grid{\rho}- \grid{\rho_*}\)\\
    \(\particle{\rho_i} \leftarrow \mathcal{I}_{g2p} (\particle{\vec{x}}, \grid{\rho_*})\) \\
    \( r \leftarrow \frac{r}{2} \) \\
    \(\grid{\rho_*} \leftarrow \grid{\rho_*} + \mathcal{I}_{p2g} (\particle{\vec{x}}, \particle{\rho_i}, r, \grid{\vec{x}})\) \\
}
\caption{Multi-scale Density Reconstruction}
\label{algo:densitySample}
\end{algorithm}






\subsection{Temporal coherency}
\begin{figure} [h!]
  \centering
  \includegraphics[width=0.4\textwidth]{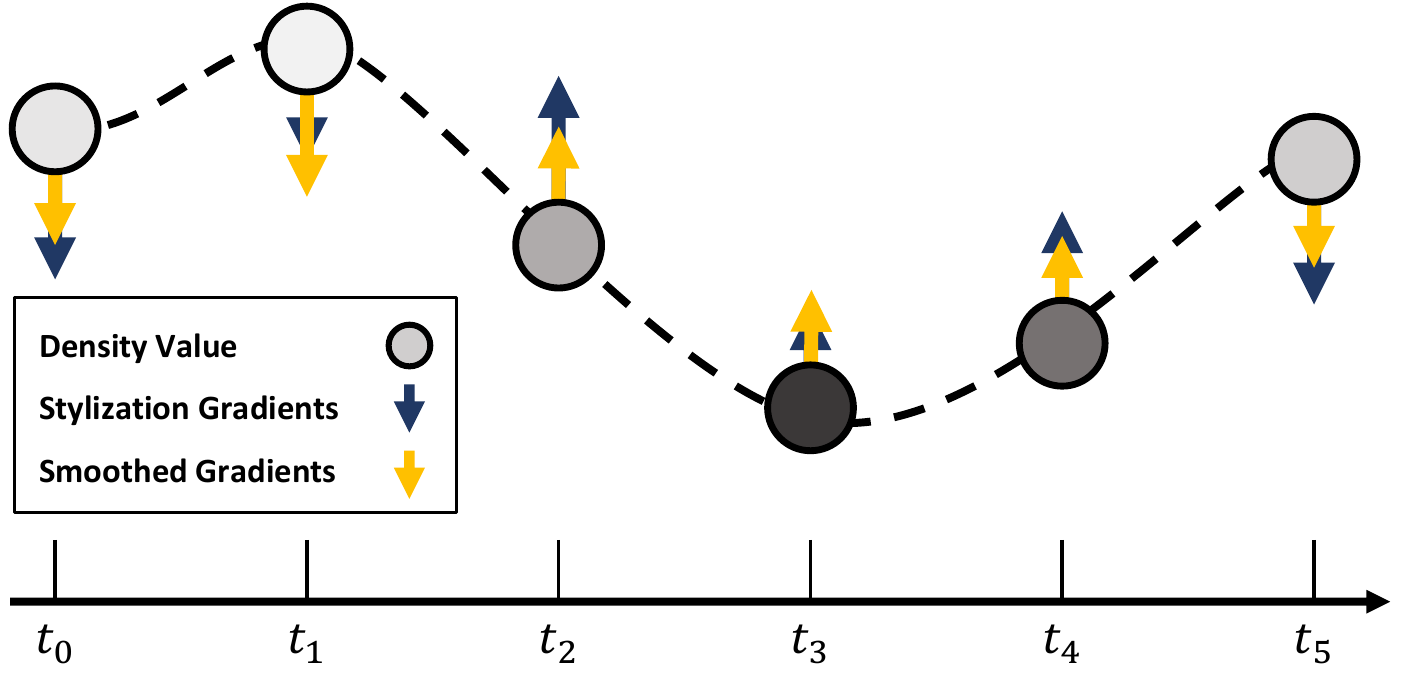}
  \caption{Particle density (circles) variation for a single particle over time. Temporal coherency is enforced by smoothing density gradients used for stylization from adjacent frames.}
  \label{fig:babyDensities}
\end{figure}

The major advantage of our Lagrangian discretization is the inexpensive enforcing of temporal coherency. Since quantities are carried individually per particle, it is intrinsically simple to track how attributes change over time. Neural style gradients are computed on the grid and need to be updated once the neighborhood of a particle changes. To ensure smooth transitions, we apply a Gaussian filter over the density changes of a particle, as shown in \Fig{babyDensities}.
Besides being sensitive to density neighborhood changes, stylization gradients are also influenced by the density carried by the particle itself (\Sec{multiscale}).



\begin{figure}[t!]
 \centering
    \includegraphics[trim=0 0 900 0,clip, width=0.435\textwidth]{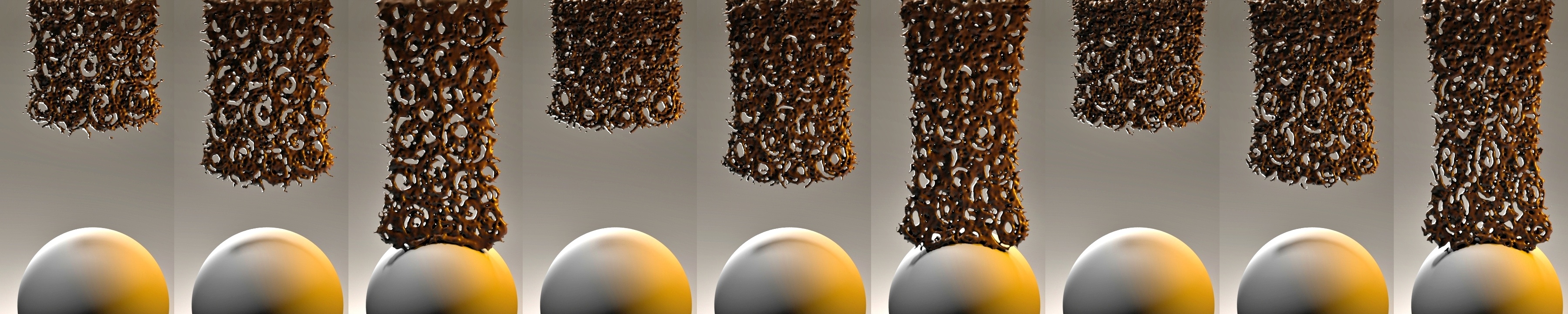}
    \caption{
    Stylization of every frame (left three images) versus keyframed stylization every 10 frames (images on the right). 
    \add{Sparse keyframing is visually similar and can be useful for quick previews.}
    }
\label{fig:keyframe}
\end{figure}

To further improve efficiency, and in contrast to TNST, we can keyframe stylizations, i.e., apply stylization to keyframes and interpolate particle attributes in-between. In practice, we reduced the stylization frames by a factor of 2 at max, but more drastic approximations could be used. Sparse keyframes still show temporally smooth transitions, but quality is degraded. Nevertheless, sparse keyframing would still be useful for generating quick previews of the simulation. The impact of sparse keyframing (every 10 frames) is shown in \Fig{keyframe}.
%



\section{Results}
We implemented the method with the \emph{tensorflow} framework and computed results on a TITAN Xp GPU (12GB). We used \emph{mantaflow} \cite{mantaflow} for smoke scene generation, a \threeD~smoke dataset from Kim et al. \shortcite{Kim2019} for comparisons with TNST, a \twoD~smoke dataset from Jamri\v{s}ka et al.~\shortcite{jamrivska2015lazyfluids} for color stylization, \emph{SPlisHSPlasH}~\cite{splishsplash,Koschier2019} for liquid simulations and \emph{Houdini} for rendering.

\begin{figure*}[h!]
\newcommand*{\mywidth}{0.158}
 \newcommand*{\triml}{350.0}
 \newcommand*{\trimr}{280.0}
 \newcommand*{\trimb}{0.0}
 \newcommand*{\trimt}{40}
 \centering
    \includegraphics[trim=\triml px \trimb px \trimr px \trimt px, clip, width=\mywidth\textwidth]{fig/results/TNST_LNST/net_gt_105.jpg} 
    \includegraphics[trim=\triml px \trimb px \trimr px \trimt px, clip, width=\mywidth\textwidth]{fig/results/TNST_LNST/net_new_105.jpg}
    \hspace{-27.7px}\includegraphics[width=0.05\linewidth]{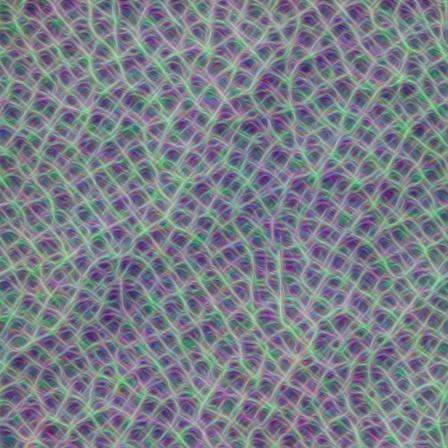}
    \includegraphics[trim=\triml px \trimb px \trimr px \trimt px, clip, width=\mywidth\textwidth]{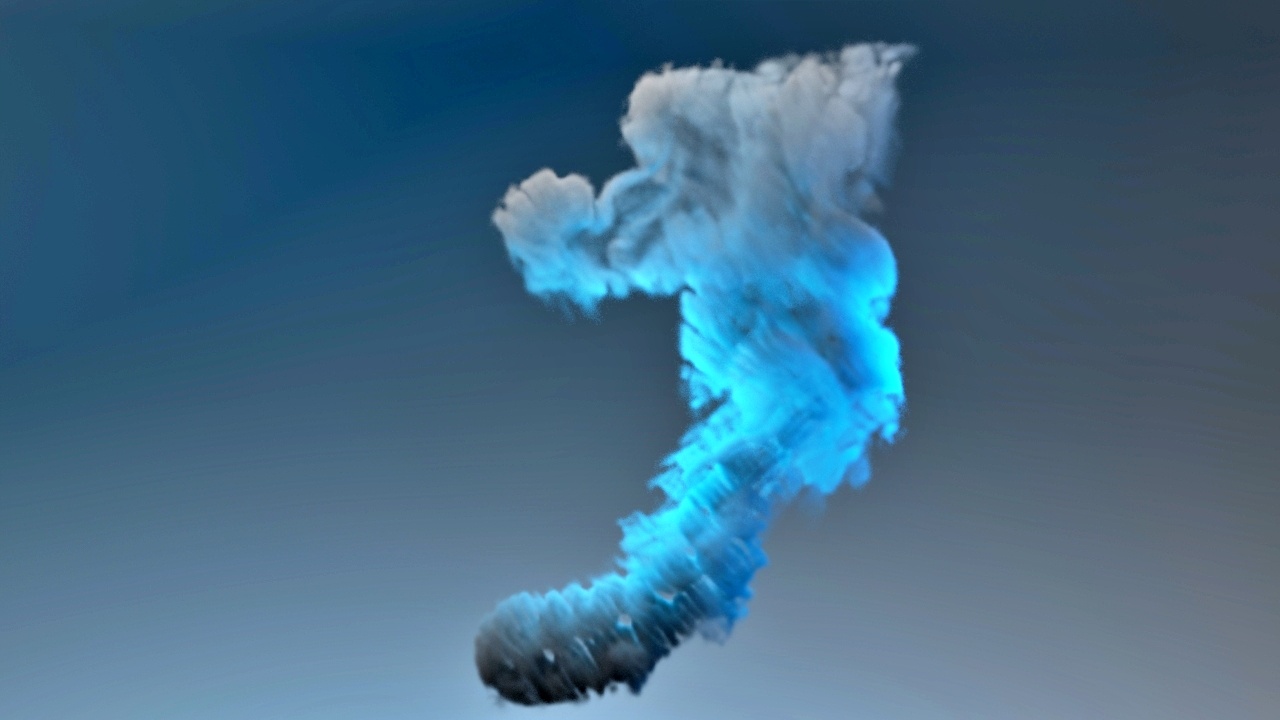}
    \hspace{-27.7px}\includegraphics[width=0.05\linewidth]{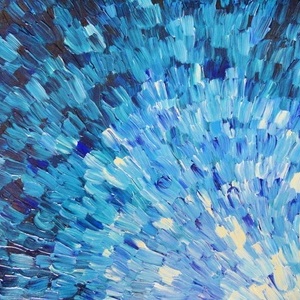}
    \includegraphics[trim=\triml px \trimb px \trimr px \trimt px, clip, width=\mywidth\textwidth]{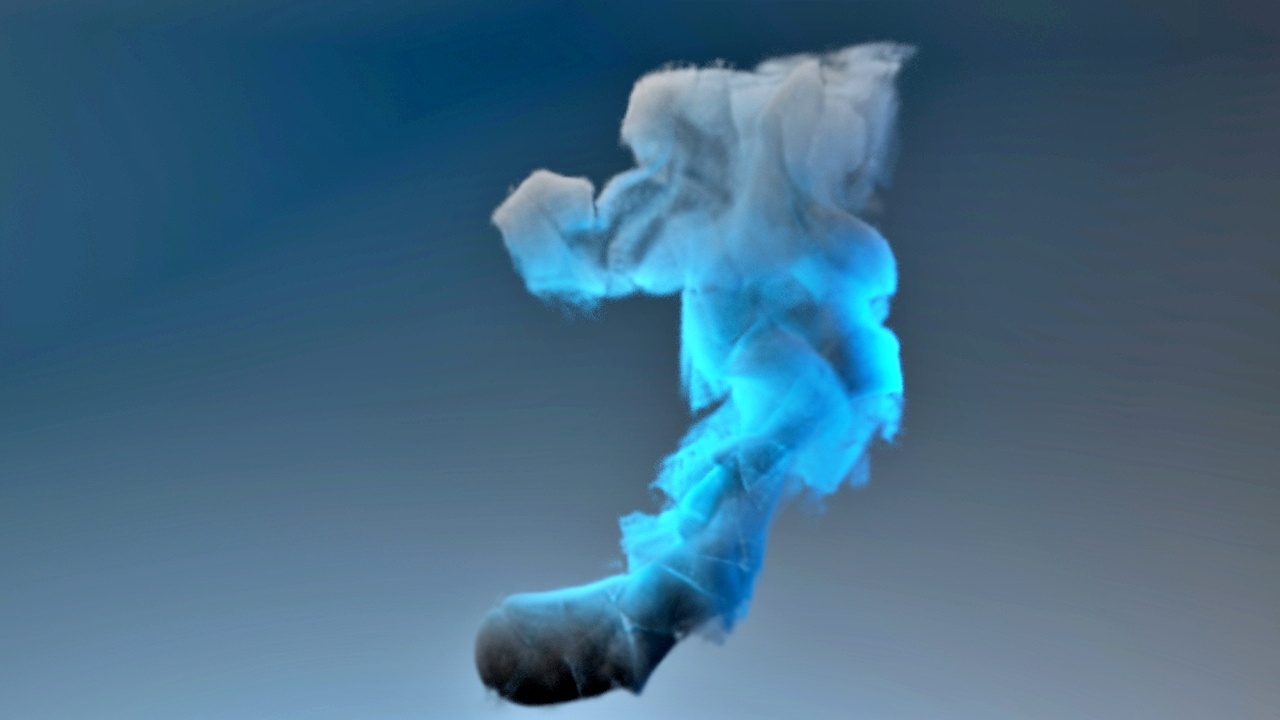}
    \hspace{-27.7px}\includegraphics[width=0.05\linewidth]{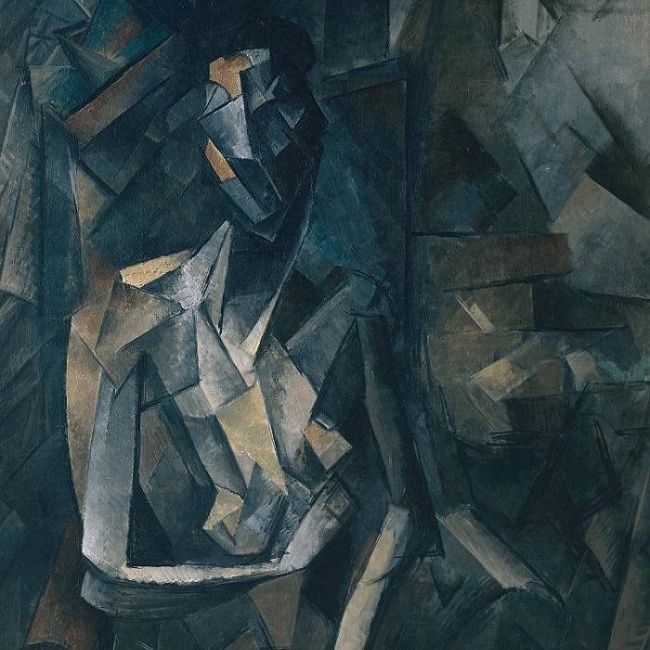}
    \includegraphics[trim=\triml px \trimb px \trimr px \trimt px, clip, width=\mywidth\textwidth]{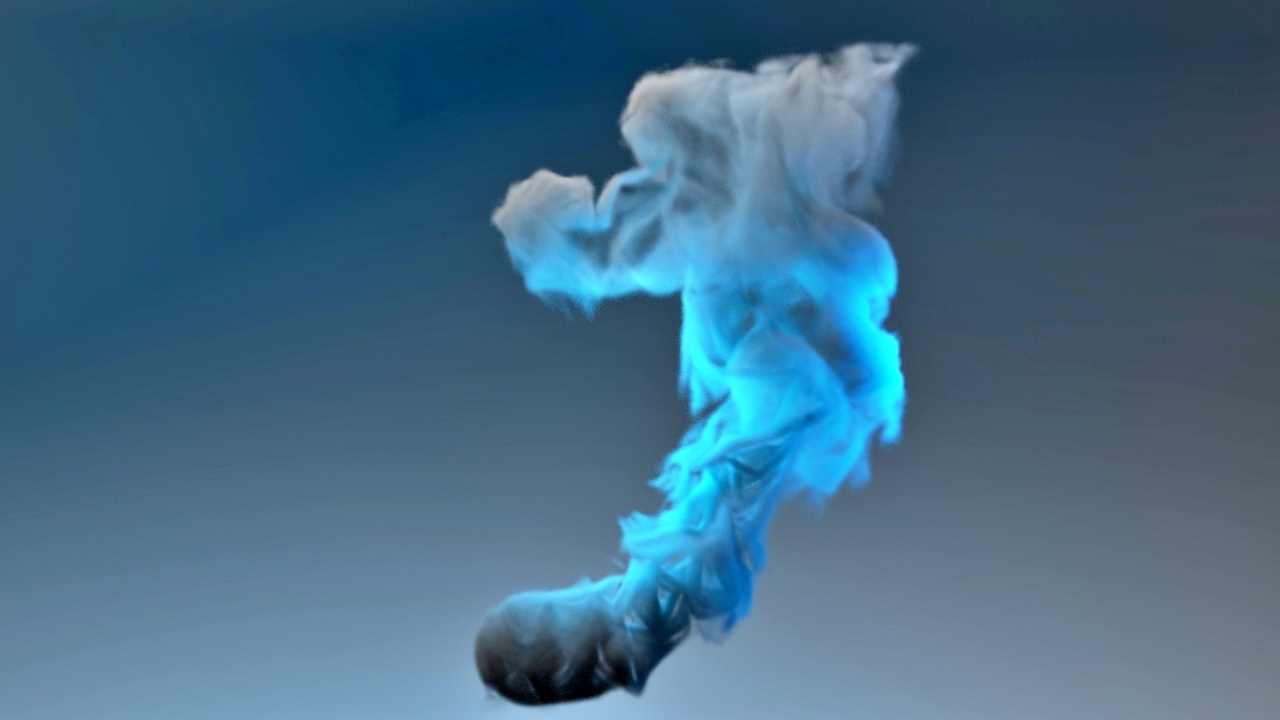}
    \hspace{-27.7px}\includegraphics[width=0.05\linewidth]{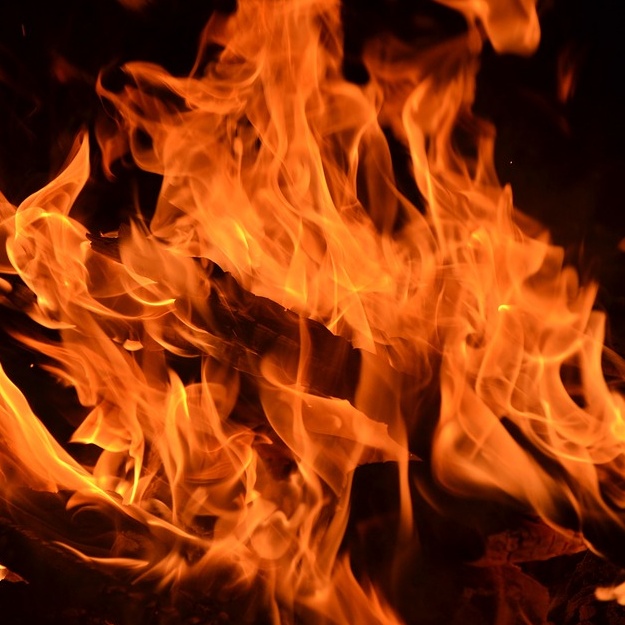}
    \\
    \hspace{\mywidth\textwidth} 
    \includegraphics[trim=\triml px \trimb px \trimr px \trimt px, clip, width=\mywidth\textwidth]{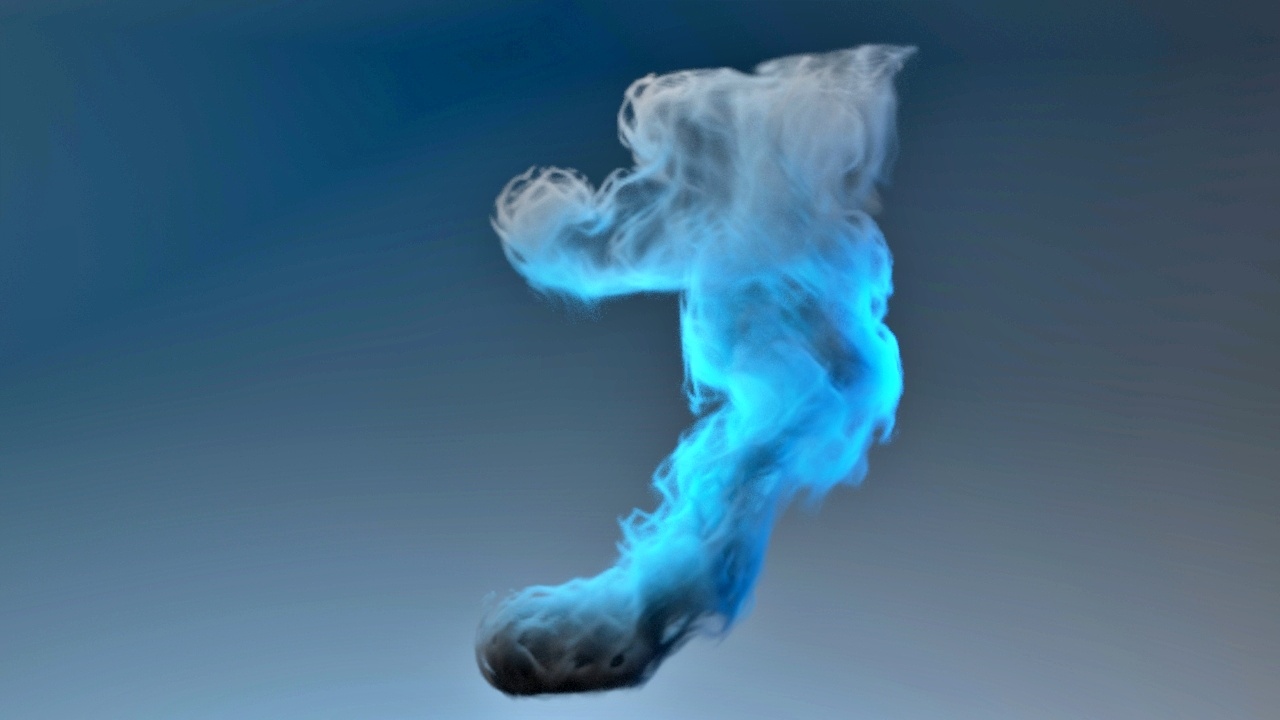}
    \includegraphics[trim=\triml px \trimb px \trimr px \trimt px, clip, width=\mywidth\textwidth]{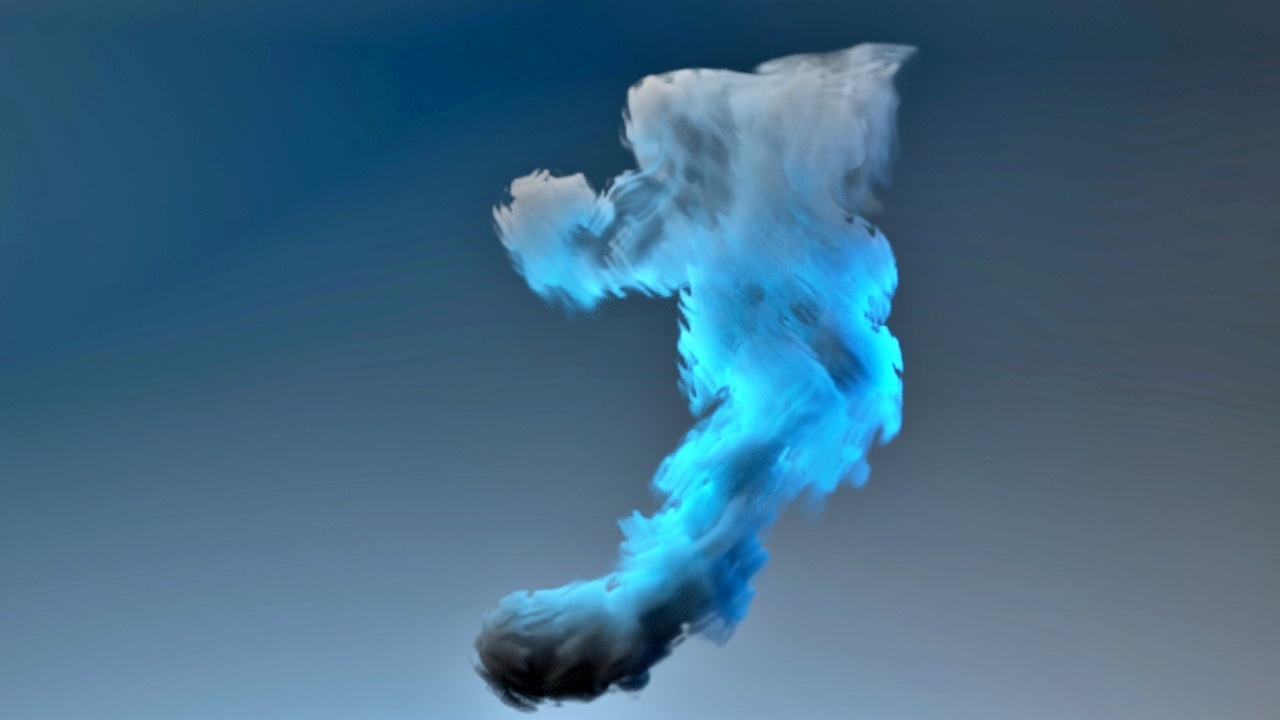}
    \includegraphics[trim=\triml px \trimb px \trimr px \trimt px, clip, width=\mywidth\textwidth]{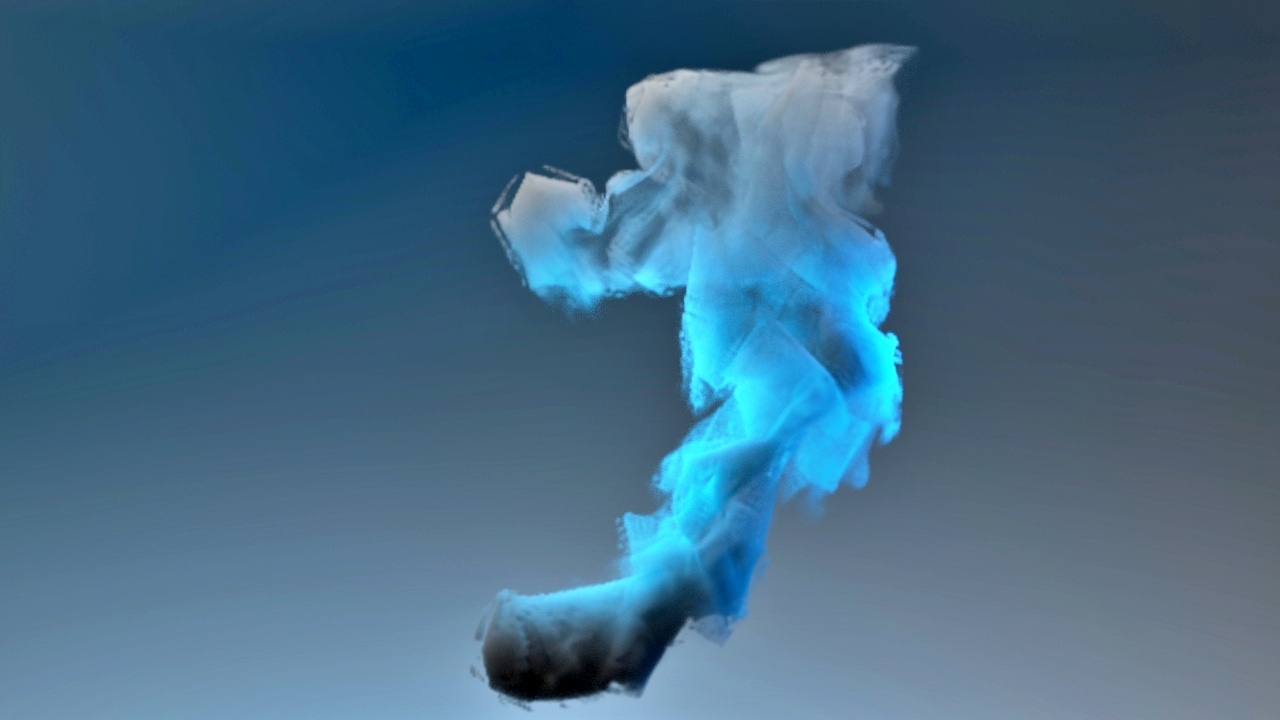}
    \includegraphics[trim=\triml px \trimb px \trimr px \trimt px, clip, width=\mywidth\textwidth]{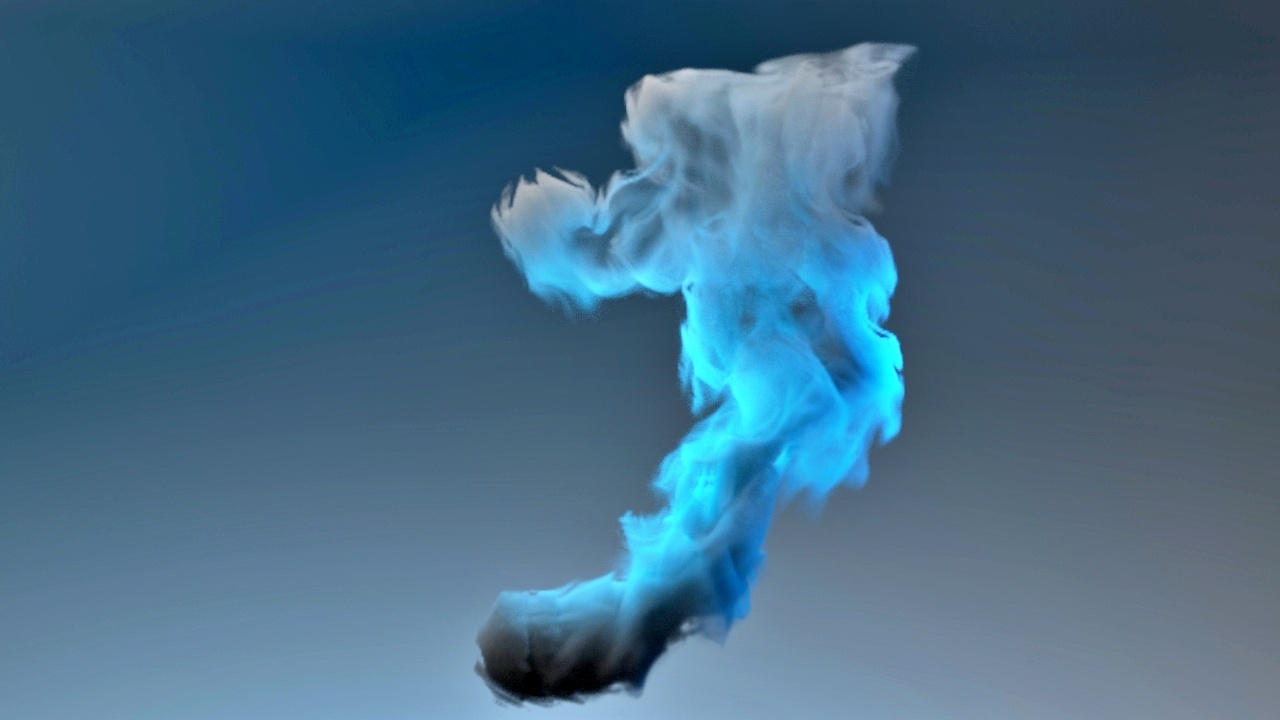}
    %
    \caption{Semantic transfer applied to the smokejet simulation of \cite{Kim2019} (leftmost column). Stylized results are shown for our \add{LNST} (top) and TNST (bottom) for semantic feature transfer \emph{net} (second column) and input images \emph{blue strokes}, \emph{Seated Nude}, and \emph{fire} (last three columns)\protect\footnotemark.}
\label{fig:tnst_lnst}
\end{figure*}

\begin{figure*}[th!]
\newcommand*{\mywidth}{0.26}
\newcommand*{\triml}{200.0}
 \newcommand*{\trimr}{230.0}
 \newcommand*{\trimb}{40.0}
 \newcommand*{\trimt}{60.0}
  \centering
\includegraphics[trim=\triml px \trimb px \trimr px \trimt px, clip,width=\mywidth\textwidth]{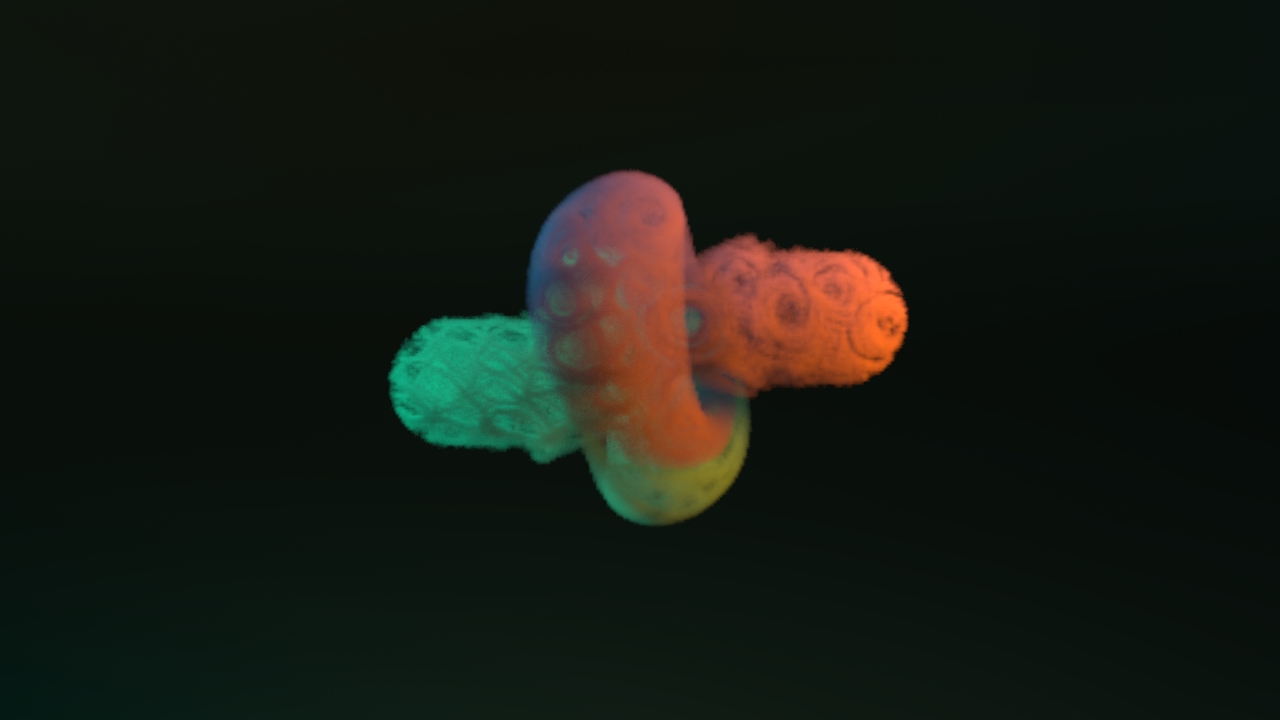}
\hspace{-53px}\includegraphics[width=0.1\linewidth]{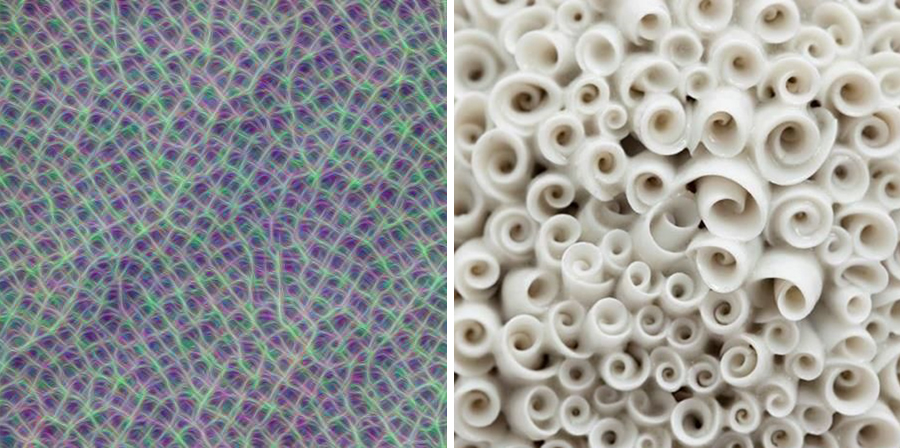}
\includegraphics[trim=\triml px \trimb px \trimr px \trimt px, clip,width=\mywidth\textwidth]{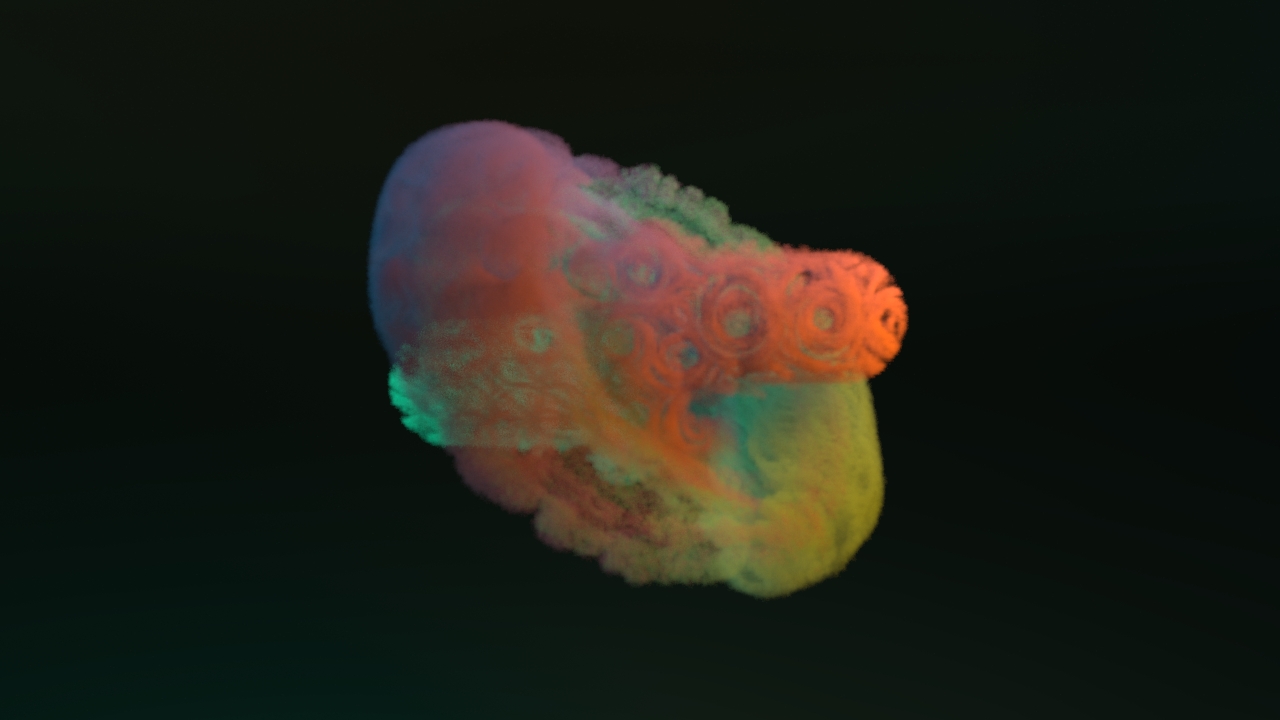}
\includegraphics[trim=\triml px \trimb px \trimr px \trimt px, clip,width=\mywidth\textwidth]{fig/results/multifluid/net_spi_0100.jpg}
  \caption{Two colliding smoke jets, which are stylized individually with the semantic feature \emph{net} and input image \emph{spirals}. The Lagrangian representation enables coherent stylization of multiple fluids even if the flow undergoes complex mixing.}
\label{fig:multifluid}
\end{figure*}

\paragraph{Performance}
Using particles for stylization eliminates the need for recursively aligning stylization velocities from subsequent frames, which notably improves the computational performance.
In combination with our sparse particle respesentation for smoke (1 particle per cell), simulations of size $200\times300\times200$ can now be stylized within an hour instead of a day (TNST). The computation time per frame is 0.66 minutes for the Smoke Jet scene shown in \Fig{tnst_lnst}, which is a speed-up of a factor of 20.41 compared to TNST. This improvement allows artists to more easily test different reference structures (input images) and hence renders neural flow stylization better applicable in production environments. \Tab{performance} gives an overview of the timings and parameters for the individual test scenes. 
Keyframing (every other frame) was applied to the Smoke Jet (\Fig{tnst_lnst}) and Double Jets (\Fig{multifluid}) examples. 
\begin{table}[h!]
\caption{Performance table.}
\vspace{-5pt}
\begin{tabular}{p{2.7cm} ccccccc}
\hline
\small{Scene} & \small{Resolution} & \# \small{Particles} & \small{Time (m/f)}
\tabularnewline \hline
\small{Moving Sphere} \footnotesize{(Fig.~\ref{fig:spheretest})} & $192\times192\times192$ & 237K & 0.8\\
\small{Smoke Jet} \footnotesize{(Fig.~\ref{fig:tnst_lnst})} & $200\times300\times200$ & 1.2M & 0.66\\
\small{Double Jets} \footnotesize{(Fig.~\ref{fig:multifluid})} & $200\times200\times200$ & 2M/2M & 0.45\\
\small{Chocolate} \footnotesize{(Fig.~\ref{fig:chocolate})} & $200\times200\times200$ & 80K & 0.05\\
\small{Colored Smoke} \footnotesize{(Fig.~\ref{fig:colorSmoke})} & $800\times800$ & 136K & 1.21\\
\small{Dam Break} \footnotesize{(Fig.~\ref{fig:color})} & $512\times1024$ & 23K & 0.58\\
\small{Double Dam} \footnotesize{(Fig.~\ref{fig:colormix})} & $512\times1024$ & 31K/8K & 0.65
\tabularnewline \hline
\end{tabular}
\label{tab:performance}
\vspace{-5pt}
\end{table}
%

\begin{figure}[t!]
 \centering
 \includegraphics[trim={0 45 40 45}, clip,
    width=0.13\textwidth]{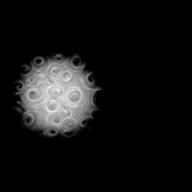}\hspace{-1px}
    \includegraphics[trim={0 45 40 45}, clip,
    width=0.13\textwidth]{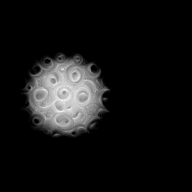}\hspace{-1px}
    \includegraphics[trim={0 45 40 45}, clip,
    width=0.13\textwidth]{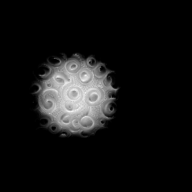}\\
\includegraphics[trim={0 45 40 45}, clip,
    width=0.13\textwidth]{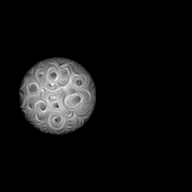}\hspace{-1px}
    \includegraphics[trim={0 45 40 45}, clip,
    width=0.13\textwidth]{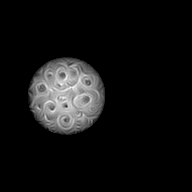}\hspace{-1px}
    \includegraphics[trim={0 45 40 45}, clip,
    width=0.13\textwidth]{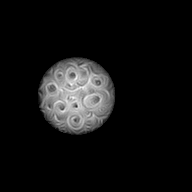}\\
\includegraphics[trim={0 45 40 45}, clip,
    width=0.13\textwidth]{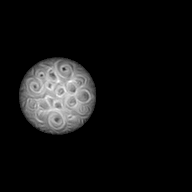}\hspace{-1px}
    \includegraphics[trim={0 45 40 45}, clip,
    width=0.13\textwidth]{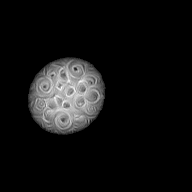}\hspace{-1px}
    \includegraphics[trim={0 45 40 45}, clip,
    width=0.13\textwidth]{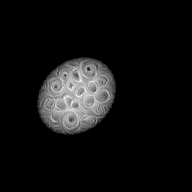}\\
    \includegraphics[trim={0 45 40 45}, clip,
    width=0.13\textwidth]{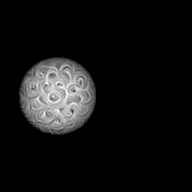}\hspace{-1px}
    \includegraphics[trim={0 45 40 45}, clip,
    width=0.13\textwidth]{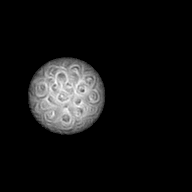}\hspace{-1px}
    \includegraphics[trim={0 45 40 45}, clip,
    width=0.13\textwidth]{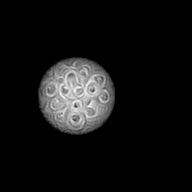}

    \caption{
    Selected frames of a stylized moving smoke sphere. From top to bottom: TNST with structures changing over time, LNST with temporally coherent structures, LNST result with applied shearing, and LNST result with noise-added density inducing style variation over time.
    }
\label{fig:spheretest}
\end{figure}

\paragraph{Time-coherency}
To illustrate the benefit of the Lagrangian formulation, we use a simple test scene where we initialize a smoke sphere with a uniform density. We then move the smoke artificially to the right, and apply the neural stylization to every frame of the sequence. We compare the results of LNST and TNST for different time instances in \Fig{spheretest}. 
The top row shows the results of TNST. It can be seen that TNST is not able to preserve constant stylized textures in regions where the density function does not change. This is due to the recursive alignment of stylization gradients, which accumulate errors especially for bigger window sizes.
The second row shows the corresponding results with LNST, demonstrating consistent stylization over time since gradients are constant. Also when applied a shearing deformation to the sphere, as shown in the third row, strucutures remain coherent. If an artist prefers to have changing structures in such situations, noise can be added to the densities carried by the particles, which in turn will induce stylization gradients as shown in the last row.

\paragraph{Smoke Stylization}
\Fig{tnst_lnst} shows a direct comparison of LNST and TNST applied to the smoke jet dataset of Kim et al. \shortcite{Kim2019}. While the resulting structures inherently depend on the underlying representation, they naturally differ and cannot be directly compared with each other. It can be observed, however, that the Lagrangian stylization may lead to more pronounced structures, well visible in the semantic transfer \emph{net} and the style transfer \emph{blue strokes}, and that boundaries are smoother, noticeable in the \emph{Seated Nude} example.

\paragraph{Multi-fluid Stylization}
Stylization of multiple fluids is naturally enabled by stylizing different sets of particles with different input images. \Fig{multifluid} shows a simulation of two smoke jets colliding, where the left one is stylized with the semantic feature \emph{net} and the right one with the style transfer of the input image \emph{spirals}. Transferred structures are retained per fluid type even if the flow undergoes complex mixing effects. 

\begin{figure}[h!]
\newcommand*{\mywidth}{0.14}
 \newcommand*{\triml}{320.0}
 \newcommand*{\trimr}{300.0}
 \newcommand*{\trimb}{0.0}
 \newcommand*{\trimt}{50}
 \centering
    \includegraphics[trim=\triml px \trimb px \trimr px \trimt px, clip, width=\mywidth\textwidth]{fig/results/liquids/gt_buckling_87.jpg} 
    \includegraphics[trim=\triml px \trimb px \trimr px \trimt px, clip, width=\mywidth\textwidth]{fig/results/liquids/buckling_spiral_87.jpg} 
    \hspace{-26.6px}\includegraphics[width=0.1\linewidth]{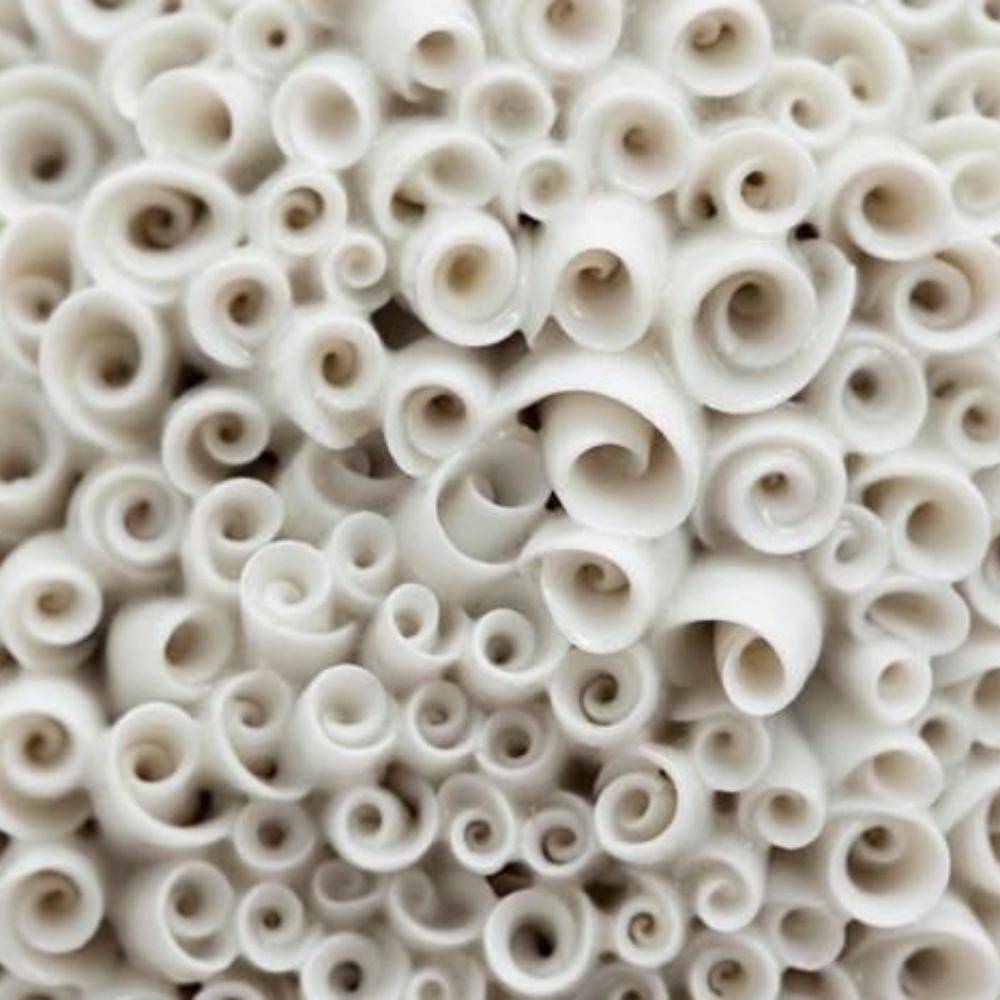}
    \includegraphics[trim=\triml px \trimb px \trimr px \trimt px, clip, width=\mywidth\textwidth]{fig/results/liquids/buckling_square_87.jpg} 
    \hspace{-26.6px}\includegraphics[width=0.1\linewidth]{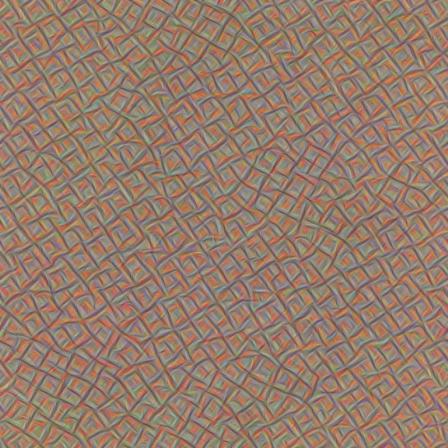}
    \caption{
 Thin sheet SPH simulation computed with \emph{SPlisHSPlasH}~\cite{splishsplash} stylized with the patterns \emph{spiral} and \emph{diagonal}.    
    }
\label{fig:chocolate}
\end{figure}

\paragraph{Stylization of Liquids}
We use a simple differentiable renderer for stylization of liquids. Unlike smoke renderer, which integrates media radiance scattered in the medium, we compute the amount of diffused light, \ie absorbed light except transmitted by its liquid volume~\cite{ihmsen2012unified}, which is given by 
\begin{equation}
\begin{split}
\tau(\vec{x}, \vec{r}) = e^{-\gamma \int_{\vec{0}}^{\vec{x}} d(\vec{r}) \; dr} \\
I_{ij} = 1 - \tau(\vec{r}_{max}, \vec{r}).
\label{eq:liquidRenderingEquation}
\end{split}
\end{equation}
\Fig{chocolate} shows the results of a stylized SPH simulation computed with \emph{SPlisHSPlasH}~\cite{splishsplash}. We applied the patterns \emph{spiral} and \emph{diagonal} to a thin sheet simulation.
\footnotetext{Image sources: \url{http://storage.googleapis.com/deepdream/visualz/tensorflow_inception/index.html},~
\url{https://github.com/byungsook/neural-flow-style}}

\begin{figure*}[t!]
\newcommand*{\mywidth}{0.31}
 \newcommand*{\triml}{0.0}
 \newcommand*{\trimr}{0.0}
 \newcommand*{\trimb}{0.0}
 \newcommand*{\trimt}{200}
 \centering
    \includegraphics[trim=\triml px \trimb px 0px \trimt px, clip, width=\mywidth\textwidth]{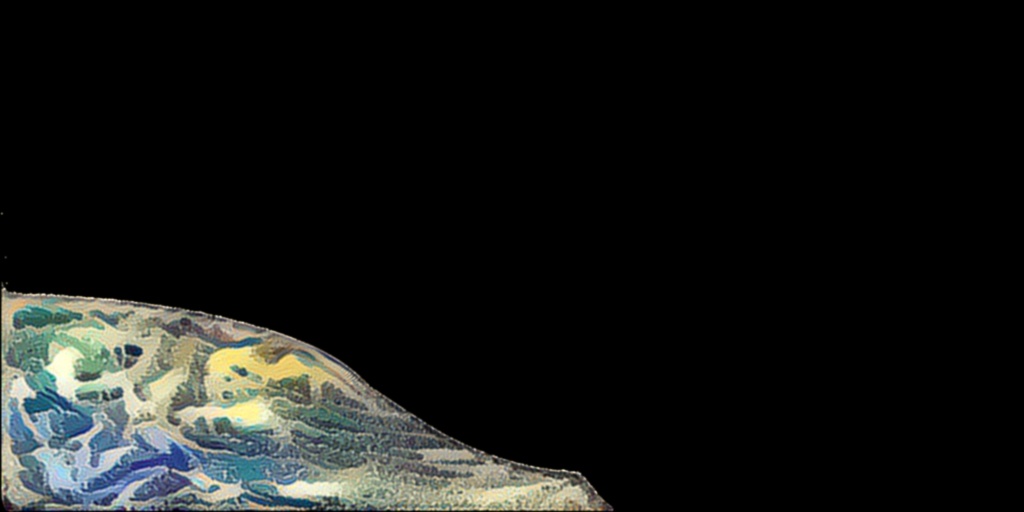} 
    \hspace{-27.7px}\includegraphics[width=0.05\linewidth]{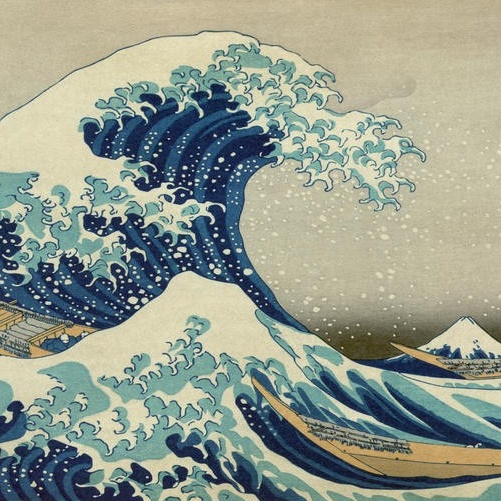}
    \includegraphics[trim=\triml px \trimb px \trimr px \trimt px, clip, width=\mywidth\textwidth]{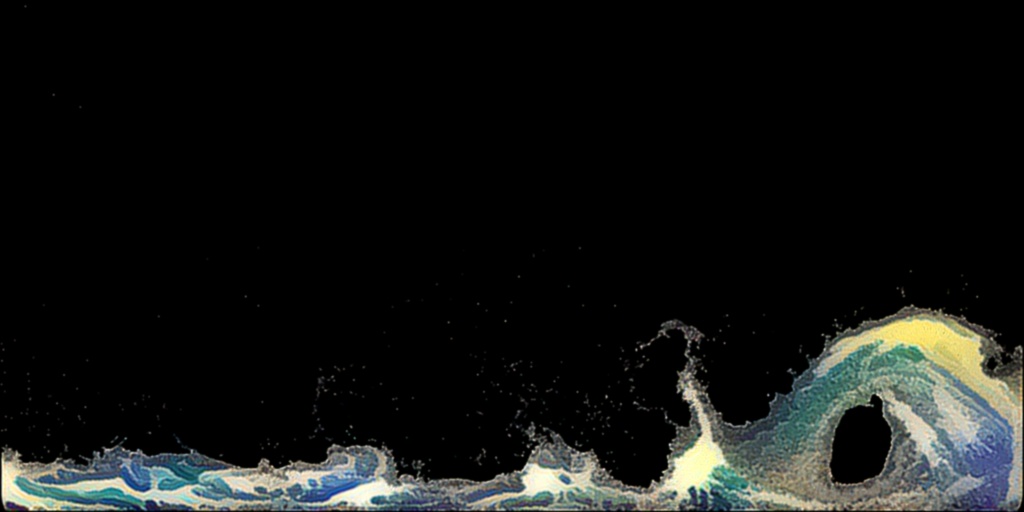} 
    \includegraphics[trim=\triml px \trimb px \trimr px \trimt px, clip, width=\mywidth\textwidth]{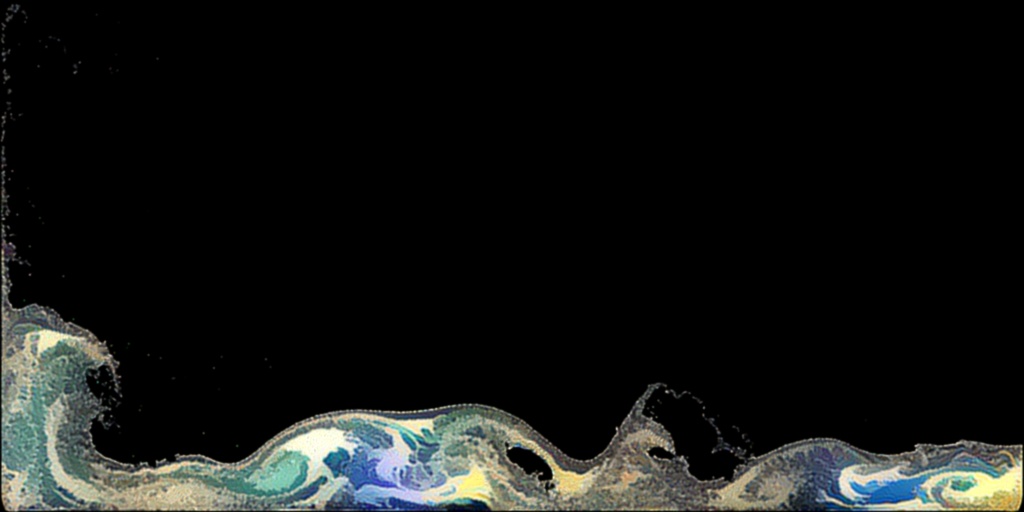} 
    \\
    \includegraphics[trim=\triml px \trimb px \trimr px \trimt px, clip, width=\mywidth\textwidth]{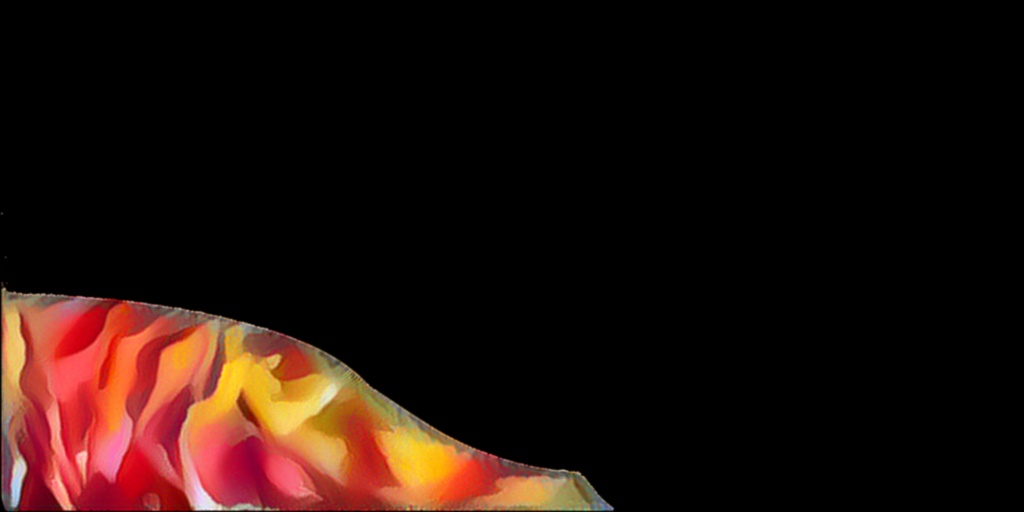} 
    \hspace{-27.7px}\includegraphics[width=0.05\linewidth]{fig/results/color/cokeffe}
    \includegraphics[trim=\triml px \trimb px \trimr px \trimt px, clip, width=\mywidth\textwidth]{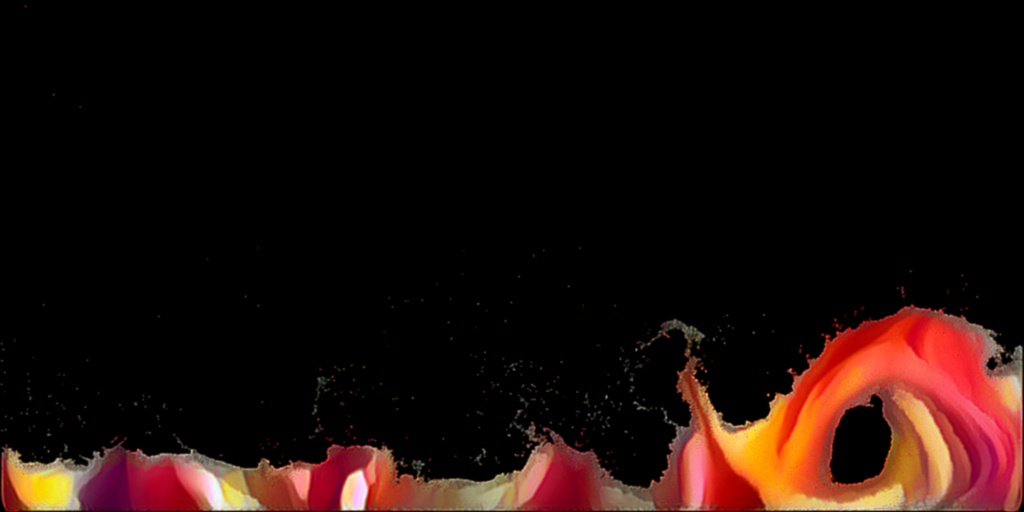} 
    \includegraphics[trim=\triml px \trimb px \trimr px \trimt px, clip, width=\mywidth\textwidth]{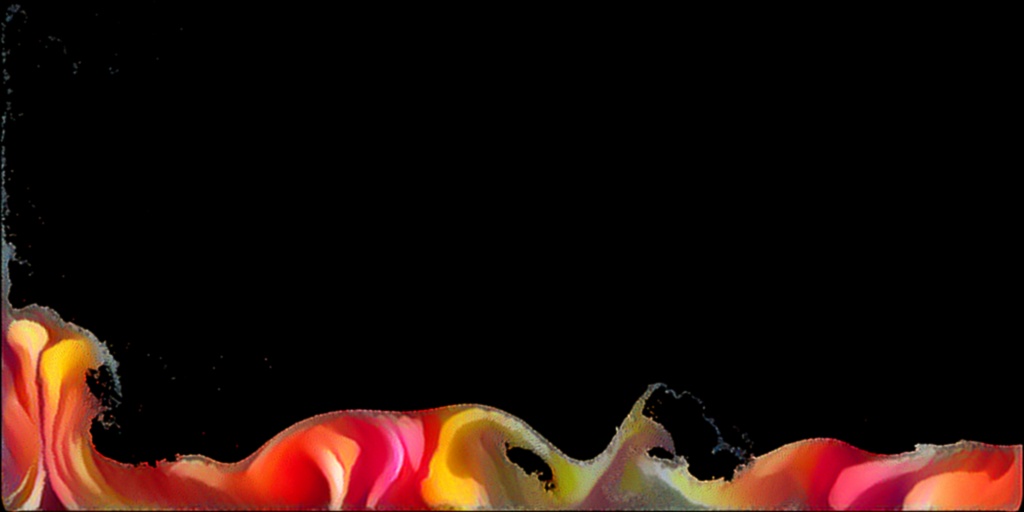} 
    \\
    \includegraphics[trim=\triml px \trimb px \trimr px \trimt px, clip, width=\mywidth\textwidth]{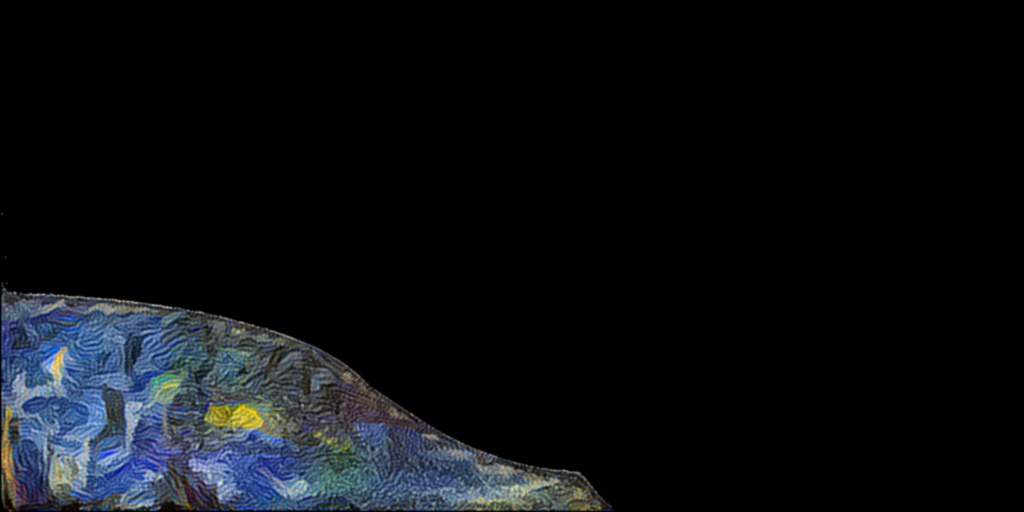} 
    \hspace{-27.7px}\includegraphics[width=0.05\linewidth]{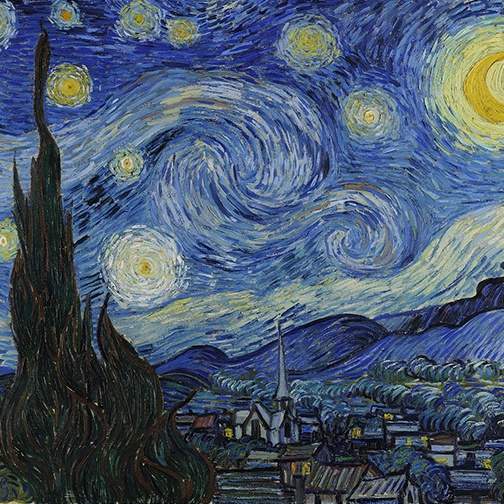}
    \includegraphics[trim=\triml px \trimb px \trimr px \trimt px, clip, width=\mywidth\textwidth]{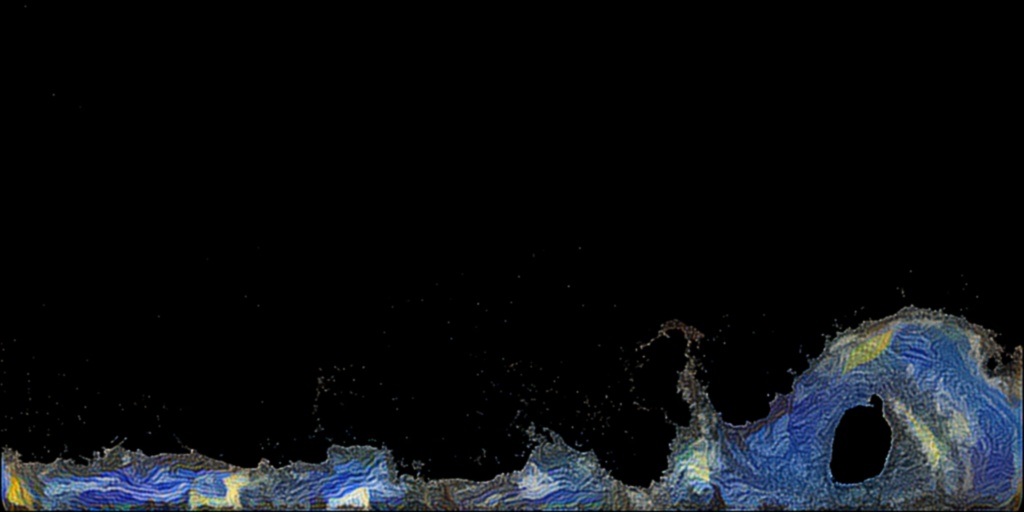} 
    \includegraphics[trim=\triml px \trimb px \trimr px \trimt px, clip, width=\mywidth\textwidth]{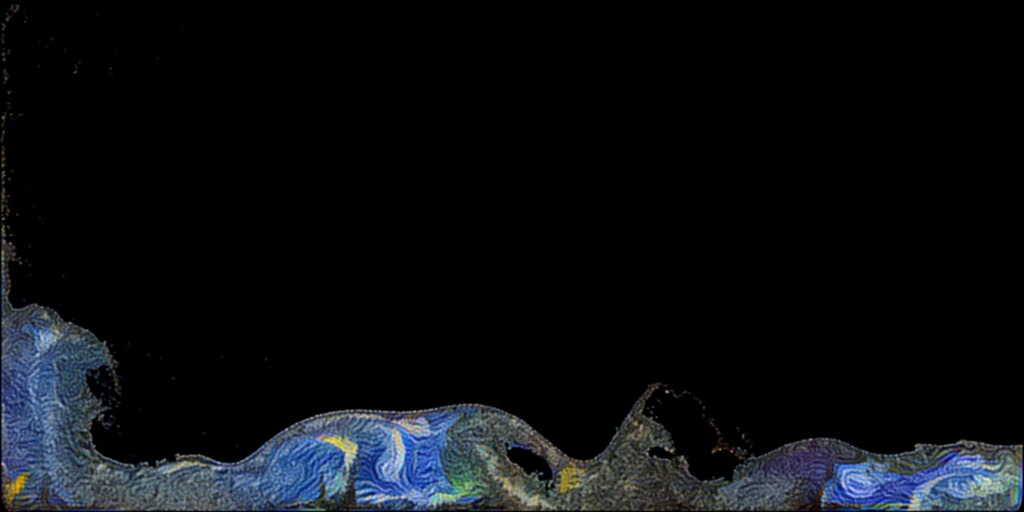} 
    %
    \caption{Lagrangian color stylization applied to a 2D particle-based liquid simulation using the input images \emph{Kanagawa Wave}, \emph{Red Canna} and \emph{Starry Night}.}
\label{fig:color}
\end{figure*}

\begin{figure*}[t!]
\newcommand*{\mywidth}{0.31}
 \newcommand*{\triml}{0.0}
 \newcommand*{\trimr}{0.0}
 \newcommand*{\trimb}{0.0}
 \newcommand*{\trimt}{150}
 \centering
    \includegraphics[trim=\triml px \trimb px 0px \trimt px, clip, width=\mywidth\textwidth]{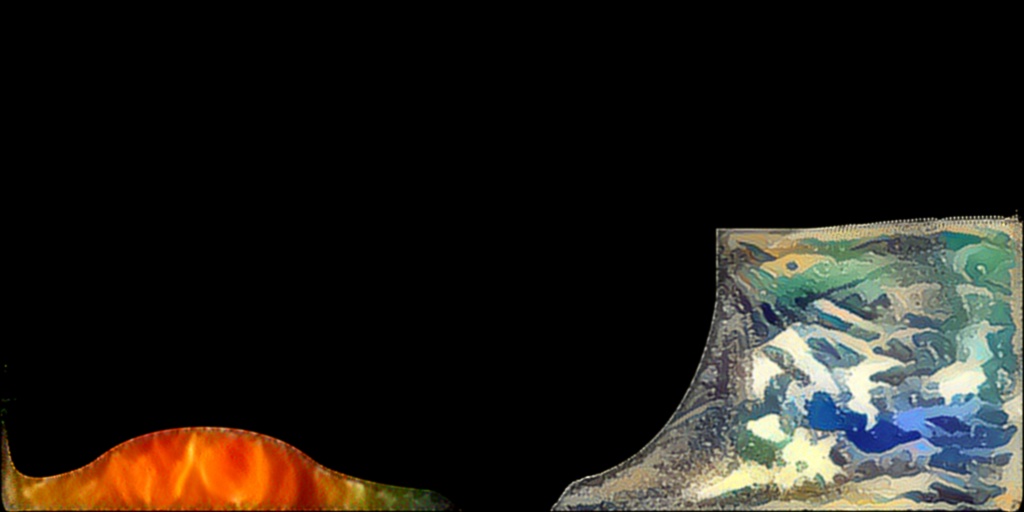} 
    \includegraphics[trim=\triml px \trimb px \trimr px \trimt px, clip, width=\mywidth\textwidth]{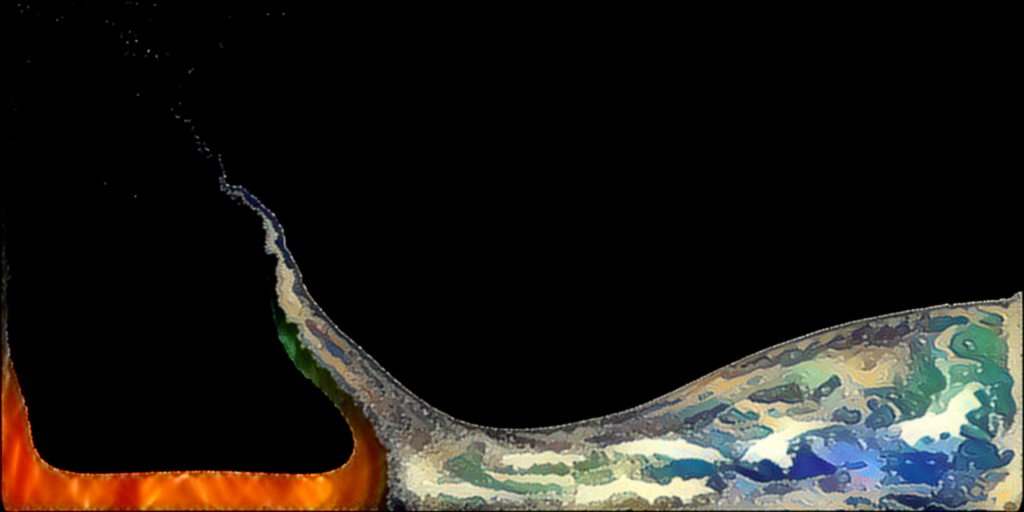} 
    \includegraphics[trim=\triml px \trimb px \trimr px \trimt px, clip, width=\mywidth\textwidth]{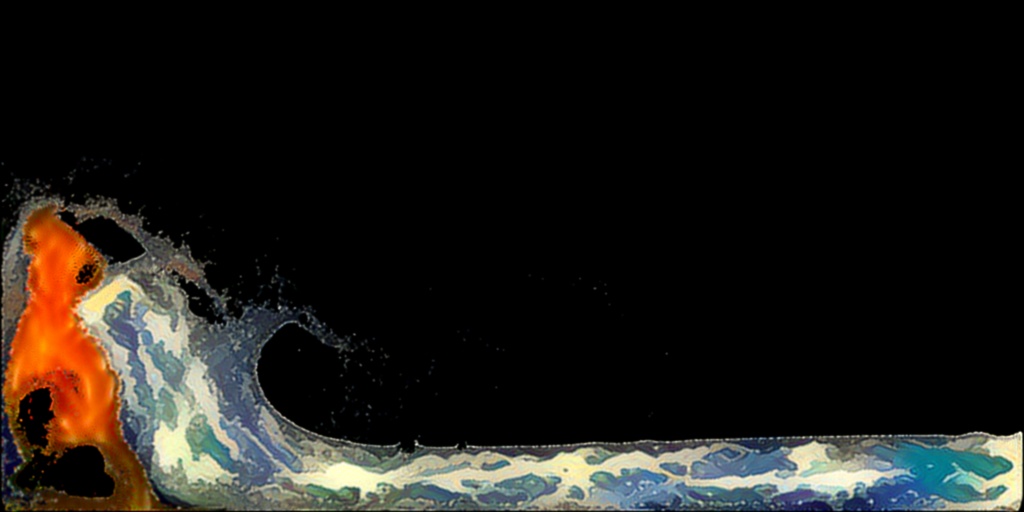} 
    \caption{Lagrangian color stylization applied to a \emph{mixed} 2D particle-based liquid simulation using the input images \emph{Kanagawa Wave} and \emph{fire}.}
\label{fig:colormix}
\end{figure*}

\paragraph{Color Transfer}
We transfer color information from input images to flow fields by storing a color value per particle and optimizing it by \Eq{LNSTloss}. This can be applied to any grid-based or particle-based smoke or liquid simulation. In \Fig{color} we applied the color stylization to a 2D dam break simulation using different example images, and in \Fig{colormix} to two liquids with distinct types (and hence color).
The accompanying videos show that local color structures change very smoothly over time, which is attributed to the improved time-coherency of the Lagrangian stylization. This is especially well visible in \Fig{colorSmoke}, where two subsequent frames are shown for TNST and LNST. In this example, we have transferred the style \emph{blue stroke} to a smoke scene. The close-up views reveal discontinuities for TNST, while LNST shows smooth transitions for color structures.


\section{Discussion and Conclusion}

We have presented a Lagrangian approach for neural flow stylization and have demonstrated benefits with respect to quality (improved temporal coherence), performance (stylization per frame in less than a minute), and art-directability (multi-fluid stylization, color transfer, liquid stylization). A key property of our approach is that it is not restricted to any particular fluid solver type (i.e., grids, particles, hybrid solvers). To enable this, we have introduced a strategy for grid-to-particle transfer (and vice versa) to efficiently update attributes and gradients, and a re-simulation that can be effectively applied to grid and particle fluid representations. This generality of our method facilitates seamless integration of neural style transfer into existing content production workflows. 

A current limitation of the method is that we use a simple differentiable renderer for liquids. While this works well for some scenarios, a dedicated differentiable renderer for liquids would improve the resulting quality and especially also support a wider range of liquid simulation setups. Similar to the smoke renderer, such a liquid renderer must be differentiable as gradients are back-propagated in the optimization. It must also be efficient as the renderer is used in each step of the optimization. Although the complexity of the renderer has a direct influence on the quality of the results, we suspect that, analogously to our smoke renderer, a lightweight renderer that can recover the core flow structures is sufficient for stylizing liquids.

We have shown that LNST enables novel effects and a high degree of art-directability, which renders flow stylization more practical in prodcution workflows. However, we have not tested the method on large-scale simulations that are typically used in such settings. While our method can handle up to 2 million particles, larger scenes are restricted by the available memory. Moreover, in practical settings the scene complexity is higher, which potentially poses challenges with respect to artist control of the stylization. 

By reducing the computation time for stylizing an entire simulation from one day with TNST to a single hour with LNST renders the method much more practical for digital artists. However, for testing different input structures, a real-time method would be desirable. Recent concepts presented on neural image stylization might be mapped to 3D simulations to further improve efficiency.

\begin{acks}
The authors would like to thank Fraser Rothnie for his artistic contributions. The work was supported by the Swiss National Science Foundation under Grant No.: 200021\_168997.
\end{acks}

\bibliographystyle{ACM-Reference-Format}
\bibliography{main}


\begin{thebibliography}{82}


\ifx \showCODEN    \undefined \def \showCODEN     #1{\unskip}     \fi
\ifx \showDOI      \undefined \def \showDOI       #1{#1}\fi
\ifx \showISBNx    \undefined \def \showISBNx     #1{\unskip}     \fi
\ifx \showISBNxiii \undefined \def \showISBNxiii  #1{\unskip}     \fi
\ifx \showISSN     \undefined \def \showISSN      #1{\unskip}     \fi
\ifx \showLCCN     \undefined \def \showLCCN      #1{\unskip}     \fi
\ifx \shownote     \undefined \def \shownote      #1{#1}          \fi
\ifx \showarticletitle \undefined \def \showarticletitle #1{#1}   \fi
\ifx \showURL      \undefined \def \showURL       {\relax}        \fi
\providecommand\bibfield[2]{#2}
\providecommand\bibinfo[2]{#2}
\providecommand\natexlab[1]{#1}
\providecommand\showeprint[2][]{arXiv:#2}

\bibitem[\protect\citeauthoryear{Ando, Th{\"{u}}rey, and Wojtan}{Ando
  et~al\mbox{.}}{2013}]%
        {Ando2013}
\bibfield{author}{\bibinfo{person}{Ryoichi Ando}, \bibinfo{person}{Nils
  Th{\"{u}}rey}, {and} \bibinfo{person}{Chris Wojtan}.}
  \bibinfo{year}{2013}\natexlab{}.
\newblock \showarticletitle{{Highly adaptive liquid simulations on tetrahedral
  meshes}}.
\newblock \bibinfo{journal}{\emph{ACM Transactions on Graphics}}
  \bibinfo{volume}{32}, \bibinfo{number}{4} (\bibinfo{date}{jul}
  \bibinfo{year}{2013}), \bibinfo{pages}{1}.
\newblock
\showISSN{07300301}


\bibitem[\protect\citeauthoryear{Ando and Tsuruno}{Ando and Tsuruno}{2011}]%
        {Ando2011}
\bibfield{author}{\bibinfo{person}{Ryoichi Ando} {and} \bibinfo{person}{Reiji
  Tsuruno}.} \bibinfo{year}{2011}\natexlab{}.
\newblock \showarticletitle{{A particle-based method for preserving fluid
  sheets}}. In \bibinfo{booktitle}{\emph{Proceedings of SCA'11}}.
  \bibinfo{pages}{7}.
\newblock
\showISBNx{9781450309233}
\urldef\tempurl%
\url{https://doi.org/10.1145/2019406.2019408}
\showDOI{\tempurl}


\bibitem[\protect\citeauthoryear{Bargteil, Sin, Michaels, Goktekin, and
  O'Brien}{Bargteil et~al\mbox{.}}{2006}]%
        {Bargteil2006}
\bibfield{author}{\bibinfo{person}{Adam~W Bargteil}, \bibinfo{person}{Funshing
  Sin}, \bibinfo{person}{Jonathan~E Michaels}, \bibinfo{person}{Tolga~G
  Goktekin}, {and} \bibinfo{person}{James~F O'Brien}.}
  \bibinfo{year}{2006}\natexlab{}.
\newblock \showarticletitle{{A Texture Synthesis Method for Liquid
  Animations}}. In \bibinfo{booktitle}{\emph{Proceedings of SCA'06}}.
  \bibinfo{pages}{345--351}.
\newblock
\showISBNx{3-905673-34-7}
\urldef\tempurl%
\url{http://dl.acm.org/citation.cfm?id=1218064.1218111}
\showURL{%
\tempurl}


\bibitem[\protect\citeauthoryear{Becker and Teschner}{Becker and
  Teschner}{2007}]%
        {Becker2007}
\bibfield{author}{\bibinfo{person}{Markus Becker} {and}
  \bibinfo{person}{Matthias Teschner}.} \bibinfo{year}{2007}\natexlab{}.
\newblock \showarticletitle{{Weakly compressible SPH for free surface flows}}.
  In \bibinfo{booktitle}{\emph{Symposium on Computer Animation}}.
  \bibinfo{pages}{1--8}.
\newblock


\bibitem[\protect\citeauthoryear{Bender}{Bender}{2016}]%
        {splishsplash}
\bibfield{author}{\bibinfo{person}{Jan Bender}.}
  \bibinfo{year}{2016}\natexlab{}.
\newblock \bibinfo{title}{{SPlisHSPlasH}}.
\newblock
\newblock
\newblock
\shownote{\emph{https://github.com/InteractiveComputerGraphics/SPlisHSPlasH}.}


\bibitem[\protect\citeauthoryear{Bender and Koschier}{Bender and
  Koschier}{2015}]%
        {Bender2015}
\bibfield{author}{\bibinfo{person}{Jan Bender} {and} \bibinfo{person}{Dan
  Koschier}.} \bibinfo{year}{2015}\natexlab{}.
\newblock \showarticletitle{{Divergence-Free Smoothed Particle Hydrodynamics}}.
  In \bibinfo{booktitle}{\emph{Symposium on Computer Animation}}.
  \bibinfo{pages}{1--9}.
\newblock


\bibitem[\protect\citeauthoryear{Bi, Kalantari, and Ramamoorthi}{Bi
  et~al\mbox{.}}{2017}]%
        {Bi2017}
\bibfield{author}{\bibinfo{person}{Sai Bi}, \bibinfo{person}{Nima~Khademi
  Kalantari}, {and} \bibinfo{person}{Ravi Ramamoorthi}.}
  \bibinfo{year}{2017}\natexlab{}.
\newblock \showarticletitle{{Patch-based optimization for image-based texture
  mapping}}.
\newblock \bibinfo{journal}{\emph{ACM ToG}} \bibinfo{volume}{36},
  \bibinfo{number}{4} (\bibinfo{date}{jul} \bibinfo{year}{2017}),
  \bibinfo{pages}{1--11}.
\newblock
\showISSN{07300301}


\bibitem[\protect\citeauthoryear{Bousseau, Neyret, Thollot, and
  Salesin}{Bousseau et~al\mbox{.}}{2007}]%
        {Bousseau2007}
\bibfield{author}{\bibinfo{person}{Adrien Bousseau}, \bibinfo{person}{Fabrice
  Neyret}, \bibinfo{person}{Jo{\"{e}}lle Thollot}, {and} \bibinfo{person}{David
  Salesin}.} \bibinfo{year}{2007}\natexlab{}.
\newblock \showarticletitle{{Video watercolorization using bidirectional
  texture advection}}.
\newblock \bibinfo{journal}{\emph{ACM ToG}} \bibinfo{volume}{26},
  \bibinfo{number}{3} (\bibinfo{year}{2007}).
\newblock
\showISSN{07300301}


\bibitem[\protect\citeauthoryear{Brackbill, Kothe, and Ruppel}{Brackbill
  et~al\mbox{.}}{1988}]%
        {Brackbill1988}
\bibfield{author}{\bibinfo{person}{J.U. Brackbill}, \bibinfo{person}{D.B.
  Kothe}, {and} \bibinfo{person}{H.M. Ruppel}.}
  \bibinfo{year}{1988}\natexlab{}.
\newblock \showarticletitle{{Flip: A low-dissipation, particle-in-cell method
  for fluid flow}}.
\newblock \bibinfo{journal}{\emph{Computer Physics Communications}}
  \bibinfo{volume}{48}, \bibinfo{number}{1} (\bibinfo{year}{1988}),
  \bibinfo{pages}{25--38}.
\newblock
\showISSN{00104655}


\bibitem[\protect\citeauthoryear{Browning, Barnes, Ritter, and
  Finkelstein}{Browning et~al\mbox{.}}{2014}]%
        {browning2014stylized}
\bibfield{author}{\bibinfo{person}{Mark Browning}, \bibinfo{person}{Connelly
  Barnes}, \bibinfo{person}{Samantha Ritter}, {and} \bibinfo{person}{Adam
  Finkelstein}.} \bibinfo{year}{2014}\natexlab{}.
\newblock \showarticletitle{{Stylized keyframe animation of fluid
  simulations}}. In \bibinfo{booktitle}{\emph{Proceedings of the Workshop on
  Non-Photorealistic Animation and Rendering}}. ACM, \bibinfo{pages}{63--70}.
\newblock


\bibitem[\protect\citeauthoryear{Christen, Kim, C.~Azevedo, and
  Solenthaler}{Christen et~al\mbox{.}}{2019}]%
        {christen19color}
\bibfield{author}{\bibinfo{person}{Fabienne Christen},
  \bibinfo{person}{Byungsoo Kim}, \bibinfo{person}{Vinicius C.~Azevedo}, {and}
  \bibinfo{person}{Barbara Solenthaler}.} \bibinfo{year}{2019}\natexlab{}.
\newblock \showarticletitle{{Neural Smoke Stylization with Color Transfer}}.
\newblock  (\bibinfo{date}{dec} \bibinfo{year}{2019}).
\newblock
\showeprint[arxiv]{1912.08757}
\urldef\tempurl%
\url{http://arxiv.org/abs/1912.08757}
\showURL{%
\tempurl}


\bibitem[\protect\citeauthoryear{Chu and Thuerey}{Chu and Thuerey}{2017}]%
        {Chu2017}
\bibfield{author}{\bibinfo{person}{Mengyu Chu} {and} \bibinfo{person}{Nils
  Thuerey}.} \bibinfo{year}{2017}\natexlab{}.
\newblock \showarticletitle{{Data-driven synthesis of smoke flows with
  CNN-based feature descriptors}}.
\newblock \bibinfo{journal}{\emph{ACM Transactions on Graphics}}
  \bibinfo{volume}{36}, \bibinfo{number}{4} (\bibinfo{date}{jul}
  \bibinfo{year}{2017}), \bibinfo{pages}{1--14}.
\newblock
\showISSN{07300301}


\bibitem[\protect\citeauthoryear{Desbrun and Gascuel}{Desbrun and
  Gascuel}{1996}]%
        {Desbrun1996}
\bibfield{author}{\bibinfo{person}{Mathieu Desbrun} {and}
  \bibinfo{person}{Marie-Paule Gascuel}.} \bibinfo{year}{1996}\natexlab{}.
\newblock \showarticletitle{{Smoothed Particles: A new paradigm for animating
  highly deformable bodies}}. In \bibinfo{booktitle}{\emph{Eurographics
  Workshop on Computer Animation and Simulation}}. \bibinfo{pages}{61--76}.
\newblock


\bibitem[\protect\citeauthoryear{Diamanti, Barnes, Paris, Shechtman, and
  Sorkine-Hornung}{Diamanti et~al\mbox{.}}{2015}]%
        {Diamanti2015}
\bibfield{author}{\bibinfo{person}{Olga Diamanti}, \bibinfo{person}{Connelly
  Barnes}, \bibinfo{person}{Sylvain Paris}, \bibinfo{person}{Eli Shechtman},
  {and} \bibinfo{person}{Olga Sorkine-Hornung}.}
  \bibinfo{year}{2015}\natexlab{}.
\newblock \showarticletitle{{Synthesis of Complex Image Appearance from Limited
  Exemplars}}.
\newblock \bibinfo{journal}{\emph{ACM Transactions on Graphics}}
  \bibinfo{volume}{34}, \bibinfo{number}{2} (\bibinfo{date}{mar}
  \bibinfo{year}{2015}), \bibinfo{pages}{1--14}.
\newblock
\showISSN{07300301}


\bibitem[\protect\citeauthoryear{Ferstl, Ando, Wojtan, Westermann, and
  Thuerey}{Ferstl et~al\mbox{.}}{2016}]%
        {Ferstl2016}
\bibfield{author}{\bibinfo{person}{Florian Ferstl}, \bibinfo{person}{Ryoichi
  Ando}, \bibinfo{person}{Chris Wojtan}, \bibinfo{person}{R{\"{u}}diger
  Westermann}, {and} \bibinfo{person}{Nils Thuerey}.}
  \bibinfo{year}{2016}\natexlab{}.
\newblock \showarticletitle{{Narrow Band FLIP for Liquid Simulations}}.
\newblock \bibinfo{journal}{\emph{CGF}} \bibinfo{volume}{35},
  \bibinfo{number}{2} (\bibinfo{year}{2016}), \bibinfo{pages}{225--232}.
\newblock
\showISSN{01677055}


\bibitem[\protect\citeauthoryear{Fi{\v{s}}er, Jamri{\v{s}}ka,
  Luk{\'{a}}{\v{c}}, Shechtman, Asente, Lu, and S{\'{y}}kora}{Fi{\v{s}}er
  et~al\mbox{.}}{2016}]%
        {Fiser2016}
\bibfield{author}{\bibinfo{person}{Jakub Fi{\v{s}}er}, \bibinfo{person}{Ondřej
  Jamri{\v{s}}ka}, \bibinfo{person}{Michal Luk{\'{a}}{\v{c}}},
  \bibinfo{person}{Eli Shechtman}, \bibinfo{person}{Paul Asente},
  \bibinfo{person}{Jingwan Lu}, {and} \bibinfo{person}{Daniel S{\'{y}}kora}.}
  \bibinfo{year}{2016}\natexlab{}.
\newblock \showarticletitle{{StyLit: illumination-guided example-based
  stylization of 3D renderings}}.
\newblock \bibinfo{journal}{\emph{ACM ToG}}  \bibinfo{volume}{35}
  (\bibinfo{year}{2016}), \bibinfo{pages}{1--11}.
\newblock
\urldef\tempurl%
\url{https://doi.org/10.1145/2897824.2925948}
\showDOI{\tempurl}


\bibitem[\protect\citeauthoryear{Fu, Guo, Gast, Jiang, and Teran}{Fu
  et~al\mbox{.}}{2017}]%
        {Fu2017}
\bibfield{author}{\bibinfo{person}{Chuyuan Fu}, \bibinfo{person}{Qi Guo},
  \bibinfo{person}{Theodore Gast}, \bibinfo{person}{Chenfanfu Jiang}, {and}
  \bibinfo{person}{Joseph Teran}.} \bibinfo{year}{2017}\natexlab{}.
\newblock \showarticletitle{{A polynomial particle-in-cell method}}.
\newblock \bibinfo{journal}{\emph{ACM ToG}} \bibinfo{volume}{36},
  \bibinfo{number}{6} (\bibinfo{date}{nov} \bibinfo{year}{2017}),
  \bibinfo{pages}{1--12}.
\newblock
\showISSN{07300301}


\bibitem[\protect\citeauthoryear{Gagnon, Dagenais, and Paquette}{Gagnon
  et~al\mbox{.}}{2016}]%
        {Gagnon2016}
\bibfield{author}{\bibinfo{person}{Jonathan Gagnon},
  \bibinfo{person}{Fran{\c{c}}ois Dagenais}, {and} \bibinfo{person}{Eric
  Paquette}.} \bibinfo{year}{2016}\natexlab{}.
\newblock \showarticletitle{{Dynamic lapped texture for fluid simulations}}.
\newblock \bibinfo{journal}{\emph{The Visual Computer}} \bibinfo{volume}{32},
  \bibinfo{number}{6-8} (\bibinfo{date}{jun} \bibinfo{year}{2016}),
  \bibinfo{pages}{901--909}.
\newblock
\showISSN{0178-2789}


\bibitem[\protect\citeauthoryear{Gagnon, Guzm{\'{a}}n, Vervondel, Dagenais,
  Mould, and Paquette}{Gagnon et~al\mbox{.}}{2019}]%
        {Gagnon2019}
\bibfield{author}{\bibinfo{person}{Jonathan Gagnon},
  \bibinfo{person}{Juli{\'{a}}n~E. Guzm{\'{a}}n}, \bibinfo{person}{Valentin
  Vervondel}, \bibinfo{person}{Fran{\c{c}}ois Dagenais}, \bibinfo{person}{David
  Mould}, {and} \bibinfo{person}{Eric Paquette}.}
  \bibinfo{year}{2019}\natexlab{}.
\newblock \showarticletitle{{Distribution Update of Deformable Patches for
  Texture Synthesis on the Free Surface of Fluids}}.
\newblock \bibinfo{journal}{\emph{CGF}} \bibinfo{volume}{38},
  \bibinfo{number}{7} (\bibinfo{year}{2019}), \bibinfo{pages}{491--500}.
\newblock
\showISSN{0167-7055}


\bibitem[\protect\citeauthoryear{Gatys, Ecker, and Bethge}{Gatys
  et~al\mbox{.}}{2015}]%
        {gatys2015neural}
\bibfield{author}{\bibinfo{person}{Leon~A Gatys}, \bibinfo{person}{Alexander~S
  Ecker}, {and} \bibinfo{person}{Matthias Bethge}.}
  \bibinfo{year}{2015}\natexlab{}.
\newblock \showarticletitle{{A neural algorithm of artistic style}}.
\newblock \bibinfo{journal}{\emph{Nature Communications}}
  (\bibinfo{year}{2015}).
\newblock


\bibitem[\protect\citeauthoryear{Gatys, Ecker, and Bethge}{Gatys
  et~al\mbox{.}}{2016}]%
        {gatys2016}
\bibfield{author}{\bibinfo{person}{Leon~A. Gatys},
  \bibinfo{person}{Alexander~S. Ecker}, {and} \bibinfo{person}{Matthias
  Bethge}.} \bibinfo{year}{2016}\natexlab{}.
\newblock \showarticletitle{{Image Style Transfer Using Convolutional Neural
  Networks}}. In \bibinfo{booktitle}{\emph{2016 IEEE CVPR}}.
  \bibinfo{pages}{2414--2423}.
\newblock
\showISBNx{978-1-4673-8851-1}


\bibitem[\protect\citeauthoryear{Gissler, Peer, Band, Bender, and
  Teschner}{Gissler et~al\mbox{.}}{2019}]%
        {Gissler2019}
\bibfield{author}{\bibinfo{person}{C Gissler}, \bibinfo{person}{A Peer},
  \bibinfo{person}{S Band}, \bibinfo{person}{J Bender}, {and}
  \bibinfo{person}{M Teschner}.} \bibinfo{year}{2019}\natexlab{}.
\newblock \showarticletitle{{Interlinked sph pressure solvers for strong
  fluid-rigid coupling}}.
\newblock \bibinfo{journal}{\emph{ACM ToG}} \bibinfo{volume}{38},
  \bibinfo{number}{1} (\bibinfo{year}{2019}), \bibinfo{pages}{5:1--5:13}.
\newblock


\bibitem[\protect\citeauthoryear{He, Liu, Li, Wang, and Wang}{He
  et~al\mbox{.}}{2012}]%
        {He2012}
\bibfield{author}{\bibinfo{person}{Xiaowei He}, \bibinfo{person}{Ning Liu},
  \bibinfo{person}{Sheng Li}, \bibinfo{person}{Hongan Wang}, {and}
  \bibinfo{person}{Guoping Wang}.} \bibinfo{year}{2012}\natexlab{}.
\newblock \showarticletitle{{Local Poisson SPH for Viscous Incompressible
  Fluids}}.
\newblock \bibinfo{journal}{\emph{CGF}}  \bibinfo{volume}{31}
  (\bibinfo{year}{2012}), \bibinfo{pages}{1948----1958}.
\newblock


\bibitem[\protect\citeauthoryear{Holl, Thuerey, and Koltun}{Holl
  et~al\mbox{.}}{2020}]%
        {Holl2020}
\bibfield{author}{\bibinfo{person}{Philipp Holl}, \bibinfo{person}{Nils
  Thuerey}, {and} \bibinfo{person}{Vladlen Koltun}.}
  \bibinfo{year}{2020}\natexlab{}.
\newblock \showarticletitle{Learning to Control PDEs with Differentiable
  Physics}. In \bibinfo{booktitle}{\emph{ICLR}}.
\newblock
\urldef\tempurl%
\url{https://openreview.net/forum?id=HyeSin4FPB}
\showURL{%
\tempurl}


\bibitem[\protect\citeauthoryear{Hu, Anderson, Li, Sun, Carr, Ragan-Kelley, and
  Durand}{Hu et~al\mbox{.}}{2020}]%
        {Hu2020DiffTaichi}
\bibfield{author}{\bibinfo{person}{Yuanming Hu}, \bibinfo{person}{Luke
  Anderson}, \bibinfo{person}{Tzu-Mao Li}, \bibinfo{person}{Qi Sun},
  \bibinfo{person}{Nathan Carr}, \bibinfo{person}{Jonathan Ragan-Kelley}, {and}
  \bibinfo{person}{Fredo Durand}.} \bibinfo{year}{2020}\natexlab{}.
\newblock \showarticletitle{DiffTaichi: Differentiable Programming for Physical
  Simulation}. In \bibinfo{booktitle}{\emph{ICLR}}.
\newblock
\urldef\tempurl%
\url{https://openreview.net/forum?id=B1eB5xSFvr}
\showURL{%
\tempurl}


\bibitem[\protect\citeauthoryear{Hu, Liu, Spielberg, Tenenbaum, Freeman, Wu,
  Rus, and Matusik}{Hu et~al\mbox{.}}{2019a}]%
        {Hu2018}
\bibfield{author}{\bibinfo{person}{Yuanming Hu}, \bibinfo{person}{Jiancheng
  Liu}, \bibinfo{person}{Andrew Spielberg}, \bibinfo{person}{Joshua~B
  Tenenbaum}, \bibinfo{person}{William~T Freeman}, \bibinfo{person}{Jiajun Wu},
  \bibinfo{person}{Daniela Rus}, {and} \bibinfo{person}{Wojciech Matusik}.}
  \bibinfo{year}{2019}\natexlab{a}.
\newblock \showarticletitle{ChainQueen: A real-time differentiable physical
  simulator for soft robotics}. In \bibinfo{booktitle}{\emph{ICRA}}.
  \bibinfo{pages}{6265--6271}.
\newblock


\bibitem[\protect\citeauthoryear{Hu, Zhang, Gao, and Jiang}{Hu
  et~al\mbox{.}}{2019b}]%
        {Hu2019}
\bibfield{author}{\bibinfo{person}{Yuanming Hu}, \bibinfo{person}{Xinxin
  Zhang}, \bibinfo{person}{Ming Gao}, {and} \bibinfo{person}{Chenfanfu Jiang}.}
  \bibinfo{year}{2019}\natexlab{b}.
\newblock \showarticletitle{{On hybrid lagrangian-eulerian simulation methods:
  practical notes and high-performance aspects}}. In
  \bibinfo{booktitle}{\emph{ACM SIGGRAPH 2019 Courses}}. \bibinfo{pages}{16}.
\newblock


\bibitem[\protect\citeauthoryear{Ihmsen, Akinci, Akinci, and Teschner}{Ihmsen
  et~al\mbox{.}}{2012}]%
        {ihmsen2012unified}
\bibfield{author}{\bibinfo{person}{Markus Ihmsen}, \bibinfo{person}{Nadir
  Akinci}, \bibinfo{person}{Gizem Akinci}, {and} \bibinfo{person}{Matthias
  Teschner}.} \bibinfo{year}{2012}\natexlab{}.
\newblock \showarticletitle{Unified spray, foam and air bubbles for
  particle-based fluids}.
\newblock \bibinfo{journal}{\emph{The Visual Computer}} \bibinfo{volume}{28},
  \bibinfo{number}{6-8} (\bibinfo{year}{2012}), \bibinfo{pages}{669--677}.
\newblock


\bibitem[\protect\citeauthoryear{Ihmsen, Cornelis, Solenthaler, Horvath, and
  Teschner}{Ihmsen et~al\mbox{.}}{2014}]%
        {Ihmsen2014}
\bibfield{author}{\bibinfo{person}{Markus Ihmsen}, \bibinfo{person}{Jens
  Cornelis}, \bibinfo{person}{Barbara Solenthaler},
  \bibinfo{person}{Christopher Horvath}, {and} \bibinfo{person}{Matthias
  Teschner}.} \bibinfo{year}{2014}\natexlab{}.
\newblock \showarticletitle{{Implicit incompressible SPH}}.
\newblock \bibinfo{journal}{\emph{IEEE TVCG}} \bibinfo{volume}{20},
  \bibinfo{number}{3} (\bibinfo{year}{2014}), \bibinfo{pages}{426--436}.
\newblock


\bibitem[\protect\citeauthoryear{Insafutdinov and Dosovitskiy}{Insafutdinov and
  Dosovitskiy}{2018}]%
        {insafutdinov2018pointclouds}
\bibfield{author}{\bibinfo{person}{Eldar Insafutdinov} {and}
  \bibinfo{person}{Alexey Dosovitskiy}.} \bibinfo{year}{2018}\natexlab{}.
\newblock \showarticletitle{Unsupervised Learning of Shape and Pose with
  Differentiable Point Clouds}. In \bibinfo{booktitle}{\emph{NeurIPS}}.
\newblock


\bibitem[\protect\citeauthoryear{Jamri\v{s}ka, Fi\v{s}er, Asente, Lu,
  Shechtman, and S\'{y}kora}{Jamri\v{s}ka et~al\mbox{.}}{2015}]%
        {jamrivska2015lazyfluids}
\bibfield{author}{\bibinfo{person}{Ond\v{r}ej Jamri\v{s}ka},
  \bibinfo{person}{Jakub Fi\v{s}er}, \bibinfo{person}{Paul Asente},
  \bibinfo{person}{Jingwan Lu}, \bibinfo{person}{Eli Shechtman}, {and}
  \bibinfo{person}{Daniel S\'{y}kora}.} \bibinfo{year}{2015}\natexlab{}.
\newblock \showarticletitle{{LazyFluids: appearance transfer for fluid
  animations}}.
\newblock \bibinfo{journal}{\emph{ACM Transactions on Graphics (TOG)}}
  \bibinfo{volume}{34}, \bibinfo{number}{4} (\bibinfo{year}{2015}),
  \bibinfo{pages}{92}.
\newblock


\bibitem[\protect\citeauthoryear{Jiang, Schroeder, Selle, Teran, and
  Stomakhin}{Jiang et~al\mbox{.}}{2015}]%
        {Jiang2015}
\bibfield{author}{\bibinfo{person}{Chenfanfu Jiang}, \bibinfo{person}{Craig
  Schroeder}, \bibinfo{person}{Andrew Selle}, \bibinfo{person}{Joseph Teran},
  {and} \bibinfo{person}{Alexey Stomakhin}.} \bibinfo{year}{2015}\natexlab{}.
\newblock \showarticletitle{{The affine particle-in-cell method}}.
\newblock \bibinfo{journal}{\emph{ACM ToG}} \bibinfo{volume}{34},
  \bibinfo{number}{4} (\bibinfo{date}{jul} \bibinfo{year}{2015}),
  \bibinfo{pages}{51:1--51:10}.
\newblock
\showISSN{07300301}


\bibitem[\protect\citeauthoryear{Jiang, Schroeder, Teran, Stomakhin, and
  Selle}{Jiang et~al\mbox{.}}{2016}]%
        {jiang2016material}
\bibfield{author}{\bibinfo{person}{Chenfanfu Jiang}, \bibinfo{person}{Craig
  Schroeder}, \bibinfo{person}{Joseph Teran}, \bibinfo{person}{Alexey
  Stomakhin}, {and} \bibinfo{person}{Andrew Selle}.}
  \bibinfo{year}{2016}\natexlab{}.
\newblock \showarticletitle{The material point method for simulating continuum
  materials}.
\newblock In \bibinfo{booktitle}{\emph{ACM SIGGRAPH 2016 Courses}}.
  \bibinfo{pages}{1--52}.
\newblock


\bibitem[\protect\citeauthoryear{Jing, Yang, Feng, Ye, Yu, and Song}{Jing
  et~al\mbox{.}}{2019}]%
        {jing1705neural}
\bibfield{author}{\bibinfo{person}{Yongcheng Jing}, \bibinfo{person}{Yezhou
  Yang}, \bibinfo{person}{Zunlei Feng}, \bibinfo{person}{Jingwen Ye},
  \bibinfo{person}{Yizhou Yu}, {and} \bibinfo{person}{Mingli Song}.}
  \bibinfo{year}{2019}\natexlab{}.
\newblock \showarticletitle{Neural style transfer: A review}.
\newblock \bibinfo{journal}{\emph{IEEE TVCG}} (\bibinfo{year}{2019}).
\newblock


\bibitem[\protect\citeauthoryear{Kato, Ushiku, and Harada}{Kato
  et~al\mbox{.}}{2018}]%
        {kato2018neural}
\bibfield{author}{\bibinfo{person}{Hiroharu Kato}, \bibinfo{person}{Yoshitaka
  Ushiku}, {and} \bibinfo{person}{Tatsuya Harada}.}
  \bibinfo{year}{2018}\natexlab{}.
\newblock \showarticletitle{{Neural 3d mesh renderer}}. In
  \bibinfo{booktitle}{\emph{Proceedings of the IEEE Conference on CVPR}}.
  \bibinfo{pages}{3907--3916}.
\newblock


\bibitem[\protect\citeauthoryear{Kim, Azevedo, Gross, and Solenthaler}{Kim
  et~al\mbox{.}}{2019a}]%
        {Kim2019}
\bibfield{author}{\bibinfo{person}{Byungsoo Kim}, \bibinfo{person}{Vinicius~C.
  Azevedo}, \bibinfo{person}{Markus Gross}, {and} \bibinfo{person}{Barbara
  Solenthaler}.} \bibinfo{year}{2019}\natexlab{a}.
\newblock \showarticletitle{{Transport-based neural style transfer for smoke
  simulations}}.
\newblock \bibinfo{journal}{\emph{ACM Transactions on Graphics}}
  \bibinfo{volume}{38}, \bibinfo{number}{6} (\bibinfo{date}{nov}
  \bibinfo{year}{2019}), \bibinfo{pages}{1--11}.
\newblock
\showISSN{07300301}
\urldef\tempurl%
\url{https://doi.org/10.1145/3355089.3356560}
\showDOI{\tempurl}


\bibitem[\protect\citeauthoryear{Kim, Azevedo, Thuerey, Kim, Gross, and
  Solenthaler}{Kim et~al\mbox{.}}{2019b}]%
        {Kim2019deep}
\bibfield{author}{\bibinfo{person}{Byungsoo Kim}, \bibinfo{person}{Vinicius~C.
  Azevedo}, \bibinfo{person}{Nils Thuerey}, \bibinfo{person}{Theodore Kim},
  \bibinfo{person}{Markus Gross}, {and} \bibinfo{person}{Barbara Solenthaler}.}
  \bibinfo{year}{2019}\natexlab{b}.
\newblock \showarticletitle{{Deep Fluids: A Generative Network for
  Parameterized Fluid Simulations}}.
\newblock \bibinfo{journal}{\emph{Computer Graphics Forum}}
  \bibinfo{volume}{38}, \bibinfo{number}{2} (\bibinfo{year}{2019}).
\newblock


\bibitem[\protect\citeauthoryear{Kim, Cha, Chang, Koo, and Ihm}{Kim
  et~al\mbox{.}}{2006}]%
        {Kim2007}
\bibfield{author}{\bibinfo{person}{Janghee Kim}, \bibinfo{person}{Deukhyun
  Cha}, \bibinfo{person}{Byungjoon Chang}, \bibinfo{person}{Bonki Koo}, {and}
  \bibinfo{person}{Insung Ihm}.} \bibinfo{year}{2006}\natexlab{}.
\newblock \showarticletitle{{Practical Animation of Turbulent Splashing
  Water}}. In \bibinfo{booktitle}{\emph{Proceedings SCA'07}}.
  \bibinfo{pages}{335--344}.
\newblock
\showISBNx{3905673347}


\bibitem[\protect\citeauthoryear{Kim, Tessendorf, and Thuerey}{Kim
  et~al\mbox{.}}{2013}]%
        {kim2013closest}
\bibfield{author}{\bibinfo{person}{Theodore Kim}, \bibinfo{person}{Jerry
  Tessendorf}, {and} \bibinfo{person}{Nils Thuerey}.}
  \bibinfo{year}{2013}\natexlab{}.
\newblock \showarticletitle{{Closest point turbulence for liquid surfaces}}.
\newblock \bibinfo{journal}{\emph{ACM Transactions on Graphics (TOG)}}
  \bibinfo{volume}{32}, \bibinfo{number}{2} (\bibinfo{year}{2013}),
  \bibinfo{pages}{15}.
\newblock


\bibitem[\protect\citeauthoryear{Kim, Th{\"{u}}rey, James, and Gross}{Kim
  et~al\mbox{.}}{2008}]%
        {kim2008wavelet}
\bibfield{author}{\bibinfo{person}{Theodore Kim}, \bibinfo{person}{Nils
  Th{\"{u}}rey}, \bibinfo{person}{Doug James}, {and} \bibinfo{person}{Markus
  Gross}.} \bibinfo{year}{2008}\natexlab{}.
\newblock \showarticletitle{{Wavelet turbulence for fluid simulation}}. In
  \bibinfo{booktitle}{\emph{ACM Transactions on Graphics (TOG)}},
  Vol.~\bibinfo{volume}{27}. ACM, \bibinfo{pages}{50}.
\newblock


\bibitem[\protect\citeauthoryear{Koschier and Bender}{Koschier and
  Bender}{2017}]%
        {Koschier2017}
\bibfield{author}{\bibinfo{person}{D Koschier} {and} \bibinfo{person}{J
  Bender}.} \bibinfo{year}{2017}\natexlab{}.
\newblock \showarticletitle{{Density maps for improved sph boundary handling}}.
  In \bibinfo{booktitle}{\emph{ACM SIGGRAPH/Eurographics Symposium on Computer
  Animation}}. \bibinfo{pages}{1--10}.
\newblock


\bibitem[\protect\citeauthoryear{Koschier, Bender, Solenthaler, and
  Teschner}{Koschier et~al\mbox{.}}{2019}]%
        {Koschier2019}
\bibfield{author}{\bibinfo{person}{Dan Koschier}, \bibinfo{person}{Jan Bender},
  \bibinfo{person}{Barbara Solenthaler}, {and} \bibinfo{person}{Matthias
  Teschner}.} \bibinfo{year}{2019}\natexlab{}.
\newblock \showarticletitle{{Smoothed Particle Hydrodynamics Techniques for the
  Physics Based Simulation of Fluids and Solids}}. In
  \bibinfo{booktitle}{\emph{Eurographics 2019 - Tutorials}}.
\newblock


\bibitem[\protect\citeauthoryear{Kwatra, Adalsteinsson, Kwatra, Carlson, and
  Lin}{Kwatra et~al\mbox{.}}{2006}]%
        {Kwatra2006}
\bibfield{author}{\bibinfo{person}{Vivek Kwatra}, \bibinfo{person}{David
  Adalsteinsson}, \bibinfo{person}{Nipun Kwatra}, \bibinfo{person}{Mark
  Carlson}, {and} \bibinfo{person}{Ming~C. Lin}.}
  \bibinfo{year}{2006}\natexlab{}.
\newblock \showarticletitle{{Texturing fluids}}. In
  \bibinfo{booktitle}{\emph{ACM SIGGRAPH '06 Sketches on}}.
  \bibinfo{pages}{63}.
\newblock


\bibitem[\protect\citeauthoryear{Kwatra, Essa, Bobick, and Kwatra}{Kwatra
  et~al\mbox{.}}{2005}]%
        {Kwatra2005}
\bibfield{author}{\bibinfo{person}{Vivek Kwatra}, \bibinfo{person}{Irfan Essa},
  \bibinfo{person}{Aaron Bobick}, {and} \bibinfo{person}{Nipun Kwatra}.}
  \bibinfo{year}{2005}\natexlab{}.
\newblock \showarticletitle{{Texture optimization for example-based
  synthesis}}. In \bibinfo{booktitle}{\emph{ACM SIGGRAPH '05}}.
  \bibinfo{pages}{795}.
\newblock


\bibitem[\protect\citeauthoryear{Ladick{\'{y}}, Jeong, Solenthaler, Pollefeys,
  and Gross}{Ladick{\'{y}} et~al\mbox{.}}{2015}]%
        {Ladicky2015}
\bibfield{author}{\bibinfo{person}{L'ubor Ladick{\'{y}}},
  \bibinfo{person}{SoHyeon Jeong}, \bibinfo{person}{Barbara Solenthaler},
  \bibinfo{person}{Marc Pollefeys}, {and} \bibinfo{person}{Markus Gross}.}
  \bibinfo{year}{2015}\natexlab{}.
\newblock \showarticletitle{{Data-driven fluid simulations using regression
  forests}}.
\newblock \bibinfo{journal}{\emph{ACM Transactions on Graphics}}
  \bibinfo{volume}{34}, \bibinfo{number}{6} (\bibinfo{date}{oct}
  \bibinfo{year}{2015}), \bibinfo{pages}{1--9}.
\newblock
\showISSN{07300301}
\urldef\tempurl%
\url{https://doi.org/10.1145/2816795.2818129}
\showDOI{\tempurl}


\bibitem[\protect\citeauthoryear{Liu and Jacobson}{Liu and Jacobson}{2019}]%
        {Liu2019}
\bibfield{author}{\bibinfo{person}{Hsueh-Ti~Derek Liu} {and}
  \bibinfo{person}{Alec Jacobson}.} \bibinfo{year}{2019}\natexlab{}.
\newblock \showarticletitle{Cubic Stylization}.
\newblock \bibinfo{journal}{\emph{ACM ToG}} (\bibinfo{year}{2019}).
\newblock


\bibitem[\protect\citeauthoryear{Liu, Tao, and Jacobson}{Liu
  et~al\mbox{.}}{2018}]%
        {liu2018paparazzi}
\bibfield{author}{\bibinfo{person}{Hsueh-Ti~Derek Liu},
  \bibinfo{person}{Michael Tao}, {and} \bibinfo{person}{Alec Jacobson}.}
  \bibinfo{year}{2018}\natexlab{}.
\newblock \showarticletitle{{Paparazzi: Surface Editing by way of Multi-View
  Image Processing}}.
\newblock \bibinfo{journal}{\emph{ACM Transactions on Graphics}}
  (\bibinfo{year}{2018}).
\newblock


\bibitem[\protect\citeauthoryear{Loper and Black}{Loper and Black}{2014}]%
        {Loper2014}
\bibfield{author}{\bibinfo{person}{Matthew~M. Loper} {and}
  \bibinfo{person}{Michael~J. Black}.} \bibinfo{year}{2014}\natexlab{}.
\newblock \showarticletitle{{OpenDR: An Approximate Differentiable Renderer}}.
\newblock \bibinfo{pages}{154--169}.
\newblock
\urldef\tempurl%
\url{https://doi.org/10.1007/978-3-319-10584-0_11}
\showDOI{\tempurl}


\bibitem[\protect\citeauthoryear{Losasso, Talton, Kwatra, and Fedkiw}{Losasso
  et~al\mbox{.}}{2008}]%
        {Losasso2008}
\bibfield{author}{\bibinfo{person}{F. Losasso}, \bibinfo{person}{J.O. Talton},
  \bibinfo{person}{N. Kwatra}, {and} \bibinfo{person}{R. Fedkiw}.}
  \bibinfo{year}{2008}\natexlab{}.
\newblock \showarticletitle{{Two-Way Coupled SPH and Particle Level Set Fluid
  Simulation}}.
\newblock \bibinfo{journal}{\emph{IEEE TVCG}} \bibinfo{volume}{14},
  \bibinfo{number}{4} (\bibinfo{date}{jul} \bibinfo{year}{2008}),
  \bibinfo{pages}{797--804}.
\newblock
\showISSN{1077-2626}


\bibitem[\protect\citeauthoryear{Macklin and Mueller}{Macklin and
  Mueller}{2013}]%
        {Macklin2013}
\bibfield{author}{\bibinfo{person}{Miles Macklin} {and}
  \bibinfo{person}{Matthias Mueller}.} \bibinfo{year}{2013}\natexlab{}.
\newblock \showarticletitle{{Position Based Fluids}}.
\newblock \bibinfo{journal}{\emph{ACM Transactions on Graphics}}
  \bibinfo{volume}{32}, \bibinfo{number}{4} (\bibinfo{year}{2013}),
  \bibinfo{pages}{104:1--104:12}.
\newblock


\bibitem[\protect\citeauthoryear{McNamara, Treuille, Popovi{\'{c}}, and
  Stam}{McNamara et~al\mbox{.}}{2004}]%
        {McNamara2004}
\bibfield{author}{\bibinfo{person}{Antoine McNamara}, \bibinfo{person}{Adrien
  Treuille}, \bibinfo{person}{Zoran Popovi{\'{c}}}, {and} \bibinfo{person}{Jos
  Stam}.} \bibinfo{year}{2004}\natexlab{}.
\newblock \showarticletitle{{Fluid control using the adjoint method}}. In
  \bibinfo{booktitle}{\emph{ACM SIGGRAPH '04}}. \bibinfo{pages}{449}.
\newblock


\bibitem[\protect\citeauthoryear{Monaghan}{Monaghan}{2005}]%
        {Monaghan2005}
\bibfield{author}{\bibinfo{person}{J~J Monaghan}.}
  \bibinfo{year}{2005}\natexlab{}.
\newblock \showarticletitle{{Smoothed Particle Hydrodynamics}}.
\newblock \bibinfo{journal}{\emph{Reports on Progress in Physics}}
  \bibinfo{volume}{68}, \bibinfo{number}{8} (\bibinfo{year}{2005}),
  \bibinfo{pages}{1703--1759}.
\newblock


\bibitem[\protect\citeauthoryear{M{\"{u}}ller, Charypar, and
  Gross}{M{\"{u}}ller et~al\mbox{.}}{2003}]%
        {Muller2003}
\bibfield{author}{\bibinfo{person}{Matthias M{\"{u}}ller},
  \bibinfo{person}{David Charypar}, {and} \bibinfo{person}{Markus Gross}.}
  \bibinfo{year}{2003}\natexlab{}.
\newblock \showarticletitle{{Particle-Based Fluid Simulation for Interactive
  Applications}}. In \bibinfo{booktitle}{\emph{Symposium on Computer
  Animation}}.
\newblock


\bibitem[\protect\citeauthoryear{Narain, Kwatra, Lee, Kim, Carlson, and
  Lin}{Narain et~al\mbox{.}}{2007}]%
        {Narain2007}
\bibfield{author}{\bibinfo{person}{Rahul Narain}, \bibinfo{person}{Vivek
  Kwatra}, \bibinfo{person}{Huai-Ping Lee}, \bibinfo{person}{Theodore Kim},
  \bibinfo{person}{Mark Carlson}, {and} \bibinfo{person}{Ming~C Lin}.}
  \bibinfo{year}{2007}\natexlab{}.
\newblock \showarticletitle{{Feature-guided Dynamic Texture Synthesis on
  Continuous Flows}}. In \bibinfo{booktitle}{\emph{Proceedings of the 18th
  EGSR}}. \bibinfo{pages}{361--370}.
\newblock


\bibitem[\protect\citeauthoryear{Narain, Sewall, Carlson, and Lin}{Narain
  et~al\mbox{.}}{2008}]%
        {Narain2008}
\bibfield{author}{\bibinfo{person}{Rahul Narain}, \bibinfo{person}{Jason
  Sewall}, \bibinfo{person}{Mark Carlson}, {and} \bibinfo{person}{Ming~C.
  Lin}.} \bibinfo{year}{2008}\natexlab{}.
\newblock \showarticletitle{{Fast animation of turbulence using energy
  transport and procedural synthesis}}.
\newblock \bibinfo{journal}{\emph{ACM Transactions on Graphics}}
  \bibinfo{volume}{27}, \bibinfo{number}{5} (\bibinfo{date}{dec}
  \bibinfo{year}{2008}), \bibinfo{pages}{1}.
\newblock
\showISSN{07300301}
\urldef\tempurl%
\url{https://doi.org/10.1145/1409060.1409119}
\showDOI{\tempurl}


\bibitem[\protect\citeauthoryear{Nielsen and Bridson}{Nielsen and
  Bridson}{2011}]%
        {Nielsen2011}
\bibfield{author}{\bibinfo{person}{Michael~B. Nielsen} {and}
  \bibinfo{person}{Robert Bridson}.} \bibinfo{year}{2011}\natexlab{}.
\newblock \showarticletitle{{Guide shapes for high resolution naturalistic
  liquid simulation}}. In \bibinfo{booktitle}{\emph{ACM SIGGRAPH '11}}.
  \bibinfo{pages}{1}.
\newblock


\bibitem[\protect\citeauthoryear{Pan, Huang, Tong, Zheng, and Bao}{Pan
  et~al\mbox{.}}{2013}]%
        {Pan2013}
\bibfield{author}{\bibinfo{person}{Zherong Pan}, \bibinfo{person}{Jin Huang},
  \bibinfo{person}{Yiying Tong}, \bibinfo{person}{Changxi Zheng}, {and}
  \bibinfo{person}{Hujun Bao}.} \bibinfo{year}{2013}\natexlab{}.
\newblock \showarticletitle{{Interactive localized liquid motion editing}}.
\newblock \bibinfo{journal}{\emph{ACM ToG}} \bibinfo{volume}{32},
  \bibinfo{number}{6} (\bibinfo{date}{nov} \bibinfo{year}{2013}),
  \bibinfo{pages}{1--10}.
\newblock
\showISSN{07300301}


\bibitem[\protect\citeauthoryear{Pan and Manocha}{Pan and Manocha}{2017}]%
        {Pan2016}
\bibfield{author}{\bibinfo{person}{Zherong Pan} {and} \bibinfo{person}{Dinesh
  Manocha}.} \bibinfo{year}{2017}\natexlab{}.
\newblock \showarticletitle{Efficient Solver for Spacetime Control of Smoke}.
\newblock \bibinfo{journal}{\emph{ACM Trans. Graph.}} \bibinfo{volume}{36},
  \bibinfo{number}{4}, Article \bibinfo{articleno}{Article 68a}
  (\bibinfo{date}{July} \bibinfo{year}{2017}), \bibinfo{numpages}{13}~pages.
\newblock
\showISSN{0730-0301}


\bibitem[\protect\citeauthoryear{Peer, Ihmsen, Cornelis, and Teschner}{Peer
  et~al\mbox{.}}{2015}]%
        {Peer2015}
\bibfield{author}{\bibinfo{person}{A. Peer}, \bibinfo{person}{M. Ihmsen},
  \bibinfo{person}{J. Cornelis}, {and} \bibinfo{person}{M. Teschner}.}
  \bibinfo{year}{2015}\natexlab{}.
\newblock \showarticletitle{{An Implicit Viscosity Formulation for SPH
  Fluids}}.
\newblock \bibinfo{journal}{\emph{ACM Transactions on Graphics}}
  \bibinfo{volume}{34}, \bibinfo{number}{4} (\bibinfo{year}{2015}),
  \bibinfo{pages}{1--10}.
\newblock


\bibitem[\protect\citeauthoryear{Pfaff, Thuerey, Selle, and Gross}{Pfaff
  et~al\mbox{.}}{2009}]%
        {Pfaff2009}
\bibfield{author}{\bibinfo{person}{Tobias Pfaff}, \bibinfo{person}{Nils
  Thuerey}, \bibinfo{person}{Andrew Selle}, {and} \bibinfo{person}{Markus
  Gross}.} \bibinfo{year}{2009}\natexlab{}.
\newblock \showarticletitle{{Synthetic turbulence using artificial boundary
  layers}}.
\newblock \bibinfo{journal}{\emph{ACM Transactions on Graphics}}
  \bibinfo{volume}{28}, \bibinfo{number}{5} (\bibinfo{date}{dec}
  \bibinfo{year}{2009}), \bibinfo{pages}{1}.
\newblock
\showISSN{07300301}


\bibitem[\protect\citeauthoryear{Raveendran, Thuerey, Wojtan, and
  Turk}{Raveendran et~al\mbox{.}}{2012}]%
        {Raveendran2012}
\bibfield{author}{\bibinfo{person}{Karthik Raveendran}, \bibinfo{person}{Nils
  Thuerey}, \bibinfo{person}{Chris Wojtan}, {and} \bibinfo{person}{Greg Turk}.}
  \bibinfo{year}{2012}\natexlab{}.
\newblock \showarticletitle{{Controlling Liquids Using Meshes}}. In
  \bibinfo{booktitle}{\emph{Proceedings of the SCA}}.
  \bibinfo{pages}{255--264}.
\newblock


\bibitem[\protect\citeauthoryear{Reinhardt, Krake, Eberhardt, and
  Weiskopf}{Reinhardt et~al\mbox{.}}{2019}]%
        {Reinhardt2019}
\bibfield{author}{\bibinfo{person}{Stefan Reinhardt}, \bibinfo{person}{Tim
  Krake}, \bibinfo{person}{Bernhard Eberhardt}, {and} \bibinfo{person}{Daniel
  Weiskopf}.} \bibinfo{year}{2019}\natexlab{}.
\newblock \showarticletitle{Consistent Shepard Interpolation for SPH-Based
  Fluid Animation}.
\newblock \bibinfo{journal}{\emph{ACM ToG}}  \bibinfo{volume}{38}
  (\bibinfo{year}{2019}).
\newblock


\bibitem[\protect\citeauthoryear{Ren, Li, Yan, Lin, Bonet, and Hu}{Ren
  et~al\mbox{.}}{2014}]%
        {Ren2014}
\bibfield{author}{\bibinfo{person}{Bo Ren}, \bibinfo{person}{Chenfeng Li},
  \bibinfo{person}{Xiao Yan}, \bibinfo{person}{Ming~C Lin},
  \bibinfo{person}{Javier Bonet}, {and} \bibinfo{person}{Shi-Min Hu}.}
  \bibinfo{year}{2014}\natexlab{}.
\newblock \showarticletitle{{Multiple-Fluid SPH Simulation Using a Mixture
  Model}}.
\newblock \bibinfo{journal}{\emph{ACM ToG}} \bibinfo{volume}{33},
  \bibinfo{number}{5} (\bibinfo{year}{2014}), \bibinfo{pages}{1--11}.
\newblock


\bibitem[\protect\citeauthoryear{Sato, Dobashi, Kim, and Nishita}{Sato
  et~al\mbox{.}}{2018}]%
        {sato2018example}
\bibfield{author}{\bibinfo{person}{Syuhei Sato}, \bibinfo{person}{Yoshinori
  Dobashi}, \bibinfo{person}{Theodore Kim}, {and} \bibinfo{person}{Tomoyuki
  Nishita}.} \bibinfo{year}{2018}\natexlab{}.
\newblock \showarticletitle{{Example-based turbulence style transfer}}.
\newblock \bibinfo{journal}{\emph{ACM Trans. Graph.}} \bibinfo{volume}{37},
  \bibinfo{number}{4} (\bibinfo{year}{2018}), \bibinfo{pages}{84}.
\newblock


\bibitem[\protect\citeauthoryear{Schechter and Bridson}{Schechter and
  Bridson}{2008}]%
        {Schechter2008}
\bibfield{author}{\bibinfo{person}{Hagit Schechter} {and}
  \bibinfo{person}{Robert Bridson}.} \bibinfo{year}{2008}\natexlab{}.
\newblock \showarticletitle{{Evolving Sub-Grid Turbulence for Smoke
  Animation.}} \bibinfo{pages}{1--7}.
\newblock
\urldef\tempurl%
\url{https://doi.org/10.2312/SCA/SCA08/001-007}
\showDOI{\tempurl}


\bibitem[\protect\citeauthoryear{Schenck and Fox}{Schenck and Fox}{2018}]%
        {Schenck2018}
\bibfield{author}{\bibinfo{person}{Connor Schenck} {and}
  \bibinfo{person}{Dieter Fox}.} \bibinfo{year}{2018}\natexlab{}.
\newblock \showarticletitle{SPNets: Differentiable Fluid Dynamics for Deep
  Neural Networks}. In \bibinfo{booktitle}{\emph{Conference on Robot
  Learning}}. \bibinfo{pages}{317--335}.
\newblock


\bibitem[\protect\citeauthoryear{Servin, Bodin, and Lacoursiere}{Servin
  et~al\mbox{.}}{2012}]%
        {Bodin2012}
\bibfield{author}{\bibinfo{person}{M. Servin}, \bibinfo{person}{K. Bodin},
  {and} \bibinfo{person}{C. Lacoursiere}.} \bibinfo{year}{2012}\natexlab{}.
\newblock \showarticletitle{Constraint Fluids}.
\newblock \bibinfo{journal}{\emph{IEEE TVCG}} \bibinfo{volume}{18},
  \bibinfo{number}{03} (\bibinfo{date}{mar} \bibinfo{year}{2012}),
  \bibinfo{pages}{516--526}.
\newblock
\showISSN{1941-0506}
\urldef\tempurl%
\url{https://doi.org/10.1109/TVCG.2011.29}
\showDOI{\tempurl}


\bibitem[\protect\citeauthoryear{Solenthaler and Pajarola}{Solenthaler and
  Pajarola}{2009}]%
        {Solenthaler2009}
\bibfield{author}{\bibinfo{person}{Barbara Solenthaler} {and}
  \bibinfo{person}{Renato Pajarola}.} \bibinfo{year}{2009}\natexlab{}.
\newblock \showarticletitle{{Predictive-corrective incompressible SPH}}.
\newblock \bibinfo{journal}{\emph{ACM Trans. Graph.}} \bibinfo{volume}{28},
  \bibinfo{number}{3} (\bibinfo{year}{2009}), \bibinfo{pages}{40:1--40:6}.
\newblock


\bibitem[\protect\citeauthoryear{Stomakhin, Schroeder, Chai, Teran, and
  Selle}{Stomakhin et~al\mbox{.}}{2013}]%
        {stomakhin2013}
\bibfield{author}{\bibinfo{person}{Alexey Stomakhin}, \bibinfo{person}{Craig
  Schroeder}, \bibinfo{person}{Lawrence Chai}, \bibinfo{person}{Joseph Teran},
  {and} \bibinfo{person}{Andrew Selle}.} \bibinfo{year}{2013}\natexlab{}.
\newblock \showarticletitle{A Material Point Method for Snow Simulation}.
\newblock \bibinfo{journal}{\emph{ACM ToG}} \bibinfo{volume}{32},
  \bibinfo{number}{4} (\bibinfo{year}{2013}).
\newblock
\showISSN{0730-0301}


\bibitem[\protect\citeauthoryear{Talmi, Mechrez, and Zelnik-Manor}{Talmi
  et~al\mbox{.}}{2017}]%
        {Talmi2016}
\bibfield{author}{\bibinfo{person}{Itamar Talmi}, \bibinfo{person}{Roey
  Mechrez}, {and} \bibinfo{person}{Lihi Zelnik-Manor}.}
  \bibinfo{year}{2017}\natexlab{}.
\newblock \showarticletitle{Template matching with deformable diversity
  similarity}. In \bibinfo{booktitle}{\emph{Proceedings of the IEEE CVPR}}.
  \bibinfo{pages}{175--183}.
\newblock


\bibitem[\protect\citeauthoryear{Thuerey}{Thuerey}{2016}]%
        {Thuerey2016}
\bibfield{author}{\bibinfo{person}{Nils Thuerey}.}
  \bibinfo{year}{2016}\natexlab{}.
\newblock \showarticletitle{{Interpolations of Smoke and Liquid Simulations}}.
\newblock \bibinfo{journal}{\emph{ACM Transactions on Graphics}}
  \bibinfo{volume}{36}, \bibinfo{number}{1} (\bibinfo{date}{sep}
  \bibinfo{year}{2016}), \bibinfo{pages}{1--16}.
\newblock
\showISSN{07300301}
\urldef\tempurl%
\url{https://doi.org/10.1145/2956233}
\showDOI{\tempurl}


\bibitem[\protect\citeauthoryear{Thuerey and Pfaff}{Thuerey and Pfaff}{2018}]%
        {mantaflow}
\bibfield{author}{\bibinfo{person}{Nils Thuerey} {and} \bibinfo{person}{Tobias
  Pfaff}.} \bibinfo{year}{2018}\natexlab{}.
\newblock \bibinfo{title}{{MantaFlow}}.
\newblock
\newblock
\newblock
\shownote{\emph{http://mantaflow.com}.}


\bibitem[\protect\citeauthoryear{Tompson, Schlachter, Sprechmann, and
  Perlin}{Tompson et~al\mbox{.}}{2017}]%
        {Tompson2016}
\bibfield{author}{\bibinfo{person}{Jonathan Tompson},
  \bibinfo{person}{Kristofer Schlachter}, \bibinfo{person}{Pablo Sprechmann},
  {and} \bibinfo{person}{Ken Perlin}.} \bibinfo{year}{2017}\natexlab{}.
\newblock \showarticletitle{Accelerating eulerian fluid simulation with
  convolutional networks}. In \bibinfo{booktitle}{\emph{Proceedings of the 34th
  ICML-Volume 70}}. JMLR. org, \bibinfo{pages}{3424--3433}.
\newblock


\bibitem[\protect\citeauthoryear{Treuille, McNamara, Popovi{\'{c}}, and
  Stam}{Treuille et~al\mbox{.}}{2003}]%
        {Treuille2003}
\bibfield{author}{\bibinfo{person}{Adrien Treuille}, \bibinfo{person}{Antoine
  McNamara}, \bibinfo{person}{Zoran Popovi{\'{c}}}, {and} \bibinfo{person}{Jos
  Stam}.} \bibinfo{year}{2003}\natexlab{}.
\newblock \showarticletitle{{Keyframe control of smoke simulations}}.
\newblock \bibinfo{journal}{\emph{ACM Transactions on Graphics}}
  \bibinfo{volume}{22}, \bibinfo{number}{3} (\bibinfo{date}{jul}
  \bibinfo{year}{2003}), \bibinfo{pages}{716}.
\newblock
\showISSN{07300301}


\bibitem[\protect\citeauthoryear{Um, Baek, and Han}{Um et~al\mbox{.}}{2014}]%
        {Um2014}
\bibfield{author}{\bibinfo{person}{Kiwon Um}, \bibinfo{person}{Seungho Baek},
  {and} \bibinfo{person}{JungHyun Han}.} \bibinfo{year}{2014}\natexlab{}.
\newblock \showarticletitle{{Advanced Hybrid Particle-Grid Method with Sub-Grid
  Particle Correction}}.
\newblock \bibinfo{journal}{\emph{CGF}} \bibinfo{volume}{33},
  \bibinfo{number}{7} (\bibinfo{date}{oct} \bibinfo{year}{2014}),
  \bibinfo{pages}{209--218}.
\newblock
\showISSN{01677055}


\bibitem[\protect\citeauthoryear{Wiewel, Becher, and Thuerey}{Wiewel
  et~al\mbox{.}}{2019}]%
        {Wiewel2018}
\bibfield{author}{\bibinfo{person}{Steffen Wiewel}, \bibinfo{person}{Moritz
  Becher}, {and} \bibinfo{person}{Nils Thuerey}.}
  \bibinfo{year}{2019}\natexlab{}.
\newblock \showarticletitle{Latent space physics: Towards learning the temporal
  evolution of fluid flow}. In \bibinfo{booktitle}{\emph{CGF}},
  Vol.~\bibinfo{volume}{38}. \bibinfo{pages}{71--82}.
\newblock


\bibitem[\protect\citeauthoryear{Xie, Franz, Chu, and Thuerey}{Xie
  et~al\mbox{.}}{2018}]%
        {xie2018tempogan}
\bibfield{author}{\bibinfo{person}{You Xie}, \bibinfo{person}{Erik Franz},
  \bibinfo{person}{Mengyu Chu}, {and} \bibinfo{person}{Nils Thuerey}.}
  \bibinfo{year}{2018}\natexlab{}.
\newblock \showarticletitle{tempogan: A temporally coherent, volumetric gan for
  super-resolution fluid flow}.
\newblock \bibinfo{journal}{\emph{ACM ToG}} \bibinfo{volume}{37},
  \bibinfo{number}{4} (\bibinfo{year}{2018}).
\newblock


\bibitem[\protect\citeauthoryear{Yan, Yang, Yumer, Guo, and Lee}{Yan
  et~al\mbox{.}}{2016}]%
        {Yan2016}
\bibfield{author}{\bibinfo{person}{Xinchen Yan}, \bibinfo{person}{Jimei Yang},
  \bibinfo{person}{Ersin Yumer}, \bibinfo{person}{Yijie Guo}, {and}
  \bibinfo{person}{Honglak Lee}.} \bibinfo{year}{2016}\natexlab{}.
\newblock \showarticletitle{Perspective transformer nets: Learning single-view
  3d object reconstruction without 3d supervision}. In
  \bibinfo{booktitle}{\emph{Advances in NIPS}}. \bibinfo{pages}{1696--1704}.
\newblock


\bibitem[\protect\citeauthoryear{Yang, Yang, and Xiao}{Yang
  et~al\mbox{.}}{2016}]%
        {Yang2016}
\bibfield{author}{\bibinfo{person}{Cheng Yang}, \bibinfo{person}{Xubo Yang},
  {and} \bibinfo{person}{Xiangyun Xiao}.} \bibinfo{year}{2016}\natexlab{}.
\newblock \showarticletitle{{Data-driven projection method in fluid
  simulation}}.
\newblock \bibinfo{journal}{\emph{CAVW}} \bibinfo{volume}{27},
  \bibinfo{number}{3-4} (\bibinfo{date}{may} \bibinfo{year}{2016}),
  \bibinfo{pages}{415--424}.
\newblock
\showISSN{15464261}


\bibitem[\protect\citeauthoryear{Yifan, Serena, Wu, {\"{O}}ztireli, and
  Sorkine-Hornung}{Yifan et~al\mbox{.}}{2019}]%
        {Yifan2019}
\bibfield{author}{\bibinfo{person}{Wang Yifan}, \bibinfo{person}{Felice
  Serena}, \bibinfo{person}{Shihao Wu}, \bibinfo{person}{Cengiz
  {\"{O}}ztireli}, {and} \bibinfo{person}{Olga Sorkine-Hornung}.}
  \bibinfo{year}{2019}\natexlab{}.
\newblock \showarticletitle{{Differentiable Surface Splatting for Point-based
  Geometry Processing}}.
\newblock  (\bibinfo{date}{jun} \bibinfo{year}{2019}).
\newblock
\urldef\tempurl%
\url{https://doi.org/10.1145/3355089.3356513}
\showDOI{\tempurl}
\showeprint[arxiv]{1906.04173}


\bibitem[\protect\citeauthoryear{Yu, Neyret, Bruneton, and Holzschuch}{Yu
  et~al\mbox{.}}{2011}]%
        {Yu2011}
\bibfield{author}{\bibinfo{person}{Q. Yu}, \bibinfo{person}{F. Neyret},
  \bibinfo{person}{E. Bruneton}, {and} \bibinfo{person}{N. Holzschuch}.}
  \bibinfo{year}{2011}\natexlab{}.
\newblock \showarticletitle{{Lagrangian Texture Advection: Preserving both
  Spectrum and Velocity Field}}.
\newblock \bibinfo{journal}{\emph{IEEE TVCG}} \bibinfo{volume}{17},
  \bibinfo{number}{11} (\bibinfo{date}{nov} \bibinfo{year}{2011}).
\newblock
\showISSN{1077-2626}


\bibitem[\protect\citeauthoryear{Zhu and Bridson}{Zhu and Bridson}{2005}]%
        {Zhu2005}
\bibfield{author}{\bibinfo{person}{Yongning Zhu} {and} \bibinfo{person}{Robert
  Bridson}.} \bibinfo{year}{2005}\natexlab{}.
\newblock \showarticletitle{{Animating sand as a fluid}}.
\newblock \bibinfo{journal}{\emph{ACM Transactions on Graphics}}
  \bibinfo{volume}{24}, \bibinfo{number}{3} (\bibinfo{date}{jul}
  \bibinfo{year}{2005}), \bibinfo{pages}{965}.
\newblock
\showISSN{07300301}
\urldef\tempurl%
\url{https://doi.org/10.1145/1073204.1073298}
\showDOI{\tempurl}


\end{thebibliography}
\end{document}